\documentclass[a4paper,twocolumn,10pt,accepted=2022-07-11]{quantumarticle}
\pdfoutput=1
\usepackage[utf8]{inputenc}
\usepackage[english]{babel}
\usepackage[T1]{fontenc}
\usepackage{amsmath}
\usepackage{hyperref}

\usepackage{amsmath,amsthm,amssymb}
\usepackage{graphicx}
\usepackage{braket}
\usepackage{comment}
\usepackage{float}
\makeatletter
\let\newfloat\newfloat@ltx
\makeatother
\usepackage{algorithm}
\usepackage[noend]{algpseudocode}

\usepackage{graphicx}% Include figure files
\usepackage{enumitem}
\usepackage{bm}% bold math
\usepackage[numbers,sort&compress]{natbib}

\usepackage{mathtools}

\usepackage[mathlines]{lineno}% Enablenumbering of text and display math
\newtheorem{Theorem}{Theorem}

\makeatletter

\begin{document}

%\preprint{APS/123-QED}

\title{TFermion: A non-Clifford gate cost assessment library of quantum phase estimation algorithms for quantum chemistry}% Force line breaks with \\

\author{Pablo A. M. Casares}
 \email{pabloamo@ucm.es}
 \orcid{0000-0001-5500-9115}
 \affiliation{Departamento de F\'isica Te\'orica, Universidad Complutense de Madrid.}%Lines break automatically or can be forced with \\
\author{Roberto Campos}%
 \email{robecamp@ucm.es}
 \orcid{0000-0002-2527-4177}
\affiliation{Departamento de F\'isica Te\'orica, Universidad Complutense de Madrid.}%\\
\affiliation{Quasar Science Resources, SL.}%
\author{Miguel A. Martin-Delgado}%
\orcid{0000-0003-2746-5062}
 \email{mardel@ucm.es}
\affiliation{Departamento de F\'isica Te\'orica, Universidad Complutense de Madrid.}
\affiliation{CCS-Center for Computational Simulation, Universidad Politécnica de Madrid.}




\begin{abstract}
Quantum Phase Estimation is one of the most useful quantum computing algorithms for quantum chemistry and as such, significant effort has been devoted to designing efficient implementations. In this article, we introduce TFermion, a library designed to estimate the T-gate cost of such algorithms, for an arbitrary molecule. As examples of usage, we estimate the T-gate cost of a few simple molecules and compare the same Taylorization algorithms using Gaussian and plane-wave basis.

\end{abstract}

\tableofcontents

%Use show keys class option if keyword display desired
\maketitle

%\tableofcontents

\section{\label{sec:Introduction}Introduction}

%% Add QFold somewhere

%% Article -> technique

%% 

Among the different applications found for quantum computing, the original aim of using quantum computers to simulate quantum systems and dynamics \cite{feynman2018simulating} still stands out as the most promising one. The reason is twofold: first, a quantum computer can encode the state of the system without needing approximations; and second, since the evolution of (closed) quantum systems is unitary, simulating it is rather natural.

Specifically, quantum computing might be particularly useful to prepare ground states of electronic Hamiltonians and find out their energies. Consequently, they can be employed in a multitude of chemical and material science problems where the ground state energy plays a key role. This includes for instance computing chemical reaction rates \cite{reiher2017elucidating,von2020quantum}, and analyzing battery properties \cite{kim2022fault,delgado2022how} or biological enzymes \cite{goings2022reliably}.%, problems where small errors in the estimated energy will have an important effect in the variable of interest. 

There exist classical computing techniques able to tackle these problems, most notably Density Functional Theory (DFT) \cite{kohanoff2006electronic}. However, they often rely on approximations, for instance, the Kohn-Sham exchange-correlation parametrized functional, which may struggle to achieve the high accuracy required in some of the problems above. For example, chemical reaction rates depend exponentially on differences in energy. In contrast, the well-known technique Quantum Phase Estimation (QPE) in principle allows achieving the high precision required by these applications. To understand how it works, remember that the Schrodinger equation dictates how a quantum system evolves according to its Hamiltonian,
\begin{equation}
    \hat{H}\ket{\psi} = i\hbar \frac{d}{dt} \ket{\psi}.
\end{equation}
If we assume for simplicity that such Hamiltonian is time independent, we can write
\begin{equation}
\begin{split}
    \psi(x,0) &= \sum_n a_n  \psi_{E_n}(x)\Rightarrow\\
    \psi(x,t) &= \sum_n a_n  e^{-iE_n t/\hbar} \psi_{E_n}(x),
\end{split}
\end{equation}
for $\psi_{E_n}(x)$ an eigenstate, and $E_n$ the corresponding eigenvalue. We are interested in the ground state energy $E_0$. Note how the eigenvalues became a phase. To recover it, we can use an inverse Quantum Fourier Transform that will encode such phase in the computational basis, from where it can be readily read out.

\begin{figure}[t]
    %\centering
    \includegraphics[width=\textwidth/2]{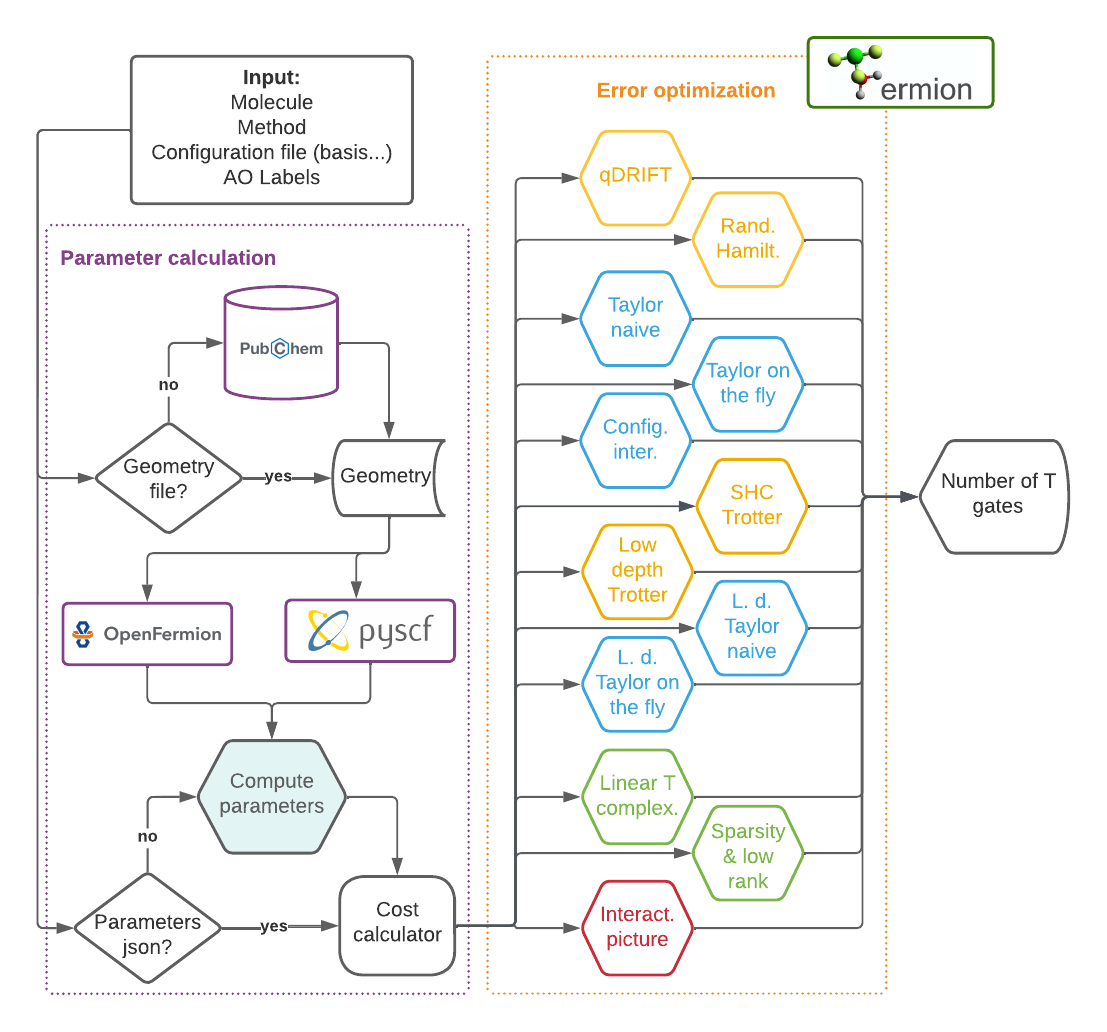}
    \caption{\label{fig:Architecture}Flowchart of the architecture of our library, divided into two parts: the first one centered on the computation of the parameters needed for the cost estimate; and a second one on using such parameters to compute the number of T-gates. Methods are colored according to the Hamiltonian simulation technique in figure \ref{fig:listmethods}.}
\end{figure}

To implement such an algorithm we need to specify how to implement the quantum Fourier transform, and also the Hamiltonian simulation $e^{-iHt}$. The former is rather straightforward and can be found in Ref. \cite{Nielsen-Chuang} for example, but the latter is more involved. Furthermore, to obtain a binary description with $b$ bits of the eigenvalue and probability of failure $p_f$, quantum phase estimation will need to implement $(e^{-iHt})^{2^j}$ for $j$ in the range $1,\ldots,b+\left\lceil \log_2\left(\frac{1}{2}+\frac{1}{2p_f}\right) \right\rceil$ \cite{cleve1998quantum}. In other words, it will require the implementation of several time segments that scales with the inverse precision, $O(1/\epsilon_{QPE})$. For this reason, it is important to be able to implement Hamiltonian simulation efficiently. 

Such a Hamiltonian might be accessed by the quantum computer in different ways. For electronic Hamiltonians, the most convenient one often is in the form of Linear Combination of Unitaries (LCU). In such framework, we decompose $H = \sum_{j} a_j H_j$, for $a_j$ some real positive coefficients, and $H_j$ the unitaries, often Pauli string-like operators. 

Given such access, there are also various methods to simulate the Hamiltonian evolution. The first way discovered was the Trotter method \cite{lloyd1996universal, Aspuru}, and soon others such as Taylor series (or Taylorization) \cite{berry2015simulating}, Qubitization \cite{low2017optimal,low2018hamiltonian} and Interaction picture simulation (or Dyson series) \cite{low2019hamiltonian,kieferova2019simulating} followed. These Hamiltonian simulation techniques, reviewed in section \ref{sec:Techniques}, are the backbone of the quantum phase estimation algorithm. Their objective is to lower as much as possible the computational cost of QPE, so large quantum systems can be simulated to high precision in reasonable amounts of time, once fault tolerant quantum computers become available.
The library we present in this article, TFermion, is an attempt to standardize and automatise the computation of the cost of several quantum phase algorithms in the literature.

However, to use quantum phase estimation, we need to prepare states with a large overlap with the ground state. This will translate into a high probability of measuring the actual ground state energy, and upon success will also project the system into the ground state. Unfortunately, it is known that preparing a representation of the ground state of a 2-body quantum Hamiltonian is Quantum Merlin Arthur (QMA) complete \cite{kempe2006complexity}, that is, a quantum computer can efficiently verify the solution, but not necessarily efficiently compute it. In other words, finding the ground state of a 2-body Hamiltonian is not in the Bounded Quantum Polynomial-time (BQP) complexity class, the class of problems a quantum computer can solve in polynomial time. Nevertheless, this does not imply either that we cannot propose algorithms to solve it as efficiently as possible \cite{ge2019faster,lin2020near}.
\begin{figure*}[htp]
    \centering
    \includegraphics[width=\textwidth]{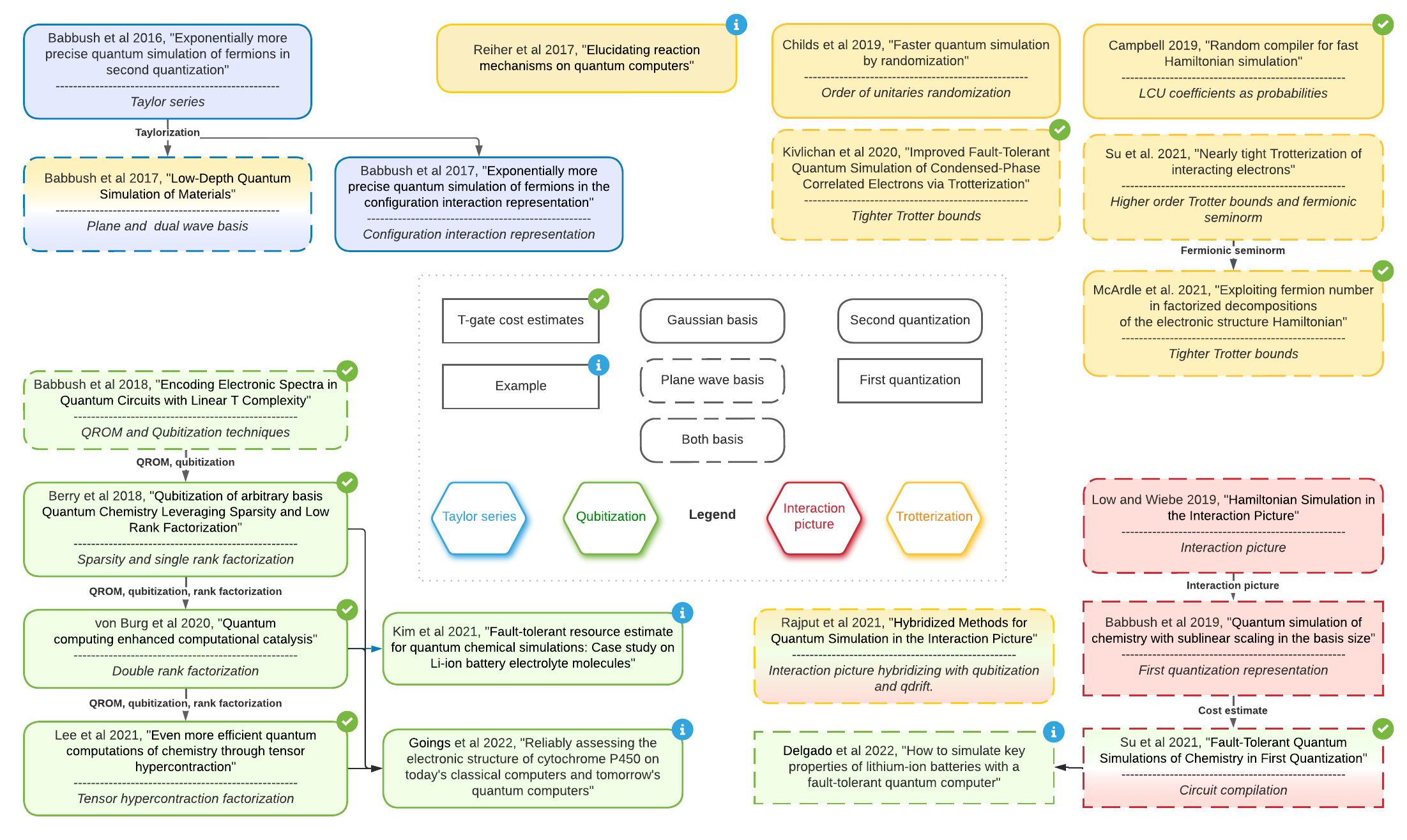}
    \caption{\label{fig:listmethods}Diagram showing some of the main techniques involved in the development of post-Trotter Quantum Phase Estimation Techniques. Not shown in the picture but of great importance are the articles crystallizing the concept of `Taylorization' \cite{berry2015simulating} and `qubitization', \cite{low2019hamiltonian}.%Blue is used to indicate Taylorization, green for QROM, qubitization and low-rank factorization, and red for the interaction picture. Dashed border indicates plane or dual wave basis, while solid line indicates Gaussian basis. Rounded borders indicate second quantization, in contrast to the first quantization; and the fill-in of the box indicates whether the article provides complexity estimates, T-gate cost estimates or an example using some method.
    }
\end{figure*}
While it is known that the general 2-body ground state preparation is QMA-complete, there is hope that the specificity of electronic Hamiltonians will make it easier to solve at least heuristically.
In fact, over the years significant effort has been devoted to the formulation of shallow-depth NISQ ansätze \cite{zhang2021computing} to prepare ground states such as the Imaginary Time Evolution ansatz \cite{mcardle2018variational} and the Variational Quantum Eigensolver (VQE) with Unitary Coupled-Cluster \cite{Peruzzo2014VQE}, adaptative \cite{Grimsley2019Adaptative}, and hardware-efficient \cite{Kandala2017Hardwareefficient} ansätze. 

Similarly, some effort has been devoted to resource estimates of particular applications \cite{reiher2017elucidating, von2020quantum, kim2022fault}, but to the best of our knowledge, no software library has been developed to allow a principled comparison between methods. This is a gap that TFermion aims to fill with the following contributions:

First, while newer algorithms often provide a specific non-Clifford gate and qubit count, older ones only give asymptotic complexity estimates (see figure \ref{fig:listmethods}). Our article aims to estimate the T-gate cost of older and some of the newer algorithms, with a molecule of choice from the software users. We believe this will be helpful to more quickly carry out research for both academics and industry. Not only that, but our software automatically performs optimization based on the different error sources to minimize the cost, and low-rank approximations \cite{berry2019qubitization}.

Second, as an example of use of our software, we address the question of whether Gaussian basis functions or plane waves are more convenient to simulate molecules, comparing the same Taylor series algorithms with a different basis.
This comparison is not definitive because the error arising from a finite-size basis is difficult to estimate. However, we can give an idea of which algorithms might be more beneficial according to some rough estimates of how many plane waves are required to simulate a system to the same precision than if one were to use Gaussian basis \cite{babbush2018low}. We furthermore provide researchers with the possibility to carry out a similar comparison but deciding the multiplicative factor for plane waves to represent a similar precision or if the comparison is not the objective, the number of plane waves too. 

In TFermion, so far we have focused on T-gates as we believe that non-Clifford gates represent a more significant bottleneck than the number of qubits. Nevertheless, in the future, we expect to add this functionality and additional algorithms to the library.
The article itself is structured as follows: first, we give an overview of the library and how it works. Then, in section \ref{sec:Techniques} we briefly explain some of the techniques for Quantum Phase Estimation and Hamiltonian simulation, including figure \ref{fig:listmethods} and table \ref{tab:algorithms} to help the reader understand the development and relation between different algorithms. In section \ref{sec:Results}, we give examples of how our software might be used, including the second contribution listed above. We then summarize the conclusions and present future work. Finally, in each appendix, we quickly describe one of the techniques studied in this paper, that can be used in combination with the original references to understand the cost estimation functions of TFermion.

%The main obstacle until now to computing the T-gate cost of the Taylorization approaches was heavy use of arithmetic subroutines and loose cost estimates. To avoid overloading the appendices with over-detailed calculations, we provide decompositions to the arithmetic or qubit phase rotation level, at which point we use the routines included in table \ref{tab:arithmetic}.

\section{\label{sec:Software}The TFermion library}

\begin{table*}[htbp]
  \centering
    \begin{tabular}{|l|l|l|}
    \hline
    Error & Mathematical definition &Where does it appear? \\
    \hline
    \hline
    $\epsilon_{QPE}$ & $\epsilon_{QPE} = \lambda 2^{-n}$, $n$ precision bits in the QPE algorithm, & Due to the Phase Estimation. \\
    & and $\lambda$ the 1-norm of the Hamiltonian. & \\
    \hline
    $\epsilon_{HS}$ & Trotter: $\left|\left|e^{-iH t/r} - \mathcal{S}_p(H; t/r)\right|\right|_2\leq W_p \left(\frac{ t}{r}\right)^{p+1}\leq\frac{\epsilon_{HS}}{r}$ \cite{mcardle2022exploiting}.  & In Hamiltonian Simulation via \\
    & Taylor: $\left|\left|\Pi_0 A\ket{0}\ket{\psi} - \ket{0}U_r \ket{\psi}\right|\right|_2\leq \frac{\epsilon_{HS}}{r}$ \cite{berry2015simulating}. & Trotter, Taylor or Dyson  \\
    & Dyson: $\left|\left|W-\mathcal{T}[e^{-i\int_0^{t/r} H(s)ds}]\right|\right|_2\leq \frac{\epsilon_{HS}}{r}$ \cite{low2019hamiltonian}. & series decomposition of $e^{-iH\tau}$. \\
    \hline
    $\epsilon_{H}$ & $\left|\int_\Omega f(\mathbf{x})d\mathbf{x} - \sum_{\mathbf{x}\in\Omega} f(\mathbf{x})(\Delta \mathbf{x})^d \right|< \epsilon_H$, with $d = dim(\Omega)$  & Error from the approximation\\
    & & of integrals by Riemannian sums. \\
    \hline
    $\epsilon_{S}$ \& $\epsilon_{SS}$ & $||U-R_z(\theta)||_2 \leq \epsilon_{SS}$ \cite{selinger2012efficient} & In the synthesis of single\\
    & (Using operator norm) & rotations $\epsilon_{SS}$ and their sum, $\epsilon_S$. \\
    \hline
    $\epsilon_{tay}$ & Defined as in Taylor's theorem. & Due to Taylor error series (and\\
    & & others) in arithmetic operations.\\
    \hline
    \end{tabular}%%
\caption{\label{tab:errors}Notation for the main sources of error that we take into account in the article and software library. Additional minor sources may appear sporadically in single articles. The norm 2 used in all cases above is the operator norm. The other algorithms used to compute arithmetic operations are the Babylon algorithm for the square root, and CORDIC algorithm for the sine or cosine. $\mathcal{S}_p(H;t/r)$ stands for the order $p$ Trotter step, and $W_p = O\left(\max_{\bm{i}} [[\ldots[H_{\gamma_{i_1}},H_{\gamma_{i_2}}], H_{\gamma_{i_3}}],\ldots],H_{\gamma_{i_{p+1}}}] \right)$ the commutator terms that bound the final error \cite{su2021nearly}.}
\end{table*}%
% Table generated by Excel2LaTeX from sheet 'Sheet2'

The first and main contribution of this article is a software library called TFermion that automatizes the estimation of T-gate cost of running a variety of Quantum Phase Estimation algorithms proposed in the literature during the last years, over arbitrary molecular geometries.

We envision several use cases of our library: 
\begin{enumerate}
    \item It could serve as a quick assessment for the feasibility of concrete QPE experiments once error-corrected quantum computers become available, such as those centered in particular scientific or industrial use cases  \cite{reiher2017elucidating,kim2022fault}.
    \item It can also help make comparisons between systems and methods. In particular, it allows comparing the impact of the chosen Hamiltonian simulation technique, or the chosen basis.
\end{enumerate}

The result provided by our library though must be interpreted as an approximation to the true value, as the final implementation will be heavily optimized, both at a hardware and software level. Our library, in contrast, aims to be more modular and system-agnostic, but we nevertheless provide built-in error optimization. It is well known that different error sources impact the final precision and gate cost in different ways. As such we have aimed at standardizing the way error sources are treated and optimized (see table \ref{tab:errors}). 

While not the main objective of our article, we also believe our work may help provide a more standardized treatment across methods, and as a consequence help better understand the choices in the Hamiltonian simulation, basis, or fermion-qubit mapping used.

One feature of our library is that it currently contains older than 1-year-old methods, and as such some excellent work \cite{von2020quantum,lee2020even,su2021fault} has not yet been included. There are two reasons: the first and most obvious one is that including new methods represents a significant amount of effort, and we believe these updates can be done later on. The second is that while for the latest methods T-gate estimates are more common, for older ones often only the complexity estimates are available. While this makes sense as the latest methods might be more useful for industrial processes, we believe that understanding well different techniques and not only the bleeding edge ones can be of significant scientific interest.

Additionally, our software was developed following a modular architecture with an easy procedure to include new methods. The process to add a newer method or updating an existing one requires two main steps: first making sure that the molecular parameters required are already calculated by some of the provided methods, or adding new ones in \texttt{molecule.py}; then create a new T-gate cost estimation function and call it from the class \texttt{Cost calculator}. The philosophy underlying this architecture is to keep TFermion updated timely and give the authors of the new methods the possibility to add their own T gate cost estimation to show practical examples of their work and make it more accessible.

The use of the library is rather straightforward: the user only needs to provide a \texttt{molecule name}, a \texttt{method} and optionally some atomic orbital labels (\texttt{ao labels}) to be used within the active space selection method AVAS \cite{sayfutyarova2017automated} to restrict the calculation and make it more efficient. This should be supplemented within a configuration file with the Gaussian basis to be used. If the method requires plane waves to be used, the system will by default approximate the number of basis functions as the thumb rule of 100 times more plane waves than Gaussian waves \cite{babbush2018low}. Alternatively, the user might provide this and other molecular parameters (eg $\lambda$, $N$...) in a JSON file under the name \texttt{[molecule name]}$\_$\texttt{[basis].json}. A flowchart of the working of the library can be seen in figure \ref{fig:Architecture}.

\begin{table}[ht!]
  \centering
  
    \begin{tabular}{|l|l|}
    \hline
    Operation & Cost \\
    \hline
    \hline
    Addition \& subtr. \cite{gidney2018halving} & $4n$  \\
    \hline
    Multiplication \cite{munoz2017t} & $21n^2$ \\
    \hline
    Division \cite{thapliyal2017quantum} & $14n^2 + 7n$ \\
    \hline
    Comparison \cite{cuccaro2004new} & $8n$ \\
    \hline
    Multi-controlled Not  \cite{barenco1995elementary} & $16(m-2)$ $m$ controls\\
    \hline
    Rotation synthesis \cite{selinger2012efficient} & $10 + 12\lceil\log_2\epsilon_{SS}^{-1}\rceil$, $SU(2)$  \\
         & $10 + 4\lceil\log_2\epsilon_{SS}^{-1}\rceil$, $R_z$ \\
    \hline
    State synthesis \cite{shende2006synthesis} & $2^{n+1}-2$ arbitrary   \\
    & rotations\\
    \hline
    \end{tabular}%%
\caption{\label{tab:arithmetic}Cost of basic arithmetic operators in T gates unless otherwise stated, omitting additive $O(1)$ factors. If the rotation synthesis is controlled, the cost will be multiplied by 2 for $R_x$, $R_y$ and $R_z$ gates, as given by Lemma 5.4 in \cite{barenco1995elementary}. Notice that $H R_z H = R_x$, while $R_y$ and $R_z$ are particular cases of the unitary $W$ in that Lemma. Finally, for general controlled rotations the cost will be thrice the synthesis cost instead of twice.}
\end{table}%

As it is shown in figure \ref{fig:Architecture}, TFermion is executed through a \texttt{main} module which receives the molecule name, the QPE method, and optionally also the ao-labels to select an active space using AVAS. It starts with the \texttt{molecule} module creating a \texttt{molecule} instance, which is passed together with the method to \texttt{cost calculator}. The latter one calls either Gaussian or Plane Waves \texttt{molecule} methods to calculate all necessary parameters. Finally, \texttt{cost calculator} minimizes the cost depending on the error sources on the selected method, and sends the result back to \texttt{main}.

TFermion manages four types of data:
\begin{itemize}
    
    \item \textbf{Molecule:} A class created to save all the molecular data, including geometric information obtained \cite{bolton2008pubchem} used to compute the electronic integrals using Pyscf \cite{sun2018pyscf}.

    \item \textbf{MolecularData:} An instance from the OpenFermion class \cite{mcclean2020openfermion}, necessary to get all parameters from the Hamiltonian and save them into instance \texttt{molecule} as attributes.

    \item \textbf{Error values:} Different QPE methods have different error sources, whose sum must not exceed a given threshold. By default we will use the \texttt{chemical accuracy} value of 0.0016 Hartrees \cite{cao2019quantum}. TFermion optimizes error values to minimize the T-gate cost output of that method without exceeding it.
    
    \item \textbf{T gate cost:} Number of T gates needed to execute the selected method, as well as the time required to synthesize the corresponding number of magic states. Calculating this value is the main goal of our library.
    
\end{itemize}

\begin{table*}[htbp]
  \centering

        \begin{tabular}{|l|l|l|l|l|}
        \hline
        \textbf{Algorithm} & \textbf{Simulation} & \textbf{Quantization} & \textbf{Basis} & \textbf{Encoding} \\
        \hline
        \hline
        Random Hamiltonian \cite{childs2019faster,campbell2019random} & Trotter & 2nd quantization & Gaussian & Jordan-Wigner \\
        \hline
        qDRIFT \cite{campbell2017shorter,campbell2019random} & Trotter-related & 2nd quantization & Gaussian & Jordan-Wigner \\
        \hline
        Taylorization `database' \cite{babbush2016exponentially} & Taylor series & 2nd quantization & Gaussian & Jordan-Wigner \\
        \hline
        Taylorization `on-the-fly' \cite{babbush2016exponentially} & Taylor series & 2nd quantization & Gaussian & Jordan-Wigner \\
        \hline
        Configuration Interaction \cite{babbush2017exponentially} & Taylor series & 1st quantization & Gaussian & Slater determinant \\
        \hline
        Low-depth `Trotter' \cite{babbush2018low} & Trotter & 2nd quantization & Plane waves & Jordan-Wigner \\
        \hline
        Low-depth `Taylor database' \cite{babbush2018low} & Taylor series & 2nd quantization & Plane waves & Jordan-Wigner \\
        \hline
        Low-depth `Taylor on-the-fly' \cite{babbush2018low} & Taylor series & 2nd quantization & Plane waves & Jordan-Wigner \\
        \hline
        Interaction picture \cite{low2018hamiltonian} & Dyson series & 2nd quantization & Plane waves & Jordan-Wigner \\
        \hline
        Sublinear scaling inter. pict. \cite{babbush2019quantum,su2021fault} & Dyson series & 1st quantization & Plane waves & Slater determinant \\
        \hline
        Sublinear scaling qubitization \cite{babbush2019quantum,su2021fault} & Qubitization & 1st quantization & Plane waves & Slater determinant \\
        \hline
        Linear T complexity \cite{babbush2018encoding}& Qubitization & 2nd quantization & Plane waves & Jordan-Wigner \\
        \hline
        Sparsity and low rank \cite{berry2019qubitization}& Qubitization & 2nd quantization & Gaussian & Jordan-Wigner \\
        \hline
        Double factorization \cite{von2020quantum} & Qubitization & 2nd quantization & Gaussian & Jordan-Wigner \\
        \hline
        Tensor hypercontraction \cite{lee2020even} & Qubitization & 2nd quantization & Gaussian & Jordan-Wigner \\
        \hline
        Hybridized method \cite{rajput2021hybridized} & Trotter \& Dyson  & 2nd quantization & Plane waves & Jordan-Wigner \\
        \hline
        \end{tabular}%
\caption{\label{tab:algorithms}Recent Hamiltonian simulation methods, named after the techniques they use, or the title of the corresponding article, explaining them for efficient Hamiltonian simulation and Quantum Phase Estimation. Notice that qDRIFT, Random Hamiltonian and Hybridize method do not specify the basis or the Fermionic encoding, but the ones we indicate seem to be the most obvious: in the case of qDRIFT and Random Hamiltonian because they are the simplest choice, while in the Hybridized method, it inherits the plane wave structure from the Interaction Picture. Recent work on Trotter Hamiltonian simulation \cite{kivlichan2020improved,su2021fault,mcardle2022exploiting,campbell2019random} has focused on bounding commutator error terms on a different basis, rather than new methods.}
\end{table*}%

Certain calculations in the library are computationally and memory intensive. The reason for this is that as the number of basis functions grows, so does the size of the one and two-body Hamiltonian terms, but does so at least quadratically. This is reflected especially in the plane wave case for molecules, where the larger number of plane waves is due to the need for significantly more basis functions. Nevertheless, an effort has been put into making the calculations relatively efficient, making use of some new techniques \cite{koridon2021orbital}.

Finally, let us briefly mention what our software does not cover yet. It only provides cost estimates for T-gate count, as it is well known that the magic state distillation required to perform the T-gate often carries the largest cost in 2 the dimensional surface code, which nevertheless exhibits a large threshold. Alternatively, there are codes in 3D, like topological color codes \cite{bombin2007topological} that avoid magic state distillation, and may provide new ways to improve this counting, but they require more qubits for similar distance codes. Furthermore, the Clifford gate count may depend on the specific chip connectivity, and for that reason, we have preferred to ignore it here. Finally, while we believe that the qubit count is important, the number of gates required may provide a more significant constraint in the long term due to the time required to perform the algorithms, as these approaches usually require on the order of $10^2$ to $10^4$ qubits for realistic targets \cite{reiher2017elucidating,kim2022fault,su2021fault}.

The cost of ground state preparation, while significant, is left for future work too. Rough estimates may be possible to obtain for moderately sized systems, using low precision QPE to project the system into the ground state \cite{berry2018improved}.

\section{\label{sec:Techniques}Quantum phase estimation techniques}

In this section, we give a quick overview of the main techniques used in the literature to perform quantum phase estimation. Quantum phase estimation requires two main ingredients: the use of a controlled Hamiltonian simulation method and sometimes an inverse Quantum Fourier Transform (QFT). While the original Quantum Phase Estimation protocol did use QFT \cite{Nielsen-Chuang,galindo2002information}, more modern versions such as Bayesian Quantum Phase Estimation avoid it \cite{wiebe2016efficient}. This latter approach has also the property of being parallelizable, implementable with minimal classical postprocessing, and requires fewer qubits. However, its cost scales as $\frac{4.7\lambda}{\epsilon_{QPE}}$ instead of the theoretical optimum of $\frac{\pi\lambda}{\epsilon_{QPE}}$ \cite{wiebe2016efficient}. Since the extra cost of the quantum Fourier transform and the qubits it requires are often negligible, we will instead assume we are using the classical version with a slightly lower cost. We will now explain the other main part, the different Hamiltonian simulation techniques.

\subsection{\label{ssec:Trotter}Trotter}

Let us assume we want to simulate $H$ for a Linear Combination of Unitaries decomposition $H = \sum_\gamma w_\gamma H_{\gamma}$. The difficulty is that since the different unitaries $H_\gamma$ do not need to commute, we cannot write $e^{-i H t} = \prod_\gamma e^{-iw_\gamma H_\gamma t}$. Instead, using the product of Hamiltonian simulation as we have just done introduces an error $O(\sum|[H_{\gamma_1},H_{\gamma_2}]|t^2)$ that depends on the commutator. 

To handle this error, within the scheme of Trotter, there are two strategies. The first one is to divide the evolution in short time segments so we can quadratically suppress the error. In other words, we implement 
\begin{equation}\label{first_order_Trotter}
    e^{-iH t} = \left(\prod_\gamma e^{-iw_\gamma H_\gamma t/r}\right)^r + O\left(\sum|[H_{\gamma_1},H_{\gamma_2}]|t^2/r\right).
\end{equation}
Alternatively, one may attempt to find higher order Trotter formulas that further suppress the error. For example, if \eqref{first_order_Trotter} is the first order formula, then
\begin{equation}\label{second_order_Trotter}
\begin{split}
    e^{-iH t} = \left(\left(\prod_{\gamma=1}^\Gamma e^{-iw_\gamma H_\gamma t/2r}\right)\left(\prod_{\gamma=\Gamma}^1 e^{-iw_\gamma H_\gamma t/2r}\right)\right)^r\\
    + O\left(\sum|[[H_{\gamma_1},H_{\gamma_2}],H_{\gamma_3}]|t^3/r^2\right)
\end{split}
\end{equation}
is the second order one.
Higher-order formulas are known, but they also become more convoluted to implement. Another possibility is to use classical randomization of the order in which each of $w_\gamma H_\gamma$ appears in the Hamiltonian, in each evolution segment \cite{childs2019faster}, or to apply Hamiltonian simulation of a random $H_\gamma$ for fixed amounts of time, with probabilities given in by $w_\gamma/\lambda$ for $\lambda = \sum w_\gamma$ \cite{campbell2019random}. The latter method is called `qDRIFT' and is explored in appendix \ref{app:qDRIFT} together with a second-order randomized Trotter simulation. Other randomized methods have been explored too \cite{wan2021randomized}.

There has also been effort devoted to tightly bounding the commutators to reduce the number of segments \cite{kivlichan2020improved,campbell2021early,su2021nearly,mcardle2022exploiting}. Of these, one with a favorable scaling number of basis functions, $O(N^3)$, is the so-called `SHC bound' for dual wave basis Hamiltonian \cite{su2021nearly,mcardle2022exploiting}. It is implemented as the method \texttt{shc\_trotter} in our library and can be found in appendix \ref{app:SHC_ Trotter}.
Finally, Trotter simulation has historically been one of the first methods to be used to estimate resource estimates, including the famous FeMoco study \cite{reiher2017elucidating}, and later ones \cite{elfving2020will}.

\subsection{\label{ssec:Taylor}Taylor series}

Methods invented after Trotterization are usually called post-Trotter, and their objective is to lower the Hamiltonian simulation error dependence, $\epsilon_{HS}$, from polynomial to polylogarithmic. Taylor series simulation or Taylorization aims to expand the evolution operator of a small time segment as a Taylor series
\begin{equation}
\begin{split}
    U_r = e^{-iHt/r}\approx \sum_{k= 0}^K \frac{1}{k!}(-iHt/r)^k =\\
    \sum_{k=0}^K \sum_{l_1,...,l_k=1}^L \frac{(-it/r)^k}{k!}a_{l_1}...a_{l_k}H_{l_1}...H_{l_k}.
\end{split}
\end{equation}
This expression is a Linear Combination of Unitaries (LCU), $U^{\text{Tay}}_{LCU} = \sum_{l=0}^L b_l U_l$. To implement it, one introduces operators
\begin{align}
    \text{Prepare}:\ket{0}\mapsto \sum_{l}\sqrt{b_l}\ket{l},\\
    \text{Select}:\ket{l}\ket{\psi}\mapsto \ket{l}U_l\ket{\psi},
\end{align}
and defines $U^{\text{Tay}}_{LCU} = (\text{Prepare}^\dagger\otimes \bm{1}) \text{Select} (\text{Prepare}\otimes \bm{1})$. Since $U_{LCU}^{\text{Tay}}$ has some failure probability in recovering $\ket{0}$ in the first register, it is customary to use (oblivious) amplitude amplification \cite{berry2015simulating}, that reduces the error to $\epsilon_{HS}/r$ in each segment.

\subsection{\label{ssec:Qubitization}Block encoding and qubitization}

Similarly, the Hamiltonian often takes the form of a linear combination of unitaries $H = \sum a_l H_l$, from which we can create as the block-encoding operator
\begin{equation}
    U_{LCU} = \begin{pmatrix}
    H/\lambda & \cdot\\
    \cdot & \cdot \\
    \end{pmatrix},
\end{equation}
with new Prepare and Select operators
\begin{align}
    \text{Prepare}:\ket{0}\mapsto \sum_{l}\sqrt{a_l}\ket{l},\\
    \text{Select}:\ket{l}\ket{\psi}\mapsto \ket{l}H_l\ket{\psi}.
\end{align}
Using them, we obtain, 
\begin{equation}
    U_{LCU}\ket{0}\ket{\psi} = \ket{0}\frac{H}{\lambda}\ket{\psi} + \sqrt{1-\frac{\|H\ket{\psi}\|}{\lambda}}\ket{(0,\psi_\lambda)^\perp}.
\end{equation}
However, as we saw this LCU implementation has some probability of failure, which requires amplitude amplification to suppress. An alternative is to construct a quantum walk operator $Q$ with the same spectrum. This is done via a procedure called qubitization \cite{low2018hamiltonian}. In the case where the corresponding $U^2=1$, as is the case for $U_{LCU} = \text{Prepare}^\dagger\cdot \text{Select}\cdot \text{Prepare}$, it can simply be implemented as \cite[Corollary 9]{low2018hamiltonian}
\begin{equation}
    Q = \underbrace{\text{Prepare}(2\ket{0}\bra{0}\otimes \bm{1}-\bm{1})\text{Prepare}^\dagger}_{R}\cdot \text{Select}.
\end{equation}
$Q$ implements a Grover rotation in each eigenspace
\begin{equation}
\begin{split}
    Q\ket{0}\ket{\psi_k}&=\cos(\theta_k)\ket{0}\ket{\psi_k}-\sin(\theta_k)\ket{(0,\psi_k)^\perp},\\
    Q\ket{(0,\psi_k)^\perp}&=\cos(\theta_k)\ket{(0,\psi_k)^\perp}+\sin(\theta_k)\ket{0}\ket{\psi_k},
\end{split}
\end{equation}
for $\cos\theta_k= \frac{E_k}{\lambda}$. In other words, $Q$ is a quantum walk operator
\begin{equation}
    Q = \bigoplus_k \begin{pmatrix}
    \frac{E_k}{\lambda} & -\sqrt{1-\frac{E_k^2}{\lambda^2}}\\
    \sqrt{1-\frac{E_k^2}{\lambda^2}} & \frac{E_k}{\lambda}
    \end{pmatrix}_k.
\end{equation}
Diagonalizing the subspace spanned by $\{\ket{0}\ket{\psi_k}, \ket{(0,\psi_k)^\perp}\}$, we might write \mbox{$Q_{LCU} = \bigoplus_k\left( e^{i\theta_k}\ket{\theta_k}\bra{\theta_k}+e^{-i\theta_k}\ket{-\theta_k}\bra{-\theta_k}\right)$.} We can use this operator to create a Chebyshev series that approximates $e^{-iHt}$ \cite{low2019hamiltonian}, with a technique called quantum signal processing \cite{low2017optimal}. However, it is more straightforward to apply phase estimation directly over $\pm \theta_k$ \cite{berry2018improved}. Then, computing $\cos(\theta_0)$ we recover the ground state energy.

Additionally, qubitization has the advantage that $R Q R = Q^\dagger$, so using this trick we can duplicate the implemented phase with almost no extra cost, so the prefactor in the cost falls from $\frac{\pi\lambda}{\epsilon_{QPE}}$ to $\frac{\pi\lambda}{2\epsilon_{QPE}}$ \cite{babbush2018encoding}. Qubitization is often used in combination with QROM and factorization techniques \cite{babbush2018encoding, berry2019qubitization, von2020quantum,lee2020even}, but has also been used in first quantization \cite{babbush2019quantum,su2021fault}. 

\subsection{\label{ssec:Dyson}Interaction picture and Dyson series}

While the qubitization method is optimal concerning the Hamiltonian simulation error, an alternative approach is to find ways to decrease the 1-norm $\lambda$ of the Hamiltonian $H$. Let us assume that $H = A+B$ such that $\|A\|\gg \|B\|$. In the interaction picture, $H_I(t) = e^{iAt}B(t) e^{-iAt}$, so in this framework, the norm of the Hamiltonian decreases to $\|B\|$, and therefore the phase estimation may be cheaper to implement.
In this picture, the Hamiltonian simulation is implemented as
\begin{equation}
    \ket{\psi(t)} = e^{-iAt}\mathcal{T}\left[e^{-i\int_0^t H(s)ds}\right] \ket{\psi(0)},
\end{equation}
where $\mathcal{T}$ denotes time ordering. While the $e^{-iAt}$ might be easy to implement if all unitary operators in LCU decomposition of $A$ commute, the time ordered exponential is more difficult to implement. This might be done with a Dyson series
\begin{equation}
\begin{split}
    U(t) = \mathcal{T}\left[e^{-i\int_0^t H(s)ds}\right] = \sum_{k = 0}^\infty (-i)^k D_k\\
    D_k = \frac{1}{k!}\int_{0}^t...\int_{0}^t\mathcal{T}[H(t_k)...H(t_1)]d^k t,
\end{split}
\end{equation}
that similarly to the Taylor series approach, bears a logarithmic complexity on $\epsilon_{HS}$, and requires to implement the simulation for short time segments and use amplitude amplification at each of them. Operator $B$ is implemented as 
\begin{equation}
    \frac{B}{\|\lambda_B\|} = \braket{0|\text{Prepare}_B^\dagger \cdot \text{Select}_B\cdot\text{Prepare}_B|0}
\end{equation}
Using this block encoding of operator $B$, we can express the block encoding of a time segment of $e^{-i(A+B)\tau}$ as \cite{su2021fault}
\begin{equation}
\begin{split}
e^{-i(A+B)\tau}\approx e^{-iA \tau}\lim_{\substack{K\rightarrow \infty\\ M\rightarrow \infty}}\sum_{k=0}^K\frac{(-i\tau)^k}{M^k k!}\sum_{m_1=0}^{M-1} \ldots \sum_{m_k=0}^{M-1}\\
\Big(e^{-i\tau(-1/2-m'_k)A/M} B e^{-i\tau(m'_k-m'_{k-1})A/M} B\ldots \\
B e^{-i\tau(m'_2-m'_{1})A/M} B e^{-i\tau(m'_1+1/2)A/M}\Big)\\
= \left(\bra{0}\text{Prepare}_B^\dagger\right)^{\otimes K}
\sum_{k=0}^K \frac{(-i\lambda_B\tau)^k}{M^k k!}\sum_{m_1,\ldots,m_k =0}^{M-1}\\
\Big(e^{-i\tau(M-1/2-m'_k)A/M} \text{Select}_B e^{-i\tau(m'_k-m'_{k-1})A/M}  \\
\text{Select}_B \ldots\text{Select}_B e^{-i\tau(m'_2-m'_{1})A/M} \text{Select}_B \\
e^{-i\tau(m'_1+1/2)A/M}\Big)\Big(\text{Prepare}_B \ket{0}\Big)^{\otimes K},
\end{split} 
\end{equation}
where $m'_1,\ldots,m'_k$ are the sorted integers from $m_1,\ldots,m_k$. This series might therefore be implemented in a similar fashion as those from Taylor series, and will similarly require amplitude amplification. The Dyson series simulation was first introduced in Refs. \cite{kieferova2019simulating,low2018hamiltonian}.

\section{\label{sec:Results}Results and an use case example: comparison between different basis functions}

In this section, we make use of our library to show usage examples. For that purpose, we will perform two tasks: (1) using the FeMoco Hamiltonian provided in the supplementary material of \cite{lee2020even}, compute the cost of performing Quantum Phase Estimation with several methods included in the library; and (2) perform T-gate estimation for a few simple molecules with a wide range of methods, making a preliminary comparison of the impact of Gaussian or plane-wave basis in the final T gate count, when using Taylorization as a Hamiltonian simulation method.

\begin{figure}[b!]
    %\centering
    \includegraphics[width=\textwidth/2]{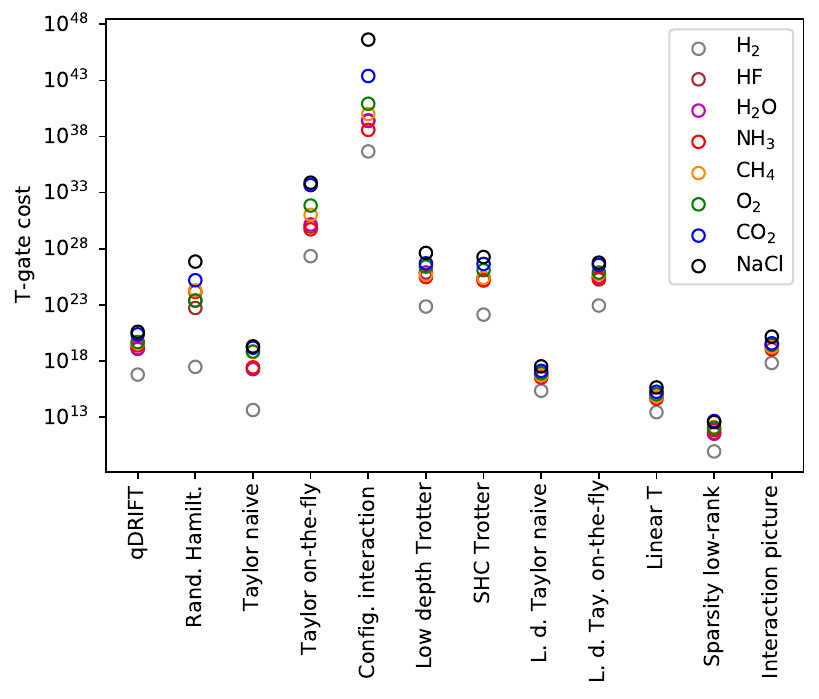}
    \caption{Representation of the results obtained for simple molecules with the results from table \ref{tab:costs}. We can see that choosing the right method greatly impacts the final cost of the Quantum Phase Estimation algorithm.}
    \label{fig:Costs}
\end{figure}

\subsection*{FeMoco estimates}

\begin{table}[t!]
\begin{tabular}{|l|l|l|}
\hline
FeMoco active space & Reiher et. \cite{reiher2017elucidating} & Li et. \cite{li2019electronic}  \\ \hline \hline
qDRIFT \cite{campbell2019random} & 7.34e+23 &  3.62e+23  \\ \hline
Rand. Hamilt. \cite{campbell2019random} & 1.32e+28 & 2.94e+28  \\ \hline
Taylor naïve \cite{babbush2016exponentially} & 1.15e+22 & 1.26e+23 \\ \hline\hline
Spars. low-rank \cite{berry2019qubitization} & 2.36e+13 & 2.17e+13 \\ \hline
w/o failure \cite{berry2019qubitization} & 4.57e+12 & 4.12e+12 \\ \hline
Results in \cite{berry2019qubitization} & 4.8e+12 & 3.9e+12 \\ \hline
\end{tabular}
\caption{\label{tab:FeMoco}Estimation of number of T-gates required to run different Quantum Phase Estimation algorithms with several algorithms. The second half of the table shows that our library gets similar results as \cite{berry2019qubitization}, where the `w/o failure' row indicates we obtain without taking into account failure probability.}
\end{table}

\begin{table}[t!]
\begin{tabular}{|l|l|l|l|l|}
\hline
$N$ & $\lambda$  & TFermion & \cite{babbush2018encoding} conditions & \cite{babbush2018encoding} results \\ \hline \hline
54 & 5 &  7.08e+08  & 2.69e+07 & 1.80e+07 \\ \hline
128 & 23 & 4.78e+09 & 2.26e+08 & 1.90e+08\\ \hline
250 & 64 & 1.96e+10 & 1.09e+09& 1.10e+09\\ \hline
1024 & 640 & 5.58e+11 & 3.88e+10 & 4.30e+10\\ \hline
\end{tabular}
\label{tab:Jellium}
\caption{Replication of the T-gate cost estimates of the \texttt{linear\_t} method with Jellium, similar to those published in table III from \cite{babbush2018encoding}. The third column includes the results with our library, while the fifth those from the original reference \cite{babbush2018encoding}. Most of the divergence can be explained because the total error budget has to be allocated between $\epsilon_{QPE}$ and $\epsilon_S$, and by considering negligible the rotation synthesis cost. To account for this, the fourth column indicates the results we get if fixed the phase estimation error to $\epsilon_{QPE} = 0.0016$ Hartree, and did not take into account the cost of gate rotation synthesis or failure probability. After this we still do not get the exact results due to other polylogarithmic contributions that the original reference did not considered; but we get quite close.}
\end{table}

Over the last years, FeMoco became a standard benchmark for quantum algorithms \cite{reiher2017elucidating}. Such a benchmark is realistic and useful because it constitutes the metal active center of an enzyme capable of converting atmospheric nitrogen and hydrogen into ammonia, bypassing the energy-intensive industrial Haber-Bosch process. As the first use case example of our library, we first extend the T-gate cost estimation for several methods. Not only this will help us understand the complexity of previous examples, but will also help check the validity of our results for the low-rank decomposition method, where previous estimates were available \cite{berry2019qubitization}.

Using the Taylorization approach \cite{babbush2016exponentially} has intermediate cost between that of Trotterization (qDRIFT and Random Hamiltonian simulation \cite{campbell2019random}) and more recent rank-decomposition and qubitization techniques \cite{berry2019qubitization}. Furthermore, the last row of table \ref{tab:FeMoco} can be compared with the published costs of $1.2\cdot 10^{12}$ and $9.8\cdot 10^{11}$ Toffoli gates for both active spaces \cite{reiher2017elucidating,li2019electronic,berry2019qubitization}. Since each Toffoli gate is equivalent to 4 T-gates, our estimation is very close to the numbers originally reported. We believe the small difference is due to a combination of factors. In the first place, the error optimization will usually give more weight to $\epsilon_{QPE}$ as it is the most costly error source. Additionally, we take into account some factors such as the Uniform subroutines and an amplitude amplification step in the preparation of uniform superpositions on registers $p$ and $q$ such that $p\leq q < N/2$ (respectively $r$ and $s$). We also take a slightly larger number of segments $r$ as described in section 3A of \cite{su2021fault}, due to the estimation of the phase of $e^{-i\tau\arccos H}$ instead of $e^{-i\tau H}$.

The FeMoco cost of other methods implemented in the library has not been computed, due to the lack of geometry-dependent parameters such as the position of the atoms in FeMoco, or because they were conceived for plane waves instead of gaussian wave functions. In any case, we believe that these results confirm the usefulness of TFermion.

%Factor de 4 debido a 2^m
%Factor de 2 debido a Prepare^\dagger
%Factor de 4 por Toffoli -> T gates

\begin{figure}[t!]
    %\centering
    \includegraphics[width=\textwidth/2]{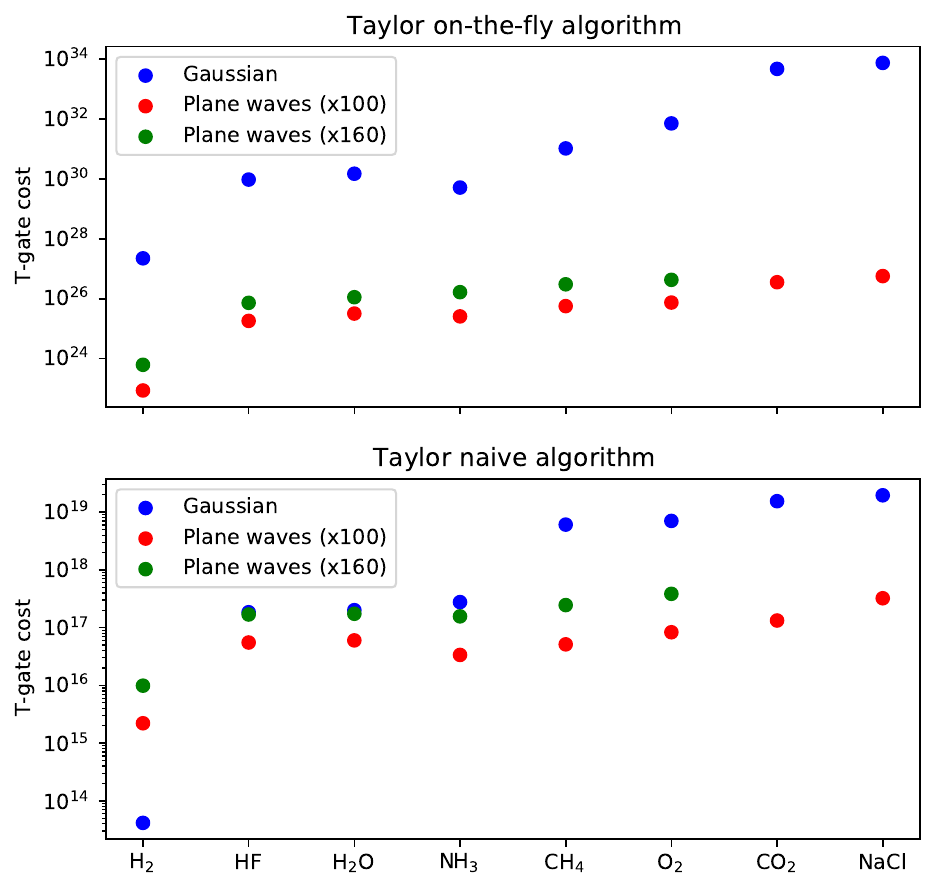}
    \caption{T-gate cost of performing the same algorithms making use of Taylorization as the main Hamiltonian simulation technique, over different molecules. The number of plane waves was chosen to be $\approx 100$ or $160$ times larger than Gaussian functions as recommended by Appendix E in \cite{babbush2018low}. The cost of computing the electronic integrals on-the-fly is larger than classically precomputing and loading them. The comparison between Gaussian and plane-wave basis should be taken with care as the error due to finite basis size was not rigorously computed and controlled.}
    \label{fig:Gaussian_vs_plane}
\end{figure}

\subsection*{Simple molecules}

Next, we run T-gate cost estimates of all the algorithms included in TFermion, with several molecules. As a use-case example, we compare the costs of similar methods on a different basis, something not previously been done in the literature. While these simple molecules can also be analyzed with classical methods, we selected these simple molecules to avoid performing active space selection on them. Of course, selecting such active space in a molecule of scientific interest will represent an important step to making the simulation efficient, but our aim here is to compare the methods rather than obtain novel results for applications of scientific or industrial interest.

The results from our calculations can be seen in table \ref{tab:costs}. We indicate the median value obtained for each entry after running the procedure $10^3$ times. We select the median instead of the average because the results have some inherent stochasticity due to the error sources optimization, but the distribution tends to be skewed to the higher values. We also do not take the lowest value to avoid numerical instability in the $\epsilon$ values that may have given rise to unrealistic lower costs.

\begin{table*}
\centering
\begin{tabular}{|l|p{1.2cm}|p{1.2cm}|p{1.2cm}|p{1.2cm}|p{1.2cm}|p{1.2cm}|p{1.2cm}|p{1.2cm}|}
\hline
Method &   H$_2$ &      HF &  H$_2$O &  NH$_3$ &  CH$_4$ &   O$_2$ &  CO$_2$ &    NaCl \\
\hline \hline
qDRIFT \cite{campbell2019random}                   & 6.2e+16 & 1.2e+19 & 1.4e+19 & 2.4e+19 & 3.9e+19 & 5.0e+19 & 2.4e+20 & 4.0e+20 \\
Rand. Hamilt. \cite{campbell2019random}            & 3.0e+17 & 5.2e+22 & 2.4e+23 & 1.4e+24 & 1.9e+24 & 2.4e+23 & 1.6e+25 & 7.1e+26 \\
Taylor naive \cite{babbush2016exponentially}       & 3.0e+13 & 1.3e+17 & 1.4e+17 & 1.9e+17 & 4.1e+18 & 4.7e+18 & 1.1e+19 & 1.4e+19 \\
Taylor on-the-fly \cite{babbush2016exponentially}  & 1.4e+27 & 5.9e+29 & 9.4e+29 & 3.3e+29 & 6.8e+30 & 4.6e+31 & 3.0e+33 & 4.8e+33 \\
Config. interaction \cite{babbush2017exponentially} & 1.6e+36 & 2.4e+39 & 2.8e+39 & 3.9e+38 & 1.0e+40 & 8.3e+40 & 2.5e+43 & 4.3e+46 \\
Low depth Trotter \cite{babbush2018low}            & 1.2e+23 & 1.3e+26 & 1.1e+26 & 5.0e+25 & 8.5e+25 & 4.4e+26 & 8.4e+26 & 6.9e+27 \\
SHC Trotter \cite{babbush2018low,mcardle2022exploiting}& 2.3e+22 & 3.6e+25 & 4.2e+25 & 2.5e+25 & 4.2e+25 & 2.0e+26 & 7.5e+26 & 3.2e+27 \\
L. d. Taylor naive \cite{babbush2018low}           & 3.1e+15 & 7.8e+16 & 8.4e+16 & 4.9e+16 & 7.6e+16 & 1.2e+17 & 1.8e+17 & 4.7e+17 \\
L. d. Tay. on-the-fly \cite{babbush2018low}        & 1.3e+23 & 2.7e+25 & 4.7e+25 & 3.7e+25 & 8.4e+25 & 1.1e+26 & 5.2e+26 & 8.5e+26 \\
Linear T \cite{babbush2018encoding}                & 3.9e+13 & 1.0e+15 & 1.1e+15 & 6.3e+14 & 9.7e+14 & 1.6e+15 & 2.6e+15 & 6.3e+15 \\
Sparsity low-rank \cite{berry2019qubitization}     & $\bm{1.2e10}$ & $\bm{4.6e11}$ & $\bm{6.0e11}$ & $\bm{1.0e12}$ & $\bm{1.8e12}$ & $\bm{1.5e12}$ & $\bm{6.3e12}$ & $\bm{5.3e12}$ \\
Interaction picture \cite{low2018hamiltonian}      & 1.4e+18 & 5.7e+19 & 5.0e+19 & 2.4e+19 & 3.6e+19 & 6.6e+19 & 8.0e+19 & 3.3e+20 \\
\hline 
\end{tabular}
\caption{T-gate cost estimates for different molecules and methods obtained using our TFermion, see Fig. \ref{fig:Costs}. The Rank decomposition technique is the most efficient between the analysed methods, closely followed by the plane wave methods using QROM and qubitization (`Linear T') or Taylorization (`Low depth Taylor naïve').\label{tab:costs}}
\end{table*}

Let us first comment on the results of some methods. The first thing that calls our attention is the large cost of the Configuration Interaction method \cite{babbush2017exponentially}. We believe this is due to a combination of three factors: the first and most important one is that the condition on the number of segments $r$ imposed by the Lemmas 1-3 in \cite{babbush2017exponentially} is a very large value, which may be understood as an upper bound rather than a real cost estimate. Secondly, our method to perform the procedure from section 4.1 was not optimized. And thirdly, it also contains a large number of arithmetic operations, similar to those in `Taylor on-the-fly'. Overall this indicates that the estimates for this method should be treated as an upper bound.

We can also observe that when using a Gaussian basis, Taylor methods are almost always more efficient than Trotter ones and that the cost of using the on-the-fly versions of Taylor is often larger than the naïve one due to the arithmetic operations. The interaction picture algorithm \cite{babbush2018low} displays a `similar' complexity as the Taylorization algorithms \cite{babbush2016exponentially}, as both operate on a Gaussian basis and decompose the evolution operator in a Taylor or Dyson series.

The most efficient algorithms among the analyzed ones are those making use of the QROM techniques, \cite{babbush2018encoding,berry2019qubitization}. Surprisingly though, the Low depth Taylor naïve \cite{babbush2018low} achieves the third-best complexity just after the rank-decomposition algorithm \cite{berry2019qubitization}, and the original article introducing the QROM \cite{babbush2018encoding}. We believe the reason for that is that the original article left unspecified the techniques that should be used to implement Prepare and Select, so we have assumed the use of modern QROM techniques \cite{babbush2018encoding}. 

To make this comparison fair, we have, as a rule of thumb, used approximately 100 times as many plane waves as Gaussian wave functions, as it has been suggested for isolated molecules \cite{babbush2018low}. The Gaussian basis used is the standard 6-31G \cite{jensen2013atomic}, but this may be changed by the user at will in the configuration file, as well as the multiplicative factor. Using the previously mentioned ratio, we can as an example of usage of our library, compute the cost of the same Taylorization methods with Gaussian and plane waves. The results are shown in figure \ref{fig:Gaussian_vs_plane}, although these results must be taken  with care as we have not controlled the error introduced by different finite basis sets.

\section{\label{sec:Conclusions}Conclusions and future work}

Over the last years significant effort has been devoted to creating efficient algorithms for Quantum Phase Estimation and Hamiltonian simulation since the estimation of ground state energy is such a central problem for quantum chemistry and a very natural application of quantum computing. 
TFermion fills a gap in standardizing and easing the use of such algorithms. It should help academics have a better understanding of algorithms for which no complexity estimates were previously available. The usefulness for the industry is
clear too, as it reduces the effort required to quickly
iterate over specific use-cases. As examples of usage,
we have run calculations with FeMoco and a range of
molecules. Among the most interesting results is the
fact that using QROM techniques in the plane wave
naïve Taylorization method \cite{babbush2018low} makes it particularly efficient, and we have seen hints that using plane-wave could be more efficient than Gaussian for the same Taylorization techniques in isolated molecules.

However, the effort is far from complete. On one hand, exciting avenues of research remain open, particularly in the use of plane waves \cite{su2021fault}. On the other, we aim to improve this library in several dimensions: (1) newer algorithms should be added; (2) our algorithms are designed for molecules instead of materials, where plane-wave methods should become very efficient; (3) TFermion only provides estimates for T-gates so the addition of other metrics such as the number of qubits would be a welcomed addition; and (4) the topic of ground state preparation is barely touched upon but should be considered a prerequisite to estimate the ground state energy.

We believe this is a particularly exciting time to explore how quantum computing can be applied to chemistry and material science. For this reason, we humbly hope that TFermion will become a useful tool to advance the field and find beneficial applications for society.

\section*{Code availability}
The code for this article can be found at
\mbox{\url{https://github.com/PabloAMC/TFermion}}.

\section*{Acknowledgements}

We want to thank the very kind explanations of Emiel Koridon of some calculations in one of his articles and beyond. Similarly, we thank answers from Nicolas Rubin and Ryan Babbush on the use of OpenFermion, Joonho Lee on the code from \cite{lee2020even}, and Antonio Hidalgo, María Jesús Morán, Nelaine Mora and Javier García on quantum chemistry.
We acknowledge financial support from the Spanish MINECO grants MINECO/FEDER Projects FIS 2017-91460-EXP, PGC2018-099169-B-I00 FIS-2018, from CAM/FEDER Project No. S2018/TCS-4342 (QUITEMAD-CM), and from Spanish MCIN with funding from European Union NextGenerationEU (PRTR-C17.I1) and Ministry of Economic Affairs Quantum ENIA project. The research of M.A.M.-D. has been partially supported by the U.S. Army Research Office through Grant No.  W911NF-14-1-0103. P. A. M. C. thanks the support of a MECD grant FPU17/03620, and R.C. the support of a CAM grant IND2019/TIC17146. 
\bibliographystyle{plainnat}
\bibliography{bibliography} 
\clearpage
\newpage

\appendix

\section{\label{app:qDRIFT}qDRIFT, a random Hamiltonian trotterization approach}

Using Hamiltonian simulation to estimate the energy of chemical configurations can be accomplished through different methods. We will present the main ones that can be chosen from in our software package in the following appendices. We first consider the \textit{Trotter-Suzuki decomposition} \cite{suzuki1990fractal,suzuki1991general,abrams1997simulation}, where the time evolution of a Hamiltonian $H= \sum_{\gamma=1}^{\Gamma} w_\gamma H_{\gamma}$, with $H_{\gamma}$ being a normalized Hermitian operator and $w_\gamma$ a non-negative Hamiltonian coefficient, is approximated by
\begin{equation}
    e^{-iHt} = e^{-it\sum_{\gamma} w_\gamma H_{\gamma}} \approx \left(\prod_{\gamma=1}^{\Gamma} e^{-iw_\gamma H_{\gamma} t/r}\right)^{r}.
    \label{Trotter_decomposition}
\end{equation}
In the limit of $r\rightarrow\infty$ the equality is exact. Notice that $H$ and $H_\gamma$ do not need to be unitary in general, only Hermitian. In contrast, $e^{-iHt}$ is unitary, and since the electronic Hamiltonian can be written in second quantization as a Linear Combination of Unitaries, for the estimation of the cost of this method we will in fact take $H_\gamma$ to be unitary, as in the rest of the described methods. In this section, we present the qDRIFT and Random Hamiltononian methods, some of the best method that uses the Trotter-Suzuki decomposition \cite{campbell2019random}. The main idea here is to reduce the complexity of the Trotter Suzuki decomposition above by randomizing the order in which the terms $e^{-iH_{\gamma}t/r}$ are applied. They suggest to simulate a single unitary $e^{-i\tau H_\gamma}$ randomly from an identical distribution, where $\tau=t\lambda/r$ is fixed, $\lambda = \sum_{\gamma=1}^\Gamma w_\gamma$, and the probability of choosing an individual unitary is weighted by the Hamiltonian coefficient $w_\gamma$. We further define $\Lambda = \max_{\gamma} w_\gamma$. This markovian method is referred to as the qDRIFT approach.

The qDRIFT algorithm achieves $O(\lambda^2 t^2/\epsilon_{HS})$ gate complexity, where $\epsilon_{HS}$ is the desired precision. This scaling stems from making the zeroth and first-order expansion terms of the qDRIFT quantum channel coincide with the channel that describes the unitary evolution. In contrast, the $2k$-th order (deterministic) Trotter methods have complexity $O(\Gamma^{2+1/2k}(\Lambda t)^{1+1/2k} / \epsilon_{HS}^{1/2k})$ \cite{campbell2019random}. As a consequence, the qDRIFT algorithm proves advantageous whenever $\lambda \ll \Lambda \Gamma$, which is the case for most electronic structure Hamiltonians, as the majority of terms $H_\gamma$ possess small coefficients $w_\gamma$ \cite{berry2019qubitization}. On the other hand, qDRIFT will most likely perform worse than higher-order Trotter expansion for large evolution times. %Notice that although qDRIFT does not depend explicitly on $\Gamma$, since $\Gamma = O(N^4)$ and $\lambda$ has some polynomial dependence on $N$, qDRIFT will not be completely independent of $\Gamma$. 

In the following, we will present the number of $T$ gates required for performing the unitary evolution of Eq.~\eqref{Trotter_decomposition} through the qDRIFT method and a second order Trotterization method, respectively. The details of this analysis are based on the supplementary material of \cite{campbell2019random} and consider the problem of estimating the ground state energy $E_0$ of a Hamiltonian $H$ using quantum phase estimation. The total number of gates $n$ of the form $e^{-i\tau H_\gamma}$ required to estimate the energy of the ground state to an additive error $\delta_E$ using qDRIFT is given by \cite{campbell2019random}
\begin{align}
n=\frac{\pi^2\lambda^2}{\epsilon_{tot}\delta_E^2}\left(\frac{1+p_f}{p_f}\right)^2,
\label{qDRIFT exponentials}
\end{align}
where $p_f$ is the failure probability inherent to the quantum phase estimation algorithm and $\epsilon_{\text{tot}}$ is the total Trotter error. Similarly, using a second-order random Trotterization, this number scales as \cite{campbell2019random}
\begin{align}
    n= 8\Gamma^2\frac{1}{\epsilon_{tot}}\left(\frac{\pi\Lambda}{2\delta_E}\right)^{3/2}\left(\frac{1+p_f}{p_f}\right)^{3/2}.
    \label{Trotter exponentials}
\end{align}
To arrive at the cost in terms of $T$-gates, we need to assess the $T$-gate cost of simulating a gate $e^{-i\tau H_\gamma}$ and then multiply it by $n$ as given by Eq.~\eqref{qDRIFT exponentials} and Eq.~\eqref{Trotter exponentials} to give an estimate for the cost of performing qDRIFT and a second-order Trotterization approach, respectively.

The difficulty here is that $H_\gamma$ will be a string of Pauli operators, so we cannot just implement the rotation in each qubit separately as it is an entangling rotation. Fortunately, we can perform each $e^{-iH_\gamma \tau}$ using Clifford gates and a single $C$-$R_z$ rotation \cite{hastings2014improving,motzoi2017linear}. This, in turn can be decomposed in two $R_z$ gates using Lemma 5.4 from \cite{barenco1995elementary}, and each rotation implemented with $\approx 10 +4 \log(\epsilon^{-1}_{SS})$ T-gates \cite{selinger2012efficient}.
 
Finally, notice that in the notation of our article, we are taking $\delta_E = 2\epsilon_{QPE}$ and $\epsilon_{tot}=\epsilon_{HS}$. Similarly $\epsilon_{SS}$ can be determined by dividing $\epsilon_S$ (which is not taken into account in \cite{campbell2019random}), by the number of unitary Pauli rotations used, $2n$.

\section{\label{app:Taylorization}Taylorization-based Hamiltonian simulation}

If in the previous appendix we explored the Trotter and Trotter-like methods for Hamiltonian simulation, from now on we would like to focus on so-called post-Trotter methods that allow avoiding having polynomial complexity in the Hamiltonian simulation precision $\epsilon_{HS}^{-1}$. We will start with a method called Taylorization \cite{babbush2016exponentially}.

\subsection{\label{app:Taylorization_method}Method explanation}

\subsubsection{`Database' algorithm}

\begin{figure}[htp]
    \centering
    \includegraphics[width=.5\textwidth]{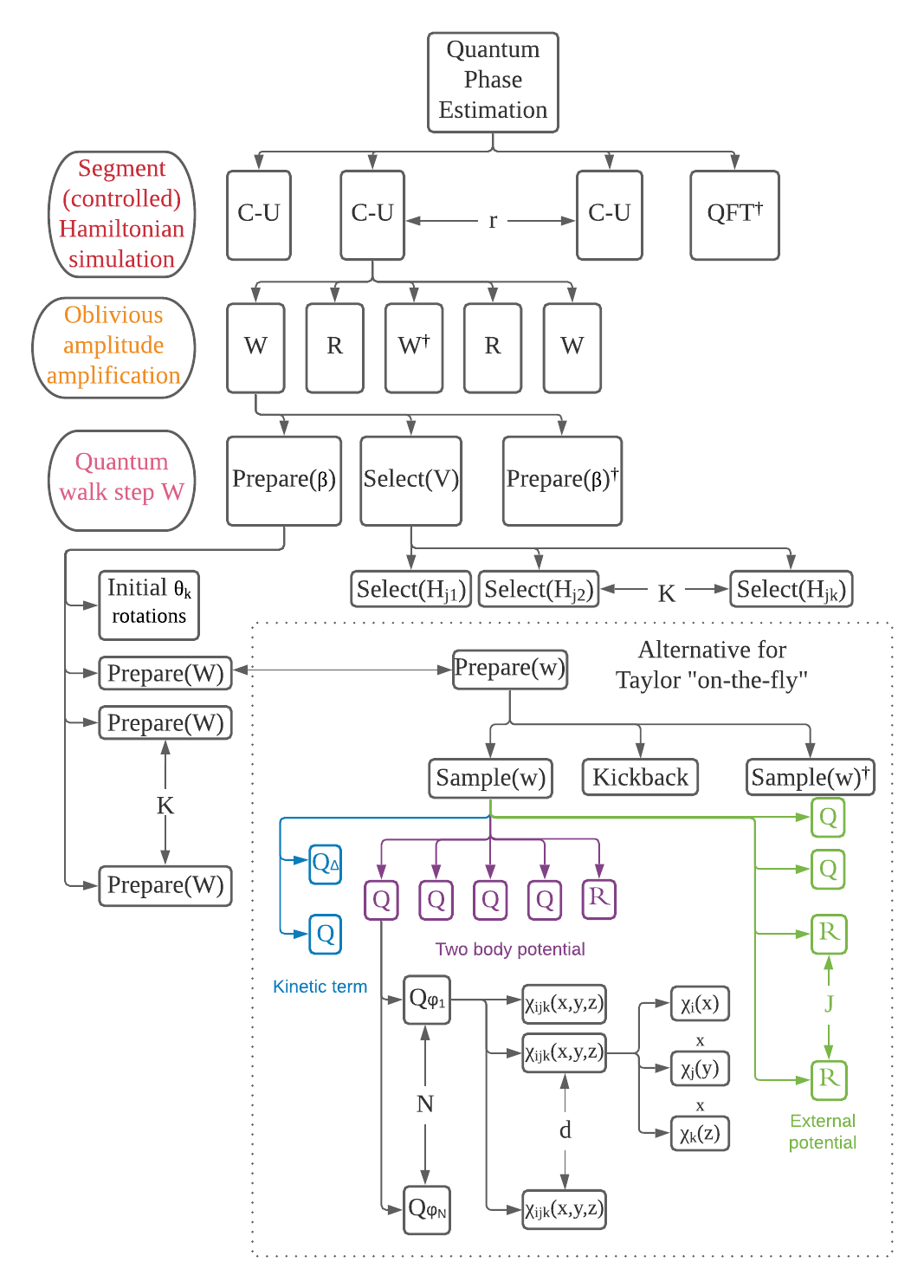}
    \caption{Abstraction level decomposition of the Taylor `database' algorithm. The x-axis represents the time steps of the algorithm, while the y-axis is the abstraction level, higher meaning more abstract. In the lower box, we also depict the substitution one does to perform the alternative Taylor `on-the-flight' algorithm. Notice that this does not show minor operations such as the computation of $\vec{\xi}$ or the multiplication in the last step of figure 4 from \cite{babbush2016exponentially}.}
    \label{fig:Taylor_database}
\end{figure}

The aim of the algorithm is to implement Hamiltonian simulation for $H = \sum_{\gamma = 1}^\Gamma w_\gamma H_\gamma$, via `Taylorization', that is, via a Taylor series:
\begin{equation}
\begin{split}
\label{Ur Taylorization phase estimation}
    &e^{-iHt/r} \approx \tilde{U}_r :=\sum_{k=0}^K \frac{(-iHt/r)^k}{k!} =\\
    &\sum_{k=0}^K \sum_{\gamma_1,...,\gamma_k =1}^\Gamma \frac{(-it/r)^k}{k!}  w_{\gamma_1}...w_{\gamma_k} H_{\gamma_1} .... H_{\gamma_k},
\end{split}
\end{equation}
with $K =  O\left( \frac{\log(r/\epsilon_{HS})}{\log \log(r/\epsilon_{HS})} \right)$. This means that in the Linear Combination of Unitaries formalism, we can write, $\tilde{U} = \sum_{j}\beta_j V_j$ with $\beta_j =  \frac{t^k}{r^k k!}w_{\gamma_1}... w_{\gamma_k}$ and $V_j = (-i)^k H_{\gamma_1} ... H_{\gamma_k}$.

Therefore we have to define how to implement Prepare($\beta$) and Select($V$), defined as 
\begin{subequations}
\begin{equation}
    \text{Prepare}(\beta)\ket{0}^{J}= \sqrt{\frac{1}{s}} \sum_j \sqrt{\beta_j} \ket{j}
\label{prepare beta}
\end{equation}
depicted in figure 1 of \cite{babbush2016exponentially}, and
\begin{equation}
    \text{Select}(V)\ket{j}\ket{\psi} = \ket{j}V_j \ket{\psi}.
\label{select V}
\end{equation}
\end{subequations}
These operators use Prepare($W$) and Select($H$) respectively:
\begin{subequations}
\begin{equation}
    \text{Prepare}(W) \ket{0}^{\otimes \lceil\log_2 \Gamma\rceil}= \sqrt{\frac{1}{\lambda}} \sum_{\gamma = 1}^\Gamma \sqrt{w_\gamma}\ket{\gamma}
\label{prepare W}
\end{equation}
with $\lambda = \sum_j |w_j| = O(N^4)$, and
\begin{equation}
    \text{Select}(H)\ket{\gamma}\ket{\psi} = \ket{\gamma}H_\gamma \ket{\psi},
\end{equation}
or in other words
\begin{equation}
    \text{Select}(H)\ket{ijkl}\ket{\psi} = \ket{ijkl}a_i^\dagger a_j^\dagger a_k a_l \ket{\psi}.\label{Select H babbush2016exponentially}
\end{equation}
\end{subequations}
To implement \eqref{Select H babbush2016exponentially} we have to transform the creation and annihilation operators according to eq. 20 and 21 from \cite{babbush2016exponentially}. This same article suggests introducing four additional qubits so that eq. 23 and 24 from \cite{babbush2016exponentially} are finally used, containing only controlled Pauli operators.

Using those operators, we define the quantum walk step implementing $\tilde{U}_r$ (figure 2 in \cite{babbush2016exponentially})
\begin{subequations}
\begin{equation}
\mathcal{W} = (\text{Prepare}(\beta) \otimes \bm{1})^\dagger\text{Select}(V)(\text{Prepare}(\beta)\otimes \bm{1})
\end{equation}
\begin{equation}
    \mathcal{W}\ket{0}^J \ket{\psi} = \frac{1}{s}\ket{0}\tilde{U}_r\ket{\psi} + \sqrt{1-\frac{1}{s^2}}\ket{\Phi}.
\end{equation}
\end{subequations}

To be able to use oblivious amplitude amplification, we need $s\approx 2$ \cite{berry2015simulating}, what can be achieved if $r = \lambda t/\ln 2$. Then $s = \sum_j |\beta_j|  = \sum_{k=0}^K \frac{1}{k!}\ln 2^k\approx 2$.

\subsubsection{`On-the-fly' algorithm}

The main difference with the `database algorithm' is that this algorithm aims to compute the integrals on-the-fly. 

One starts observing that the Hamiltonian is constant in time, but at the same time it can be expressed as a spatial integral over a given region $\mathcal{Z}$, given that it decays exponentially outside it

\begin{equation}
    H = \int_\mathcal{Z} \mathcal{H}(\vec{z}) d\vec{z}\approx \frac{\mathcal{V}}{\mu}\sum_{\rho = 1}^\mu \mathcal{H}(\vec{z}).
\end{equation}

As done in previous appendices, we divide the Hamiltonian evolution into segments $U_r$,
\begin{equation}
    U_r \approx \sum_{k=0}^K \frac{(-it/r)^k}{k!}\int_{\mathcal{Z}}\mathcal{H}(\vec{z}_1)... \mathcal{H}(\vec{z}_k) d\vec{\bm{z}}.
\end{equation}
If we substitute the integrals by Riemannian sums, $\mathcal{H}(\vec{z}) = \sum_{\gamma=1}^\Gamma w_\gamma(\vec{z}) H_\gamma$, 
\begin{equation}
\begin{split}
    &U_r \approx \sum_{k=0}^K \frac{(-it\mathcal{V})^k}{r^k \mu^k k!}\cdot\\
    &\cdot\sum_{\gamma_1,...,\gamma_k=1}^\Gamma \sum_{\rho_1,...,\rho_k=1}^\mu w_{\gamma_1}(\vec{z}_{\rho_1})...w_{\gamma_k}(\vec{z}_{\rho_k})H_{\gamma_1}...H_{\gamma_k}
\end{split}
\label{Riemannian sum Ur Taylorization}
\end{equation}

Now, the question is how to prepare $w_{\gamma_i}(\vec{z}_{\rho_i})$ in the amplitudes. What the article does is first assume we have a method sample($w$) such that
\begin{equation}
    \text{sample}(w)\ket{\gamma}\ket{\rho}\ket{0}^{\otimes \lceil\log_2 M\rceil} = \ket{\gamma}\ket{\rho}\ket{\tilde{w}_{\gamma}(\vec{z}_\rho)}
\end{equation}
with $\tilde{w}_{\gamma}(\vec{z}_\rho)$ an approximation of $w_{\gamma}(\vec{z}_\rho)$. Then the preparation procedure of the amplitudes consists of calculating the coefficients $w_{\gamma,m}(\vec{z}_\rho) \in \{\pm 1\}$ of a superposition such that $w_{\gamma}(\vec{z})\approx \zeta \sum_{m=1}^M w_{\gamma,m}(\vec{z})$; $\zeta =\Theta\left( \frac{\epsilon_H}{\Gamma \mathcal{V} t} \right)$. To do that, defining $\ket{l} = \ket{\gamma}\ket{m}\ket{\rho}$, one performs $Kickback$:
\begin{equation}
    \ket{l}\ket{\tilde{w}_{\gamma}(\vec{z}_\rho)}\rightarrow \left\{ \begin{array}{lcc}
\ket{l}\ket{\tilde{w}_{\gamma}(\vec{z}_\rho)}  & \tilde{w}_{\gamma}(\vec{z}_\rho) > (2m-M)\zeta \\
i\ket{l}\ket{\tilde{w}_{\gamma}(\vec{z}_\rho)} &  \tilde{w}_{\gamma}(\vec{z}_\rho) \leq (2m-M)\zeta  \end{array} \right.
\label{w gamma m amplitude preparation}
\end{equation}
before uncomputing sample$(w)$.

In summary, to prepare the amplitudes, one calculates sample($w$) in the basis, performs \eqref{w gamma m amplitude preparation} in a superposition of $\ket{m}$, and uncomputes the register prepared by sample($w$). We will call such procedure Prepare($w$):
\begin{equation}
    \text{Prepare}(w)\ket{0}^{\otimes \lceil\log_2 L\rceil} = \sqrt{\frac{1}{\lambda'}}\sum_{l = 1}^L \sqrt{\frac{\zeta\mathcal{V}}{\mu}w_{\gamma,m}(\vec{z}_\rho)}\ket{l},
\end{equation}
where $\lambda' = L \frac{\zeta\mathcal{V}}{\mu}= \Theta (\Gamma \mathcal{V}\max_{\vec{z},\gamma}|w_{\gamma}(\vec{z})|)$; $L = \Theta(\Gamma \mu M)$ and $M = \Theta (\max_{\vec{z},\gamma}|w_{\gamma}(\vec{z})| / \zeta)$. Additionally, due to equation 66 from \cite{babbush2016exponentially} we know that
\begin{equation}
    \mathcal{V}\max_{\vec{z},\gamma}(|w_{\gamma}(\vec{z})|) = 2^6 \varphi_{\max}^4 x_{\max}^5,
\label{Vol_max_w_gamma taylor_on-the-fly}
\end{equation}
where the $2^6$ is due to there being a hypercube with $(2x_{\max}/\delta x)^6$ terms.

This means that this alternative algorithm is similar to the `database' one, but substitutes Prepare($W$) with Prepare($w$) that we just explained. The preparation over $\ket{k}$ is similar to the one depicted in figure 1 of \cite{babbush2016exponentially}, except that $\lambda$ gets substituted by $\lambda'$.

The final, important detail we have to explain is how to perform the sample($w$) routine. We want to calculate 
\begin{subequations}
\begin{equation}
\begin{split}
    w_\gamma(\vec{z})&= h_{ijkl}(\vec{x},\vec{y})= \frac{\varphi_i^\dagger(\vec{x})\varphi_j^\dagger(\vec{y})\varphi_l(\vec{x})\varphi_k(\vec{y})}{|\vec{x}-\vec{y}|}\\
    &= \varphi_i^\dagger(\vec{x})\varphi_j^\dagger(\vec{x}-\vec{\xi})\varphi_l(\vec{x})\varphi_k(\vec{x}-\vec{\xi})|\vec{\xi}| \sin(\theta),
\end{split}
\end{equation}
with $\vec{\xi} = \vec{x}-\vec{y}$ and $\theta$ the polar angle of $\vec{\xi}$; as well as
\begin{equation}
\begin{split}
    w_\gamma(\vec{z})&= h_{ik}(\vec{x})\\
    &= \varphi_i^\dagger(\vec{x})\left(-\sum_{j=0,1,2}\frac{\nabla^2_j}{2} - \sum_{j=0,...,J} \frac{Z_j}{|\vec{R}_j -\vec{x}|}\right)\varphi_k(\vec{x})\\
    &= - \varphi_i^\dagger(\vec{x})\frac{\nabla^2}{2}\varphi_k(\vec{x})\\
    &- \sum_j Z_j|\vec{\xi_j}|\sin(\theta_j)\varphi_i^\dagger(\vec{R}_j-\vec{\xi}_j)\varphi_k(\vec{R}_j-\vec{\xi}_j)
\end{split}
\end{equation}
again transforming to polar coordinates in the external potential, $\vec{\xi_j} = \vec{R}_j - \vec{x}$.
\label{w_gamma(z) babbush2016exponentially}
\end{subequations}
We need a subroutine $Q$ to calculate the integrals. 
\begin{equation}
\begin{split}
    Q &= \prod_{j=1}^N \ket{j}\bra{j}\otimes Q_{\varphi_j},\\
    Q_{\varphi_j}&\ket{\rho}\ket{0}^{\otimes \lceil\log_2 M} = \ket{\rho }\ket{\varphi_j(\vec{z}_\rho)}.
\end{split}
\end{equation}
From the previous equation, one can see that the complexity of $Q$ is $N$ times the complexity of $Q_{\varphi_j}$. Notice that we will have to integrate over the space volume $\mathcal{V}$, summing over its discretization.

\subsection{\label{app:Taylorization_cost}How to compute its cost}

\subsubsection{`Database' algorithm}
We will use figure \ref{fig:Taylor_database} as the main guide to compute the cost of the different abstraction levels. The first thing we have to take is the simulation time required, fixed by the error in the Phase Estimation algorithm, $\epsilon_{QPE}$. One takes the number of segments $\tilde{U}_r$ to be
\begin{equation}
    r = \frac{\lambda t}{\ln 2} = \frac{\pi \lambda}{\epsilon_{QPE}\ln2}.
\end{equation}

Another important parameter is the value of $K$, that controls the number of Prepare($W$) in Prepare($\beta$) and Select($H$) in Select($V$), which we can take from \cite{low2018hamiltonian} to be
\begin{equation}
    K = \left\lceil -1 + \frac{2\log(2r/\epsilon_{HS})}{\log(\log(2r/\epsilon_{HS}) +1)}\right\rceil.
\end{equation}

The final aspects to take into account are:
\begin{enumerate}
    \item \textbf{$\theta_k$ initial rotations}. This can be done using $K-1$ controlled $R_y$ rotations.

    \item \textbf{Prepare(W)} The cost of an arbitrary state preparation for can be estimated as $2^{\lceil \log_2 N^4 \rceil+1}$ arbitrary rotations, using the protocol from \cite{shende2006synthesis}, as it is preferable to encode $\ket{ijkl}$ instead of a continuous register that later on gets converted to that. This will be the most expensive part of the algorithm.
    
    \item \textbf{Select(H)} First we have to specify how to create the circuit for each operator $a_{j,q}$ (analogously $a_{j,q}^\dagger$). For that we iterate over $n\in\{1,...,N\}$. If $j = n$ we apply a $\sigma_x$ or $\pm i\sigma_y$ as dictated by $\ket{q}$, if $j < n$ then we apply $\sigma_z$.
    
    The equality case can be performed via multi-controller Pauli operators. For each creation/annihilation operator, there will be $4N$ options due to the possible values of $\ket{j}\ket{q}$. We have to control on one qubit of register $\ket{k}$ encoded in unary to take into account the amplitude term corresponding to $\frac{(t/r)^k}{k!}$, on $\ket{j}$ with $\lceil \log_2 N\rceil$ qubits, and on $\ket{q}$; we will need to resort to multi controlled gate decomposition.
    
    To avoid the comparison in the case of $n<j$ we can create an accumulator. That is, when $n = j$ we switch an ancilla from $\ket{1}\rightarrow\ket{0}$, and controlled on such ancilla (and the unary register $\ket{k}$), at each step we perform $\sigma_z$ on the $n-$th register of $\ket{\psi}$. This means $N$ Toffolis and $N$ multi-controlled (on $\lceil \log_2 N\rceil$ qubits) Not gates due to the equality comparison.

\end{enumerate}

\subsubsection{`On-the-fly' algorithm}

To compute the cost of the `on-the-fly' variation of this algorithm, the key step is substituting the Prepare($W$) operator by something less expensive. The way we do this is by computing the one and two body integrals on the fly, by creating a sign-weighted superposition in register $\ket{\rho}$. Such superposition will use $\lceil \log_2 \mu \rceil$ qubits and can take values from $0$ to $\mu-1$ where
\begin{equation}
\begin{split}
    \mu &\approx \left(\frac{2r\times 6K}{\epsilon_{H}}(4\varphi'_{\max} + \varphi_{\max}/x_{\max}) \varphi_{\max}^3 x_{\max}^6\right)^6\\
    &=\Theta \left(\left(\frac{N^4t}{\epsilon_{H}}(\varphi'_{\max} + \varphi_{\max}/x_{\max}) \varphi_{\max}^3 x_{\max}^6\right)^6\right)
    \label{mu definition babbush2016exponentially}
\end{split}
\end{equation}
as can be seen from equations 73 and 74, and the text in the paragraph before equation 61, from \cite{babbush2016exponentially}.
Although this is a large number, it will only appear logarithmically in the number of qubits in the $\ket{\rho}$ register as explained in \eqref{Riemannian sum Ur Taylorization}, so does not represent a too large complexity overhead. Notice that from equation 60 in \cite{babbush2016exponentially}, $r= \frac{\lambda' t}{\ln 2} = \frac{t}{\ln 2}\Gamma\mathcal{V}\max_{\vec{z},\gamma}(|w_{\gamma}(\vec{z})|)$, and the factor of $4$ in front of $\varphi'_{\max}$ appear because we were deriving $\varphi_{\max}^4$; whereas the $2$ appears because if we assume a hypercube, there should be $(2x_{\max}/\delta x_{\max})^6$ blocks in the discretization. Additionally, we can choose the coordinate system centered around the orbital such that $x_{\max} = O(\log(Nt/\epsilon_H)) = C \log(Nt/\epsilon_H)$, $C$ a constant given by the software package users. $\varphi_{\max}$ will not depend on $N$.
Similarly, since $\zeta$ is $\epsilon_H$ divided by the number of integral terms calculated in the process,
\begin{equation}
    M = \frac{\max_{\vec{z},\gamma}(|w_{\gamma}(\vec{z})|)}{\zeta} = \frac{6Kr\Gamma \mathcal{V}\max_{\vec{z},\gamma}(|w_{\gamma}(\vec{z})|)}{\epsilon_H},
\end{equation}
 where we can use the expressions from \eqref{Vol_max_w_gamma taylor_on-the-fly}.

The final contribution we should take into account is that of the arithmetic operations required to calculate $\varphi_j(\vec{z}_\rho)$, which will also depend on the basis function we are using. 

For that we will be using quantum addition \cite{gidney2018halving}, multiplication \cite{munoz2017t} and integer division \cite{thapliyal2017quantum}. The respective T-gate costs are $4n+O(1)$, $21n^2-14$ and $14n^2+7n+7$, where $n$ is the number of digits, $n = \lceil\log_2\mu\rceil/3$, as there are three coordinates. Additionally, performing comparison between two numbers \cite{cuccaro2004new} can be done using $2n$ Toffoli gates if each of the inputs to compare is length $n$, so $8n$ T-gates.

To calculate the number of operations needed, we have to first remember that we are using a Gaussian basis set. In such basis, we expand the wave function as $\phi = \sum_{i=1}^M c_i \chi_i$. Each $\chi_j(x,y,z) = (x-X)^k(y-Y)^l(z-Z)^m e^{-\zeta_i (\mathbf{r}-\mathbf{R})^2}$, where $(X,Y,Z)$ indicate the center of the atom, and $k+l+m$ is the angular momentum (eg. $k+l+m = 1$ means p-type basis etc. We assume that we only use up to $d$ basis). The orbitals are usually contracted $\kappa_j = \sum_{i=1}^d d_{ij} \chi_i$ and $\phi = \sum_{j=1}^N c_j \kappa_j$. Each $\kappa_j$ is one of the $N$ basis functions that we use. More information on the topic of Gaussian basis sets might be found in a recent review \cite{jensen2013atomic}.

In any case, to calculate each basis function $\kappa_j=\varphi_j$ we have to do the following:
\begin{enumerate}
    \item Calculate $(x-X)$, $(y-Y)$, and $(z-Z)$, using $12n+O(1)$ T gates.
    \item Calculate $(\mathbf{r}-\mathbf{R})^2 = (x-X)^2 + (y-Y)^2 + (z-Z)^2$, with cost $3(21n^2-14)$ for the multiplications, that is the leading cost. The sums mean $8n+O(1)$ additional cost.
    \item Calculate the exponential $\zeta_i (\mathbf{r}-\mathbf{R})^2$ with a single multiplication, at T-gate cost $(21n^2-14)$. \item $e^{-\zeta_i(\mathbf{r}-\mathbf{R})^2}$ via a Taylor series. Expanding to order $o$ means $o-1$ multiplications and divisions, and $o$ sums. 
    \item The error in the previous expansion can be bounded as $\max \left(\zeta_i(\mathbf{r}-\mathbf{R})^2\right)^o/o!$
    \item To construct $\chi_j(x,y,z)$ we need 3 multiplications, so the cost is  $\approx 3(21n^2-14)$.
    \item Each $\kappa_j$ will be a sum of weighted exponentials, so the previous cost should be multiplied by $d$, the number of terms in such sum. 
\end{enumerate}

The number of terms $d$ in each $\kappa_j$ depends on the basis used, but it can be seen in tables 1-4 from \cite{jensen2013atomic} that the number of primitive basis sets $\chi_i$ that form each $\kappa_j$ does not exceed 6 functions in the case of segmented basis sets (sparse $d_{ij}$), so we will take $d = 6$. However, if the basis set is general-contracted, $d_{ij}$ is dense and the number might be much greater.

Once we have computed $\kappa_j = \varphi_j(\vec{x})$, we want to compute $\tilde{w}_{\gamma}(\vec{z})$:
\begin{itemize}
    \item Whenever we have to compute $\vec{\xi}_j$ or $\vec{\xi}_j$, the cost is $12n + O(1)$ T-gates.
    \item Performing $\mathcal{R}\ket{\vec{\xi}}\ket{0}\mapsto
    \ket{\vec{\xi}}\ket{|\vec{\xi}|\sin\theta}$, and similarly for $\vec{\xi}_j$. To do that, observe that $|\vec{\xi}|\sin\theta = \sqrt{\vec{x}_x^2 + \vec{x}_y^2}$, so we need two multiplications at cost $2(21n^2-14)$, one sum at T-gate cost $4n + O(1)$, and a square root calculation. We compute the square root using the Babylonian method, which only involves a sum and a division per order.
    \item $\nabla^2 \chi_k(x) = \left(4x^2 -2 + 4k - (1+k)/x^2\right)\chi_k(x)$. If we call the parenthesis $a_k(x)$, then $\nabla^2\chi_{ijk}(x,y,z) = (a_i(x)+a_j(y)+a_k(z))\chi_{ijk}$. Computing $a_i(x)$ can be done using $4$ sums, 1 multiplication ($x^2$ term) and 1 division. This is because multiplying by 4 is free, just shifting bit positions. This has to be multiplied by 3 to take into account the three coordinates in the Laplacian, and done before the combination of the $d$ functions into a single $\kappa_j = \varphi_j$.
\end{itemize}
In a similar fashion can $Q_\Delta$ be computed, for the sake of a name for outputting $\nabla^2 \varphi$. 

Overall, the cost of Sample($w$) is 
\begin{itemize}
    \item Two-body term: 4Q +$\mathcal{R}$ +  4 multiplication + computation of $\vec{\xi}$.

    \item Kinetic term: $Q + Q_\Delta$ + multiplication.
    
    \item External potential term: 2Q +$J\times\mathcal{R}$ + $J$ multiplications by $Z_j$ and $J-1$ sums + $J$ computations of $\vec{\xi}_j$.
\end{itemize}
Remember that in the previous calculations we are taking $n = \lceil \log_2 \mu \rceil/3$. 

The cost of the rotation $Kickback$ between the two applications of Sample($w$) can be seen as a controlled rotation on the result of a comparison with $\lceil \log_2 \mu \rceil$ qubits. This requires one sum, one multiplication, and one comparison, which should be done twice to uncompute the result once the rotation has happened.  From the previous, the cost of the `on-the-fly' version of algorithm \cite{babbush2016exponentially} can be computed using figure \ref{fig:Taylor_database}.

\subsection{\label{app:Taylorization_adaptation}How to adapt the Hamiltonian simulation to control the direction of the time evolution}

Quantum Phase Estimation requires being able to control the time direction of the Hamiltonian evolution of a segment. We do that by slightly modifying the Select($V$) operator: if we want to simulate $e^{-iHt/r}$, for $k = 4j + 1$  we apply a C-S$^\dagger$ operation (to apply $-i$ phase) and C-S if $k = 4j+3$, while if we instead want to simulate $e^{iHt/r}$ additionally apply C-X in those situations to flip the sign. Here the Control bits are the value of $k$ and the control qubits in Quantum Phase Estimation.

Adapting the Hamiltonian simulation method for Phase Estimation operation then amounts to two multi-controlled Not gates, with $K/2+1$ controls because $k$ is encoded in unary and we are using Bayesian Phase Estimation with a single control ancilla.

\section{\label{app:Configuration_interation} Configuration interaction and first quantization}

\subsection{\label{app:Configuration_interaction_method}Method explanation}

In the previous section, we saw how to use Taylorization as a Hamiltonian simulation method in second quantization. Here, we explain the approach of \cite{babbush2017exponentially}, which relies on the same approach but in first quantization, in a formulation called Configuration Interaction. The general structure of the algorithm will consequently be similar.

In the Configuration Interaction representation one writes $\ket{\alpha} = \ket{\alpha_0,...,\alpha_{\eta-1}}$, where each $\alpha_i$ indicates an occupied orbital. The determinant of the corresponding wave functions is an antisymmetric function called Slater determinant and represents the state of the system

\begin{equation}
\begin{split}
    &\braket{\vec{r}_0,...\vec{r}_{\eta-1}|\alpha} =\\ &\frac{1}{\sqrt{\eta!}}\left|
    \begin{pmatrix}
    \varphi_{\alpha_0}(\vec{r}_0) & \cdots & \varphi_{\alpha_{\eta-1}}(\vec{r}_0)\\
 \vdots& &\vdots\\
     \varphi_{\alpha_{0}}(\vec{r}_{\eta-1}) & \cdots & \varphi_{\alpha_{\eta-1}}(\vec{r}_{\eta-1})
    \end{pmatrix}
    \right|.
\end{split}
\end{equation}

An important aspect of this method is that it can only be applied with local basis functions, such as Gaussian orbitals, but not the plane-wave basis. The reason is that at one point one has to bound the error by approximating Hamiltonian integrals from Riemannian sums, and bounding the error is only possible if we are restricted to a local volume of space. To make it work with molecular orbitals appearing in the Hartree-Fock procedure, one can use the operator $U = \exp{\left(-\sum_{ij}\kappa_{ij}a_i^\dagger a_j\right)}$ that changes the basis and may be applied using $\tilde{O}(N^2)$ gates \cite{wecker2015solving}. $\kappa$ here is an antihermitian matrix that is obtained by the self-consistent Hartree Fock procedure.

Expressing the Configuration Interaction Hamiltonian as a linear combination of unitaries is not efficient. On the other hand, though, it can be expressed as a sparse matrix, called Configuration Interaction (CI), whose elements are a sum of integrals. 

The Slater-Condon rules indicate how to compute those matrix elements, based on one- and two-body integrals \cite{babbush2017exponentially}. Because of them, the sparsity of the Configuration Interaction matrix is
\begin{equation}
\begin{split}
    d &= \binom{\eta}{2}\binom{N-\eta}{2} + \binom{\eta}{1}\binom{N-\eta}{1} + 1\\
    &= \frac{\eta^4}{4}- \frac{\eta^3N}{2} + \frac{\eta^2N^2}{2} + O(\eta^2 N + \eta N^2)\in O(\eta^2N^2).
\end{split}
\end{equation}

After decomposing the Configuration Interaction matrix in 1-sparse operators, we approximate its integrals as a Riemannian sum of self inverse operators. Finally, we construct  $\text{Select}(\mathcal{H})$, that applies such self inverse operators
\begin{equation}
    \text{Select}(\mathcal{H})\ket{l}\ket{\rho}\ket{\psi} =  \ket{l}\ket{\rho}\mathcal{H}_{l,\rho}\ket{\psi} 
    \label{Select CI}
\end{equation}
and allows to evolve the system under the Hamiltonian. The steps are the following:

\begin{figure}[htp]
    \centering
    \includegraphics[width=.5\textwidth]{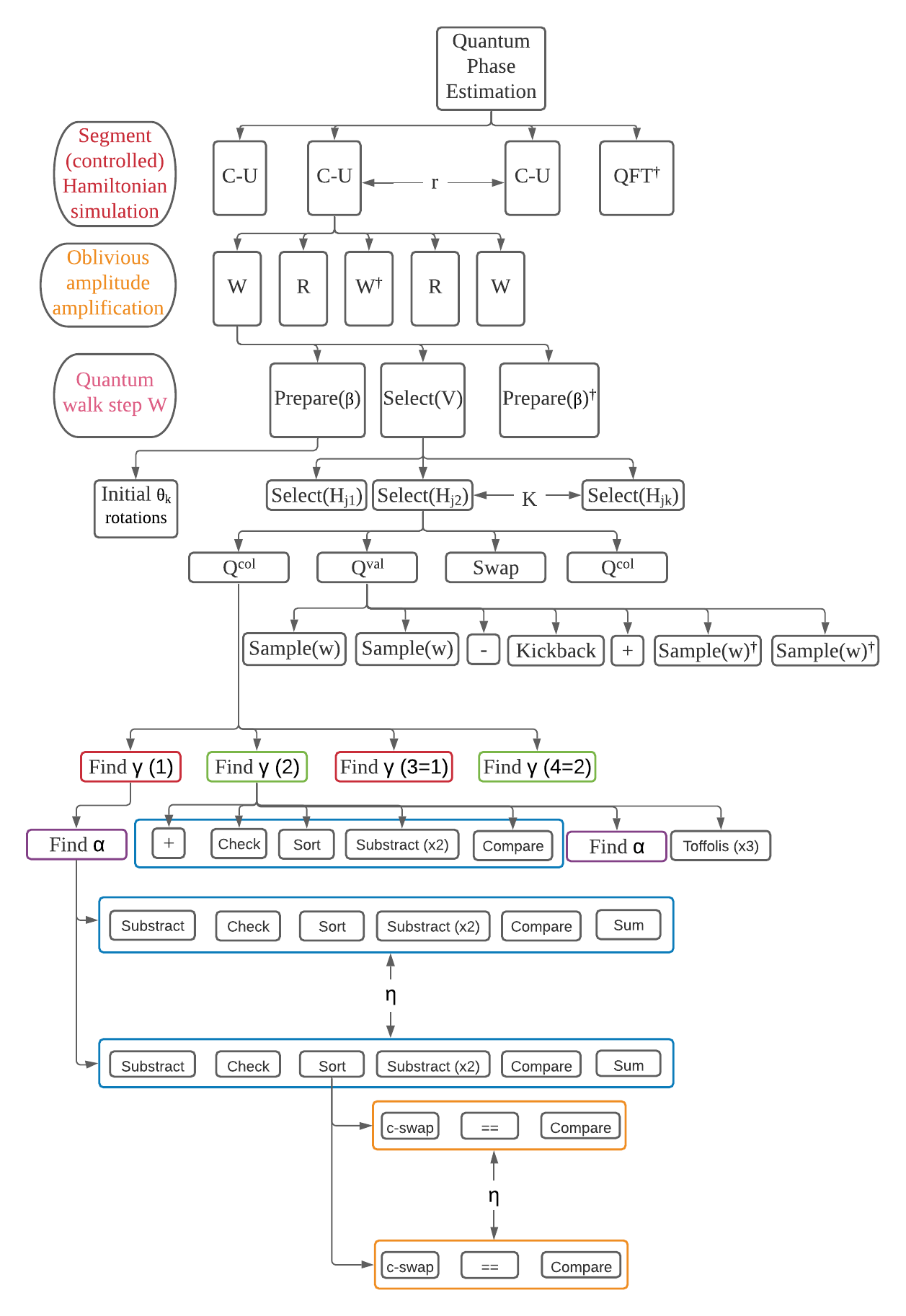}
    \caption{Abstraction level decomposition of the Configuration Interaction procedure \cite{babbush2017exponentially}. The Sample operation shown is the same as in figure \ref{fig:Taylor_database}.}
    \label{fig:Configuration_interaction}
\end{figure}

\begin{enumerate}
\item \textbf{Decompose the  Hamiltonian into 1-sparse operators}. Such operators will be indexed by 2 4-tuples $(a_1,b_1, i,p)$ and $(a_2,b_2,j,q)$ that denote the differing orbitals. This tuples will be used to perform the operator 
    \begin{equation}
        Q^{col}: \ket{\gamma}\ket{\alpha}\ket{0}^{\eta\lceil \log_2 N \rceil}\mapsto \ket{\gamma}\ket{\alpha}\ket{\beta},
        \label{Qcol}
    \end{equation}
    within the Select operator \eqref{Select CI}.
    The specific algorithms for this procedure can be found in appendix A of the article of reference for this appendix \cite{babbush2017exponentially}. These procedures require, between other things, the ability to order a list of orbitals, which we explain in Algorithm \ref{AlgorithmSort}. 
    
    \item \textbf{Decompose each 1-sparse operator into $h_{ij}$ and $h_{ijkl}$}. The Slater Condon rules sometimes requires the sum over $\eta$ integrals. Here we decompose the previous sum such that only at most two integrals are summed for each term. This decomposition can be seen in section 4.2 of the original article \cite{babbush2017exponentially}. It will allow us to write the Hamiltonian as $H = \sum_{\gamma} H_\gamma$, with $\Gamma = \eta +\eta(\eta-1)/2 + (N-1)\eta^2 + (N-1)^2\eta(\eta-1)/2$.
    
     \item \textbf{Discretising the integrals into Riemannian sums}.
    
    Each Hamiltonian term from the previous equation might be represented as $H_{\gamma}^{\alpha\beta} = \int \aleph^{\alpha\beta}_\gamma (\vec{z}) d\vec{z}$. Since the domain of each integral might be different, we write
    $H^{\alpha\beta}_{\gamma}\approx \sum_{\rho=1}^\mu \aleph^{\alpha\beta}_{\gamma \rho}$.
    Here is where we need the requirement that the orbitals are local.

\item \textbf{Decomposition into self-inverse operators}. Finally, we decompose in a sum of $M\in \Theta(\max_{\gamma,\rho}||\aleph_{\gamma,\rho}||_{\max}/\zeta)$ self-inverse operators, using a similar strategy as in the previous section \ref{app:Taylorization} \cite{babbush2016exponentially}. Operators will be indexed by $\rho$ and $l = (\gamma,m,s)$, where $m$ controls whether a phase $i$ is added in the Kickback, and $s$ is sign. $\rho$ controls the Riemmanian sum. The final decomposition can be written as $H = \zeta \sum_{l = 1}^L \sum_{\rho= 1}^\mu \mathcal{H}_{l,\rho}$. Using this we can perform
    \begin{equation}
        Q^{val}\ket{l}\ket{\rho}\ket{\alpha}\ket{\beta} = \mathcal{H}^{\alpha\beta}_{l,\rho}\ket{l}\ket{\rho}\ket{\alpha}\ket{\beta},
        \label{Qval}
    \end{equation}
which also appears in the Select operator.
\end{enumerate}
In conclusion, one time segment of the Taylorized Hamiltonian evolution will be 
\begin{equation}
    U_r\approx \sum_{k=0}^K \frac{(-it\zeta)^k}{r^k k!}\sum_{l_1,...,l_k = 0}^L \sum_{\rho_1,...,\rho_k = 0}^\mu \mathcal{H}_{l_1,\rho_1}...\mathcal{H}_{l_k,\rho_k},
\end{equation}
where $\ket{l} = \ket{\gamma, m, s}$. The role of Prepare will be restricted to the preparation of $\theta$ angles for $\frac{(-it\zeta)^k}{r^k k!}$.

To compute the algorithm cost, we will need constants $\alpha$, $\gamma_1$ and $\gamma_2$ to comply with equations 28, 29 and 30 from \cite{babbush2017exponentially}, and will bound the error from computing the Hamiltonian integrals as Riemannian sums:
\begin{itemize}
\item For each $l$ there is a vector $c_l$ such that if $||\vec{r}-\vec{c}_l||\geq x_{\max}$ then
        \begin{equation}
        |\varphi_l(\vec{r})|\leq \varphi_{\max}\exp\left(-\frac{\alpha}{x_{\max}}||\vec{r}-\vec{c}_l||\right) \label{alpha definition babbush2017exponentially}
        \end{equation}
        
        \item For each $l$, $\varphi_l$ is twice differentiable and there exists $\gamma_1$ and $\gamma_2$ such that
        \begin{subequations}
        \label{gammas definition babbush2017exponentially}
        \begin{equation}
            ||\nabla \varphi_l(\vec{r})||\leq \gamma_1 \frac{\varphi_{\max}}{x_{\max}}
        \end{equation}
        and
        \begin{equation}
            ||\nabla^2 \varphi_l(\vec{r})||\leq \gamma_2 \frac{\varphi_{\max}}{x^2_{\max}}
        \end{equation}
        \end{subequations}
\end{itemize}

\subsection{\label{app:Configuration_interation_cost}How to compute its cost}

We will use figure \ref{fig:Configuration_interaction} as a guide to computing the cost of the algorithm.
There are three key differences with the cost calculated in the previous appendix. First, some parameters change. These are notably $r$, the number of time segments, and $M$, which indicates the size of register $\ket{m}$ and as a consequence influences the cost. The other two aspects that change are that we need to compute the cost of $Q^{val}$ and $Q^{col}$ in figure \ref{fig:Configuration_interaction}.

Let us start computing $r$, the number of segments. $r = \zeta L \mu t / \ln(2)$ (according to the paragraph before equation 68 in \cite{babbush2017exponentially}), with $L = 2(M\Gamma)$ (the $2$ because of register $s$ in $\ket{l} = \ket{\gamma}\ket{m}\ket{s}$). The product $\mu \max_{\gamma,\rho}||\aleph_{\rho,\gamma}|| = \mu M \zeta$ can optimized from Lemmas 1-3 in the original article \cite{babbush2017exponentially}, so
\begin{equation}
    r = 2\Gamma t (\mu M \zeta)/\ln(2),
\end{equation}
with $t = \pi/\epsilon_{QPE}$ and 
\begin{equation}
\begin{split}
    \Gamma &= \binom{\eta}{2}\binom{N-\eta}{2} + \binom{\eta}{1}\binom{N-\eta}{1} + 1\\
    &= \frac{\eta^4}{4}- \frac{\eta^3N}{2} + \frac{\eta^2N^2}{2} + O(\eta^2 N + \eta N^2)\in O(\eta^2N^2).
\end{split}
\end{equation}
To compute $M$, similarly as in the previous appendix
\begin{equation}
    M = \Theta \left(\frac{\max_{\gamma,\rho}||\aleph_{\rho,\gamma}||}{\zeta}\right),
\end{equation}
and in the previous appendix we saw that $\zeta$ is the error that we allow, modelled as the error budget for this error source $\epsilon_H$, divided by the number of times we called the decomposition, $\Gamma\mathcal{V}r$. The reason why $\mathcal{V}$ appeared in place of $\mu$ is because instead of writing 
\begin{subequations}
\begin{equation} 
H_\gamma = \sum_{\rho} w_{\gamma}(\vec{z}_\rho) 
\end{equation}
we were taking
\begin{equation} 
H_\gamma = \frac{\mathcal{V}}{\mu}\sum_{\rho} w_{\gamma}(\vec{z}_\rho ),
\end{equation}
so the precision must be scaled correspondingly.
\end{subequations}
In this case however,
\begin{equation} 
H_\gamma = \sum_{\rho} \aleph_{\gamma}(\vec{z}_\rho),
\end{equation}
integrating the cell volume as a multiplicative constant in $\aleph_{\gamma}(\vec{z}_\rho)$, so the error has to be appropriately scaled by $\mathcal{V}/\mu$.
Similarly, this time,
\begin{equation}
    \zeta = \frac{\epsilon_H}{3\cdot 2Kr(\#\gamma) (\#\rho)} = \frac{\epsilon_H}{6Kr\Gamma \mu}.
\end{equation}
Since $\max_{\gamma,\rho}||\aleph_{\rho,\gamma}||$ is bounded from Lemmas 1, 2 and 3 in \cite{babbush2017exponentially}, we can compute $M$.
These lemmas will also depend on $\delta$, taken to be the individual error in each of the integrals. Therefore, we should take (see paragraph before eq. 74 in \cite{babbush2017exponentially}):
\begin{equation}
    \delta = \frac{\epsilon_H}{6K r},\quad \zeta = \frac{\delta}{\Gamma \mu}
\end{equation}
where $6K$ is the number of times these integrals are used in each segment, indicated figure \ref{fig:Configuration_interaction}. We can see that $\delta$ depends on $r$, which depends on $\mu M \zeta$, which from the previously mentioned lemmas depends on $\delta$. We solve this by computing $r$ such that $\mu$ times equations 39, 43 and 47 in \cite{babbush2017exponentially} become approximate equalities to $\mu M \zeta$. This way we obtain a close result to if we had used $\delta = \epsilon_H / (6K\Gamma t)$.

Now let us turn to two main operators involved in the algorithm, $Q^{val}$ in \eqref{Qval} and $Q^{col}$ in \eqref{Qcol}.
To compute the cost of $Q^{val}$ the procedure is the same as we did in the previous appendix \ref{app:Taylorization}. In this case, however, we will have to compute up to 2 basis functions. To do so we iterate over the different possibilities of $\gamma$ to decompose in $h_{ij}$ and $h_{ijkl}$.
\begin{enumerate}[label=\alph*.]
    \item $p=0=q$. This point requires calculating $\eta$ terms of type $h_{\chi_i \chi_i}$, and $\eta(\eta-1)/2$ terms $(h_{\chi_i \chi_j \chi_i \chi_j} - h_{\chi_i \chi_j \chi_j \chi_i})$.
    \item $p=0$, $q \neq 0$. In this case there are $(N-1)\eta(\eta-1)$. terms of the form $h_{k\chi_i l \chi_i} - h_{k\chi_i \chi_i l}$, and $(N-1)\eta$ for the terms of the form $h_{kl}$.
    \item $p\neq0$, $q = 0$. No integrals are needed.
    \item $p\neq 0$, $q \neq 0$. All of the integrals in this last point are of the form $h_{ijkl}-h_{ijlk}$. There are $(N-1)^2\eta(\eta-1)/2$ of them.
\end{enumerate}
From this and the previous appendix \ref{app:Taylorization}, the cost of $Q^{val}$ can be readily calculated.

Computing $Q^{col}$ requires implementing the procedure `Find Alphas' and a more general one indicated in cases 2 and 4 in appendix A, that we will call `Find Gammas' \cite{babbush2017exponentially}. Both `Find Alphas' and `Find Gammas' require a sorting algorithm that has the peculiarity that only up to one item might be out of order, and we know its position. For that reason, we have described a possible sorting algorithm \ref{AlgorithmSort}. To compute the cost, one should also make use of the basic operations described in table \ref{tab:arithmetic}.

\begin{algorithm}
\begin{algorithmic}[1]
\Procedure{Order}{$\ket{\beta}\ket{p}\ket{j}$}%\Comment{The g.c.d. of a and b}
    \State Calculate unordered $\ket{\tilde{\alpha}}_1$ subtracting $\ket{p}$ from $\ket{\beta_j}$.
    \State Use Cnots to create two 'basis' copies of $\ket{\tilde{\alpha}}_1$, called $\ket{\tilde{\alpha}}_1$ and $\ket{\tilde{\alpha}}_2$
    \For{$i\in \text{reversed}(\text{range}(j))$}%\Comment{We have the answer if r is 0}
        \If $\ket{\tilde{\alpha}_i}_1 == \ket{\tilde{\alpha}_{i+1}}_1$ then
            \State\textbf{return} Invalid \Comment{If this is activated, reverse the entire computation. Thus cost $\times 2$.}
        \EndIf
        \State $\ket{0}_a\gets (\ket{\tilde{\alpha}_i}_1 > \ket{\tilde{\alpha}_{j}}_1$)
        \State Controlled on $\ket{\cdot}_a$ swap $\ket{\tilde{\alpha}_i}_2$ and  $\ket{\tilde{\alpha}_{i+1}}_2$
        \State Uncompute $\ket{\cdot}_a$
    \EndFor
    \State Uncompute $\ket{\tilde{\alpha}}_1$
    \State \textbf{return} $\ket{\beta}\ket{p}\ket{j}\ket{\tilde{\alpha}}_2$ %\Comment{Complexity = 2 $\eta$ comparison and c-swaps}
\EndProcedure
\end{algorithmic}
\caption{Algorithm to order the orbitals $\ket{\tilde{\alpha}}$ generated from $\ket{\beta}$, shift $\ket{p}$ and position $\ket{j}$}\label{AlgorithmSort}
\end{algorithm}

Using this and figure \ref{fig:Configuration_interaction} it is relatively straightforward to compute the cost of the present algorithm. Notice however that the initial Hartree-Fock rotation $U = \exp\left(-\sum_{ij}\kappa_{kj}a_i^\dagger a_j\right)$ has not yet been implemented in the cost estimation, but it is not a dominant factor. 

\subsection{\label{app:CI_adaptation}How to adapt the Hamiltonian simulation to control the direction of the time evolution}

Adapting the Hamiltonian simulation for its use in Quantum Phase Estimation can be done as in appendix \ref{app:Taylorization_adaptation}. The cost can be therefore calculated in the same way.

\section{\label{app:QROM}Introducing the QROM}
\subsection{\label{app:QROM_method}Method explanation}
\begin{figure}[htp]
    \centering
    \includegraphics[width=.5\textwidth]{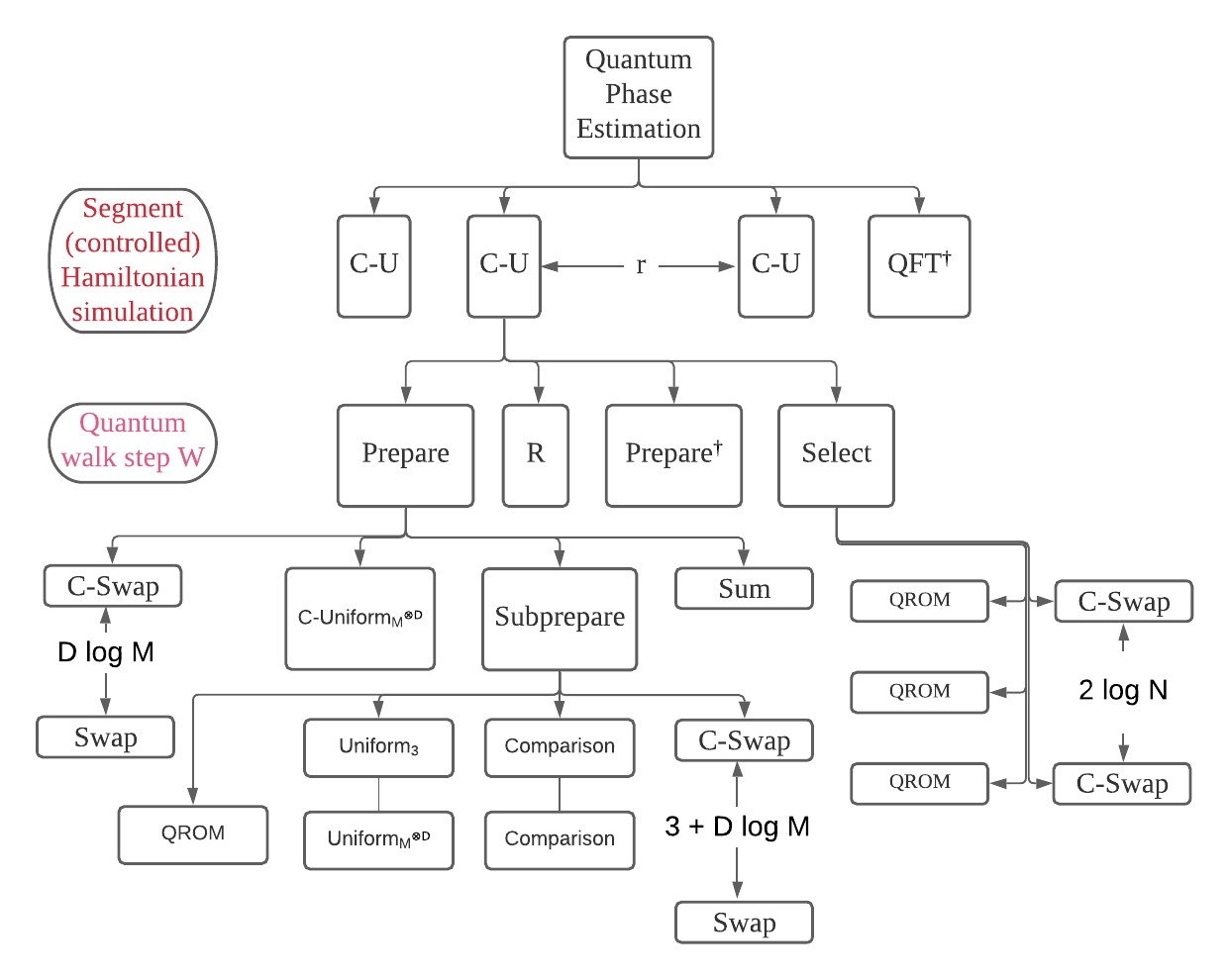}
    \caption{Abstraction level decomposition of the procedure \cite{babbush2018encoding}.}
    \label{fig:QROM}
\end{figure}

One of the key innovations used in this method is that if instead of simulating $\mathcal{W}(H)=e^{\pm iH\tau}$ one chooses $\mathcal{W}(H)=e^{\pm i\arccos(H/\lambda)}$, one can eliminate the Taylor series error completely \cite{babbush2018encoding}, as we already explained in section \ref{ssec:Qubitization}. This idea had been previously introduced \cite{poulin2018quantum,berry2018improved}, and has the consequence that instead of phase estimating the ground state energy $E_0$ one phase estimates $\arccos(E_0)$. 
We can simulate $\mathcal{W}(H)$ with the standard quantum walk $(\text{Prepare}^\dagger\otimes \bm{1}) \text{Select}(\text{Prepare}^\dagger\otimes \bm{1})$. Notice that in contrast to \cite{su2021fault} we are using $\arccos$ instead of $\arcsin$ because, since $\arccos\theta + \arcsin\theta = \pi/2$, the change amounts to a global phase and sign change, and we want to use similar notation everywhere.

Therefore, in this appendix, we aim to explain the implementations of the Prepare and Select operators, and the key innovation of article \cite{babbush2018encoding}, the proposal of an efficient QROM that will play an important role in both Prepare and Select. We will start with the latter. The role of the QROM is to iterate over all possible inputs preparing the corresponding outputs.

How can we construct such a unary iterator? The easiest way is just to use as control all the index qubits for each of the $L$ values the indices can take. But this is clearly wasteful since we are often repeating the same controls over consecutive values in the indices. Therefore, \cite{babbush2018encoding} proposes using auxiliary qubits to hierarchically save the combinations of controls, giving rise to circuits similar to their figure 5, called the "sawtooth" circuit. This circuit, in contrast to the original, can be simplified avoiding the wasteful repetition of AND gates that we indicated previously. As shown in their figure 6, allows for converting their figure 5 to their figure 7, requiring only $(L-1)$ AND gates. Since each AND can be constructed from 4 T gates, the unary iterator requires $4L-4$ T gates.

A variation over the previous iterator is the accumulator. Instead of directly applying the chosen gates to the target qubits, one defines an accumulator qubit, which is at state $\ket{0}$ until we control on the selected value of the indices and stays $\ket{1}$ until the end of the iterator, at which point it can be uncomputed since at the end the accumulator will be at disentangled state $\ket{1}$. A picture of this variant is figure 8 in \cite{babbush2018encoding}. This accumulator is specially useful because it will allow us to apply the Majorana fermion operator $\ket{l}\ket{\psi}\rightarrow \ket{l}\left(\frac{a_l
^\dagger - a_l}{i}\ket{\psi} \right) = \ket{l} Y_l Z_{l-1}...Z_0 \ket{\psi}$, as can be seen in figure 9 in \cite{babbush2018encoding}.

This QROM is useful to perform the Prepare circuit. However, we will not prepare
\begin{equation}
    \ket{0}^{\lceil\log_2\Gamma\rceil}\mapsto \sum_{ \gamma = 0}^{\Gamma-1} \sqrt{\frac{w_\gamma}{\lambda}},
\end{equation}
but rather
\begin{equation}
    \ket{\mathcal{L}} = \sum_{ \gamma = 0}^{\Gamma-1} \sqrt{\frac{w_\gamma}{\lambda}}\ket{\gamma}\ket{\text{temp}_\gamma},
\end{equation}
with $\ket{\text{temp}_\gamma}$ a junk register entangled with $\ket{\gamma}$. The way to ensure that this entanglement does not interfere with other computations is to ensure that the same qubits are fed into the uncomputation and that the reflection $\mathcal{R_L} = (2\ket{\mathcal{L}}\bra{\mathcal{L}} - \bm{1})$ that appears in the quantum walk step $\mathcal{W} = \mathcal{R_L}\cdot \text{Select}$ is done only over state $\ket{0}$. Here, we will be looking for an algorithm that performs the following transformation:
\begin{equation}
    \ket{0}^{\otimes (1 + 2\mu + 2\lceil\log_2\Gamma\rceil)} \rightarrow \sum_{\gamma}^{\Gamma-1}\sqrt{\tilde{\rho}_l}\ket{\gamma}\ket{\text{temp}_\gamma},
\end{equation}
with $\tilde{\rho}_\gamma$ a $\mu$-bits binary approximation to $w_\gamma/\lambda$. For this, one chooses 
\begin{equation}
\begin{split}
\mu = \biggl\lceil  \log_2 \left( \frac{2\sqrt{2}\lambda}{\Delta E}\right) &+ \log_2 \left(1 + \frac{\Delta E^2}{8\lambda^2}\right) \\& - \log_2 \left(1-\frac{||H||}{\lambda} \right)\biggr\rceil .
\end{split}
\label{mu babbush2018}
\end{equation} 
as given in equation 36 from \cite{babbush2018encoding}. Since the Hamiltonian is frustrated, the quotient in the last logarithm is upper bounded away from 1, and thus the last term is $O(1)$. Similarly, since $\Delta E < \lambda$, the second term can be upper bounded by $\log_2 (1 + 1/8)$.

We will prepare this new $\ket{\mathcal{L}}$ indirectly, using a circuit that they depicted in figure 11 and called Subprepare. We start from the uniform superposition $\sum \ket{\gamma}$ and have two registers that depend on $\gamma$, $\ket{\text{keep}_\gamma}$ and $\ket{\text{alt}_\gamma}$. $\ket{\text{keep}_\gamma}$ will dictate the probability that we coherently exchange $\ket{\gamma}$ and $\ket{\text{alt}_\gamma}$. The objective is to find keep$_\gamma$ and alt$_\gamma$ such that in the end, we obtain the correct amplitudes. The details of the procedure can be found in section 3D in the main reference for this appendix, and it is the inverse procedure of the depicted one in their figure 13 \cite{babbush2018encoding}.

The Hamiltonian basis explored in this technique is plane waves, with the same structure that we saw in eq. \eqref{JW Dual plane wave Hamiltonian} \cite{babbush2018encoding}. The article suggests that to make the basis set as compact as possible, one may choose Gausslet basis sets, that combine some of the features of plane waves and of Gaussian waves \cite{white2017hybrid, white2019multisliced}. They represent however a very complex basis set, so for the time being we have not implemented it yet, working in dual waves instead.

The following question we need to answer is how to index the terms of the Hamiltonian. We will have registers $\ket{p}$ and $\ket{q}$ which in binary encode the orbitals without taking into account the spin, while $\ket{\alpha}$, and $\ket{\beta}$ will take that into account. Thus, $\ket{p}$ and $\ket{q}$ will encode numbers from $0$ to $N/2-1$ ($N$ the number of spin-orbitals) and will need $\lceil\log_2 N\rceil - 1$ qubits each. Then we will have two one-qubit registers $\ket{U}$ and $\ket{V}$, that will decide what term in the Hamiltonian to apply. Finally $\ket{\theta}$ will be used to apply a phase $(-1)^\theta$. Overall, we have the following Select operator
\begin{equation}
\begin{split}
    &\text{Select}\ket{\theta, U, V, p, \alpha, q, \beta} \ket{\psi}=\\
    &(-1)^{\theta}\ket{\theta, U, V, p, \alpha, q, \beta}\\
    &\otimes
     \begin{cases}
       Z_{p,\alpha} &\quad U \land \neg V \land ((p,\alpha)= (q,\beta))\\
       Z_{p,\alpha}Z_{q,\beta} &\quad \neg U \land V \land ((p,\alpha) \neq (q,\beta))\\
       X_{p,\alpha}\vec{Z} X_{q,\alpha}&\quad \neg U \land \neg V \land (p< q) \land (\alpha = \beta) \\
      Y_{p,\alpha}\vec{Z} Y_{q,\alpha}&\quad \neg U \land \neg V \land (p> q) \land (\alpha = \beta) \\
      \text{Undefined} & \quad\text{Otherwise}
     \end{cases}.
     \label{Select operator babbush18}
\end{split}
\end{equation}

As an aside notice that $p$ and $q$ are three dimensional vectors whose elements take integer values in the range $[0, (N/2)^{1/3} -1]$, so we need to map $(p,\sigma)$ to an integer index representing a qubit. The mapping is, for a $D$ dimensional system ($D=3$)
\begin{equation}
    M = (N/2)^{1/D}, \qquad f(p,\sigma) = \delta_{\sigma, \uparrow} M^D + \sum_{j=0}^{D-1} p_j M^j.
\end{equation}

Similarly, the Prepare operator performs 
\begin{equation}
    \begin{split}
        &\text{Prepare}: \ket{0}^{\otimes (3+2\lceil \log_2 N \rceil)} \mapsto\\ &\sum_{p,\sigma}\tilde{U}(p)\ket{\theta_p}\ket{1}_U\ket{0}_V\ket{p,\sigma,p,\sigma}\\
&+\sum_{p\neq q,\sigma}\tilde{T}(p-q)\ket{\theta^{(0)}_{p-q}}\ket{0}_U\ket{0}_V\ket{p,\sigma,q,\sigma}\\
&+\sum_{(p,\alpha)\neq (q,\beta)}\tilde{V}(p-q)\ket{\theta^{(1)}_{p-q}}\ket{0}_U\ket{0}_V\ket{p,\sigma,q,\sigma},
    \end{split}
\end{equation}
with coefficients
\begin{equation}
    \begin{split}
        \tilde{U}(p) = \sqrt{\frac{|T(0)+ U(p) + \sum_q V(p-q)|}{2\lambda}}\\
        \tilde{T}(p) =\sqrt{\frac{|T(p)|}{\lambda}}; \qquad
        \tilde{V}(p) =\sqrt{\frac{|V(p)|}{4\lambda}}\\
    \end{split}
\end{equation}
and
\begin{equation}
    \begin{split}
        \theta_p = \frac{1- sign(-T(0)-  U(p) - \sum_q V(p-q))}{2}\\
        \theta_p^{(0)} =\frac{1-sign(T(p))}{2};\qquad
        \theta_p^{(1)} =\frac{1-sign(V(p))}{2}.
    \end{split}
\end{equation}

To implement Prepare, first, we prepare a unitary operator called Subprepare, which acts as
\begin{equation}
    \begin{split}
        \ket{0}^{\otimes (2 + \log_2 N)}\mapsto\\
        \sum_{d=0}^{N-1}
\left(\tilde{U}(d)\ket{\theta_d}\ket{1}_U \ket{0}_T +  \tilde{T}(d)\ket{\theta_d^{(0)} } \ket{0}_U\ket{0}_V \right.\\ +\left.\tilde{V}(d)\ket{\theta_d^{(1)} }\ket{0}_U\ket{1}_V\right)\ket{d}.
    \end{split}
\end{equation}

The construction of Select, Subprepare and Prepare can be seen in fig.  14, 15 and 16 from \cite{babbush2018encoding}. Taking this into account, the total cost will be $r(2\cdot \text{Prepare}+\text{Select} + R)$, where $R$ stands for the reflection in each step.

\subsection{\label{app:QROM_cost}How to compute its cost}

The circuit implementing the Select operator is depicted in the above-mentioned figure 14 \cite{babbush2018encoding}. It will require the use of 3 QROM applications of size $O(N)$, and $2\lceil \log_2 N \rceil$ controlled swaps (Fredking gates) each requiring one T gate. So, the total T-gate cost is $12N + 8\lceil \log_2 N \rceil - 14$.

Subprepare is the main building block for Prepare, and it is depicted in figure 15 in \cite{babbush2018encoding}. It uses one QROM, with AND complexity $3M^D -1 = 3N/2 - 1$, so T complexity $6N-4$. The 3 is due to the three possible combinations that can appear in $\ket{U}$ and $\ket{V}$, whereas $M^D$ is due to register $\ket{p}$ having $D\lceil\log_2 M\rceil = \lceil\log_2 N/2\rceil$ qubits. Appart from the QROM, Subprepare contains $3 + \lceil\log_2 N/2\rceil = 2 + \lceil\log_2 N\rceil$ controlled swaps (each requiring a Toffoli gate or $4$ T gates); two comparison test between 2 $\mu$-sized registers; and finally operators Uniform$_M^{\otimes D}$ and Uniform$_3$. 

The Uniform operators prepare an uniform superposition over the first $L$ basis states, and is analyzed in figure 12 in \cite{babbush2018encoding}. Since in particular we are using Uniform$_3$ and Uniform$_M^{\otimes D}$, this will require $8\lceil\log_2 L\rceil +O( \log_2 \epsilon^{-1}_{SS})= 8\lceil\log_2 3\rceil+O(\log \epsilon^{-1}_{SS})$ T gates in the first case, and $8D\log M +O(\log \epsilon^{-1}_{SS})=8\lceil\log_2 N\rceil - 8 +O(\log \epsilon^{-1}_{SS})$ in the second. The $O(\log \epsilon^{-1}_{SS})$ term stands for 2 rotations $R_z$ in each Uniform operator. Overall, Subprepare requires $6N + 12\lceil \log_2 N \rceil + 10\mu +16\lceil\log_2 \epsilon^{-1}_{SS}\rceil$ T gates.

The Prepare operator can be seen in figure 16 in \cite{babbush2018encoding}. It requires another Uniform$_M^{\otimes D}$, at cost $8\lceil \log_2 N \rceil + 8\lceil \log_2 \epsilon^{-1}_{SS}\rceil$; $D\lceil\log_2 M\rceil = \lceil\log_2 N\rceil -1 $ swaps with $4$ times as many T gates; 2 multicontrolled Not gates with $\lceil \log_2 N \rceil$ controls each, which can be implemented using $16\lceil \log_2 N \rceil$ T gates \cite{barenco1995elementary}; and one sum over $D\lceil\log_2 M\rceil$ qubits. 

With the previous, we have everything we need to calculate the total T gate cost accurately.

\section{\label{app:Plane_wave}Plane and dual wave basis}
\subsection{\label{app:Plane_waves_method}Method explanation}
 When looking for a basis of functions to perform chemical calculations, one is primarily looking for a basis that \cite{babbush2018low}
\begin{enumerate}
    \item Leads to a small number of terms in the Hamiltonian.
    \item Allows for simple preparation of initial state.
\end{enumerate}
On the Gaussian basis, initial states are easy to prepare using the Hartree-Fock procedure. However, the Hamiltonian may have up to $O(N^4)$ terms.

One idea to avoid having so many terms in the Hamiltonian is to use the plane waves and dual wave basis. The plane wave basis functions have the form
\begin{equation}
\begin{split}
    \varphi_{\bm{\nu}} (\bm{r}) = \sqrt{\frac{1}{\Omega}} e^{i\bm{k_{\nu}}\cdot \bm{r}}, \quad \bm{k_{\nu}} = \frac{2\pi \bm{\nu}}{\Omega^{1/3}},\\
    \bm{\nu}\in [-N^{-1/3},N^{1/3}]^3 \in \mathbb{Z}^3.
\end{split}
\end{equation}

In the plane wave basis, the Hamiltonian will take the form \cite{babbush2018low}
\begin{equation}
\begin{split}
    &H =\underbrace{+\frac{2\pi}{\Omega}\sum_{\substack{(p,\sigma)\neq (q,\sigma')\\
    \nu\neq 0}}\frac{c_{p,\sigma}^\dagger c_{q,\sigma'}^\dagger c_{q+\nu,\sigma'} c_{p-\nu,\sigma}}{k_\nu^2}}_{V}\\
&\underbrace{+\frac{1}{2}\sum_{p,\sigma}k_p^2 c_{p,\sigma}^\dagger c_{p,\sigma}}_{T}
\underbrace{-\frac{4\pi}{\Omega}\sum_{\substack{p\neq q;\\ j,\sigma}}\left(\zeta_j \frac{e^{ik_{q-p}\cdot R_j}}{k_{p-q}^2}\right)c_{p,\sigma}^\dagger c_{q,\sigma}}_{U},
\label{Plane wave Hamiltonian}
\end{split}
\end{equation}
$p$, $q\in [-N^{-1/3},N^{1/3}]^3$ indexing the momentum. Notice that in this basis the kinetic operator $T$ is diagonal, a property that we will use abundantly.

Fourier transforming \eqref{Plane wave Hamiltonian}, we get the dual plane wave Hamiltonian,
\begin{equation}
    \begin{split}
        H &=\underbrace{\frac{1}{2N}\sum_{p,q,\nu,\sigma}k_\nu^2 \cos[k_\nu\cdot r_{q-p}] a_{p,\sigma}^\dagger a_{q,\sigma}}_{T}\\
        &\underbrace{-\frac{4\pi}{\Omega}\sum_{p, j,\sigma,\nu\neq 0}\left( \frac{\zeta_j\cos[k_{\nu}\cdot (R_j-r_p)]}{k_{\nu}^2}\right)n_{p,\sigma}}_{U}\\
        &\underbrace{+\frac{2\pi}{\Omega}\sum_{(p,\sigma)\neq (q,\sigma'); \nu\neq 0}\frac{\cos[k_\nu \cdot r_{p-q}]}{k_\nu^2}n_{p,\sigma}n_{q,\sigma'}}_{V},
    \end{split}
    \label{Dual plane wave Hamiltonian}
\end{equation}
with $n_p = a_p^\dagger a_p$, $a_p$ and $a_p^\dagger$ the Fourier transformed annihilation and creation operators, and $\bm{r_p} = \bm{p}(\Omega/N)^{1/3}$. We can see that in this basis the potential terms become diagonal, and since the term $V$ only has $\Theta(N^2)$ terms, the number of terms in the Hamiltonian is $O(N^2)$.

In Jordan Wigner mapping, \eqref{Dual plane wave Hamiltonian} can be represented as
\begin{equation}
    \begin{split}
        &H =\frac{\pi}{2\Omega}\sum_{\substack{(p,\sigma)\neq (q,\sigma')\\ \nu \neq 0}}\frac{\cos[k_\nu\cdot r_{p-q}]}{k_\nu^2}Z_{p,\sigma}Z_{q,\sigma'} \\
        &\sum_{\substack{p,\sigma\\ \nu \neq 0}} \left(\frac{\pi}{\Omega k_\nu^2} - \frac{k_\nu^2}{4N}+ \frac{2\pi}{\Omega}\sum_j \frac{\zeta_j\cos[k_{\nu}\cdot (R_j-r_p)]}{k_{\nu}^2}\right)Z_{p,\sigma}\\
        &+\frac{1}{4N}\sum_{\substack{p\neq q\\ \nu,\sigma }}k_\nu^2\cos[k_\nu\cdot r_{q-p}](X_{p,\sigma}Z_{p+1,\sigma}...Z_{q-1,\sigma}X_{q,\sigma}\\
        &+Y_{p,\sigma}Z_{p+1,\sigma}...Z_{q-1,\sigma}Y_{q,\sigma}) + \sum_{\nu\neq 0}\left(\frac{k_\nu^2}{2}- \frac{\pi N}{\Omega k_\nu^2}\right)I.
    \end{split}
    \label{JW Dual plane wave Hamiltonian}
\end{equation}

Depending on the situation, to simulate the Hamiltonian in the most efficient way possible we will jump back and forth between dual and primal representations depending on the operator of the Hamiltonian
\begin{equation}\label{diagonal dual plane wave Hamiltonian}
\begin{split}
    H &=\underbrace{FFFT^\dagger\left(\frac{1}{2}\sum_{\nu,\sigma}k_\nu^2 a_{\nu,\sigma}^\dagger a_{\nu,\sigma}\right) FFFT}_{T}\\
    &\underbrace{-\frac{4\pi}{\Omega}\sum_{p, j,\sigma,\nu\neq 0}\left( \frac{\zeta_j\cos[k_{\nu}\cdot (R_j-r_p)]}{k_{\nu}^2}\right)n_{p,\sigma}}_{U}\\
    &\underbrace{+\frac{2\pi}{\Omega}\sum_{(p,\sigma)\neq (q,\sigma'); \nu\neq 0}\frac{\cos[k_\nu \cdot r_{p-q}]}{k_\nu^2}n_{p,\sigma}n_{q,\sigma'}}_{V}
\end{split}    
\end{equation}
where all the terms are diagonal. To implement this Hamiltonian, we need to perform a Fermionic Fast Fourier Transform (FFFT) \cite{ferris2014fourier}, an adaptation of the classical Fast Fourier Transform. We cannot use here the Quantum Fourier Transform because we are using the Jordan-Wigner mapping that encodes the value of the qubits not in the amplitudes but the basis.

\subsubsection{Trotterization algorithm}

The most basic way to use the plane wave approach is to use \eqref{diagonal dual plane wave Hamiltonian} to simulate a segment of the Hamiltonian simulation procedure

\begin{equation}
\begin{split}
    e^{-iH\delta_t} &\approx
    e^{-i(U+V)\delta_t/2}\cdot\\
    &FFFT^\dagger e^{-i(\delta_t/2)\sum_{\nu,\sigma}k_{\nu}^2 a_{\nu,\sigma}^\dagger a_{\nu,\sigma}}FFFT\\
    &\cdot e^{-i(U+V)\delta_t/2} + O(\delta_t^3),
    \label{Trotter segment Hamiltonian simulation in plane waves}
\end{split}
\end{equation}
with $U$ and $V$ given in \eqref{diagonal dual plane wave Hamiltonian}.
This formulation allows us to perform Hamiltonian simulation and Quantum Phase Estimation. The FFFT will be explained later on in this appendix.

\subsubsection{Taylorization `database' algorithm}

Alternatively, we may use the Taylorization procedures from appendix \ref{app:Taylorization}. Let us start with the `database' algorithm. To carry it out we need to define how to perform the Prepare($W$) and Select($H$) operators.

Select($H$) is virtually the same as the same preparation method as we describe in appendix \ref{app:QROM} \cite{babbush2018encoding}, except that in this case we use the notation of $p$ odd or even for up and down spin values:
\begin{equation}
\begin{split}
    &\text{Select}(H)\ket{p, q, b} \ket{\psi}= \ket{ p, q, b} \otimes\\
&\begin{cases}
Z_{p}\ket{\psi} &\quad p=q\\
Z_{p}Z_{q}\ket{\psi} &\quad (b=0)\land (p\neq q)\\
X_{p}\vec{Z} X_{q}\ket{\psi}&\quad (b=1)\land (p> q) \land (p\oplus q = 0) \\
Y_{p}\vec{Z} Y_{q}\ket{\psi}&\quad (b=1)\land (p< q) \land (p\oplus q = 0) \\
\ket{\psi}& \quad (b=1) \land (p\oplus q = 1)
\end{cases}.
\end{split}
\label{Select(H) in plane waves}
\end{equation}
where $\oplus$ indicates sum modulus 2; and can therefore be implemented at cost $12N + 8\lceil \log_2 N \rceil + O(1)$ T gates. 

Since the Prepare($W$) method is not specified in the main reference for this appendix \cite{babbush2018low}, we will also use the method from \cite{babbush2018encoding}.

\subsubsection{Taylorization `on-the-fly' algorithm}

In appendix K of \cite{babbush2018low} it is explained how to use the `on-the-fly algorithm' in this context, which is similar to what we explained in appendix \ref{app:Taylorization} \cite{babbush2016exponentially}.

The amplitudes we want to prepare, $W_{p,q,b}$, can be divided in a sum
\begin{equation}
    W_{p,q,b} = \sum_{\nu\neq 0} W_{p,q,b,\nu},
\end{equation}
where
\begin{equation}
\begin{split}
    &W_{p,q,b} =\\
&\begin{cases}
      \begin{array}{r@{}}
\sum_{\nu\neq 0}\left(\frac{\pi}{2\Omega k_\nu^2} - \frac{k_\nu^2}{8N} +\right.\\
\left.\frac{\pi}{\Omega}\sum_j \zeta_j \frac{\cos [k_\nu \cdot (R_j-r_p)]}{k_{\nu}^2}\right)
      \end{array} & \quad p = q\\
\frac{\pi}{4\Omega}\sum_{\nu \neq 0}\frac{\cos[k_\nu \cdot r_{p-q}]}{k_{\nu}^2} & \quad (b= 0)\land (p\neq q)\\
\frac{1}{4N}\sum_{\nu}k_{\nu}^2 \cos [k_{\nu}\cdot r_{p-q}] & \quad (b=1)\land (p\oplus q = 0 )\\
1 & \quad (b=1)\land (p\oplus q = 1 ).
\end{cases}
\end{split}\label{W Taylor babbush2018low}
\end{equation}
If we have to sum over a large number of atoms $J$, we may also decompose each of the terms in the $j$ sum independently.

Since it is easy to apply phases but not to change the amplitudes of a given state, \cite{babbush2018low} proposes further dividing each
\begin{equation}
\begin{split}
    W_{p,q,b,\nu} \approx \zeta\sum_{m=0}^{M-1}W_{p,q,b,\nu,m}; \quad  W_{p,q,b,\nu,m}\in \{\pm 1\};\\
    \zeta = \Theta \left(\frac{\epsilon}{\Gamma t}\right); \quad M \in \Theta\left(\frac{\max_{p,q,b,\nu}|W_{p,q,b,\nu}|}{\zeta} \right).
\end{split}
\end{equation}

To perform the logic of the on-the-fly algorithm we first have to perform the calculations for the coefficients, which means we need costly arithmetic operations:
\begin{equation}
    \text{Sample}(W)\ket{p,q,b,\nu}\ket{0}^{\otimes \lceil \log_2 N\rceil}\mapsto \ket{p,q,b,\nu}\ket{\tilde{W}_{p,q,b,\nu}},
\end{equation}
with $W_{p,q,b,\nu}$ a binary approximation to $\tilde{W}_{p,q,b,\nu}$.

The complexity will be $O(N^3 + \log_2\epsilon^{-1}_{M})$, where the $\epsilon_M$ appears due to the use of Subprepare techniques from \cite{babbush2018encoding}.

\subsection{\label{app:Plane_wave_cost}How to compute its cost}

\subsubsection{Trotterization algorithm}

In this subsection we aim to explain the cost of performing Trotterization using this approach. To do so, we have to compute the cost of the FFFT operator, as well as the number of single qubit rotations in the exponentials and the number of segments required, $r$.

Let us start with the computation of the cost of FFFT. From \cite{ferris2014fourier} it can be seen that the number of gates required to perform an $m$-mode Fourier Transform are are $(m/2)\lceil\log_2 (m/2)\rceil$ single qubit rotations and $(m/2)\lceil\log_2 m\rceil$ $F_2$ gates. The matrix representation of $F_2$ in the Jordan-Wigner representation is 
\begin{equation}
\begin{split}
    F_2&=\begin{pmatrix}
    1 & 0 & 0 & 0\\
    0 & 2^{-1/2} & 2^{-1/2} & 0\\
    0 & 2^{-1/2} & -2^{-1/2} & 0\\
    0 & 0 & 0 & -1
    \end{pmatrix}\\
    &=\begin{pmatrix}
    1 & 0 & 0 & 0\\
    0 & 2^{-1/2} & 2^{-1/2} & 0\\
    0 & 2^{-1/2} & -2^{-1/2} & 0\\
    0 & 0 & 0 & 1
    \end{pmatrix} \cdot
    \begin{pmatrix}
    1 & 0 & 0 & 0\\
    0 & 1 & 0 & 0\\
    0 & 0 & 1 & 0\\
    0 & 0 & 0 & -1
    \end{pmatrix}.
\end{split}
\end{equation}
Therefore, we can see that $F_2$ is the product of a matrix that we will call $W$ with a Control-Z. The gate $W$ works as a Hadamard in the subspace spanned by $\{\ket{01},\ket{10}\}$. Any gate with the structure of a unitary gate $U$ in that subspace can be constructed as $C-U$ between two C-Nots in the opposite direction. In this case, $U$ is the Hadamard gate, and the controlled-Hadamard gate can be performed using $R_y(\pi/4)$, a C-Not, and $R_y(-\pi/4)$. Therefore, in total $F_2$ requires two T gates in the Jordan-Wigner representation. 

Overall, the FFFT requires $(N/2)\log_2(N/2) = (N/2)(\log_2 N -1)$ single qubit z-rotations and $(N/2)\log_2(N)$ $F_2$ gates, as can be seen from figure 1b from \cite{ferris2014fourier}.

The next step is computing the cost of the exponential rotations in \eqref{Trotter segment Hamiltonian simulation in plane waves}.
There are $8N$ terms in $U$, $8N(8N-1)/2$ terms in $V$ and $8N$ terms in $T$ in \eqref{diagonal dual plane wave Hamiltonian}, so the same number of $R_z$ rotations for operators $T$ and $U$. Notice that in the simulation of $e^{-iV \tau}$ we will need Clifford gates and a single $C-R_z$ rotation per term \cite{hastings2014improving,motzoi2017linear}, as it was the case in appendix \ref{app:qDRIFT}.

Finally we want to compute the number of time segments in the Trotter decomposition $r$. Using the equations 5 and 6 from \cite{poulin2014trotter} we can see that the error in each time step is bounded by 
\begin{equation}
    2\left([T,[T,U+V]] + [(U+V),[T, (U+V)]]\right)\delta_t^3.
\end{equation}
This, in turn, can be bounded \cite{babbush2018low} by 
\begin{equation}
\begin{split}\label{loose trotter bounds}
    2 ( &\max_{\psi}|\braket{\psi|T|\psi}|^2\cdot \max_{\psi}|\braket{\psi|U+V|\psi}| +  \\ & \max_{\psi}|\braket{\psi|T|\psi}|\cdot \max_{\psi}|\braket{\psi|U+V|\psi}|^2 )\delta_t^3.
\end{split}
\end{equation}
Since there are $r := t/\delta_t$ terms, the Trotter error is
\begin{equation}
\begin{split}
    \frac{\epsilon_{HS}}{r} \leq 2 ( \max_{\psi}|\braket{\psi|T|\psi}|^2\cdot \max_{\psi}|\braket{\psi|U+V|\psi}| +  \\  \max_{\psi}|\braket{\psi|T|\psi}|\cdot \max_{\psi}|\braket{\psi|U+V|\psi}|^2 ) \left(\frac{t}{r}\right)^3 .
    \label{babbush2018low trotter error}
\end{split}
\end{equation}
Asymptotically, this means we will take
\begin{equation}
    r = \Theta\left( \frac{\eta^2 N^{5/6}t^{3/2}}{\Omega^{5/6}\sqrt{\epsilon_{HS}}}\sqrt{1 +\frac{\eta\Omega^{1/3}}{N^{1/3}}} \right).
\end{equation}
We can find bounds for the expected values of $U$, $V$ and $T$, in appendix F \cite{babbush2018low}. 
\begin{subequations}
From equation F1
\begin{equation}
\begin{split}
    \max_{\psi}|\braket{\psi|V|\psi}|&\leq \frac{2\pi \eta^2}{\Omega}\sum_{\nu\neq 0}\frac{1}{k_\nu^2}\\
    &= \frac{\eta^2}{2\pi \Omega^{1/3}}\sum_{(\nu_x,\nu_y, \nu_z)\neq (0,0,0)}\frac{1}{\nu_x^2 + \nu_y^2+ \nu_z^2},
\end{split} \label{bound V babbush2018low}
\end{equation}
from F8
\begin{equation}
\begin{split}
    \max_{\psi}|\braket{\psi|U|\psi}|&\leq \frac{4\pi \eta}{\Omega}\left(\sum_j \zeta_j\right)\sum_{\nu\neq 0}\frac{1}{k_\nu^2}\\
    &= \frac{\eta^2}{\pi \Omega^{1/3}}\sum_{(\nu_x,\nu_y, \nu_z)\neq (0,0,0)}\frac{1}{\nu_x^2 + \nu_y^2+ \nu_z^2},
\end{split} \label{bound U babbush2018low}
\end{equation}
and from F10
\begin{equation}
    \max_{\psi}|\braket{\psi|T|\psi}| \leq \frac{2\pi^2 \eta}{\Omega^{2/3}}\nu_{\max}^2.
    \label{bound T babbush2018low}
\end{equation}
\end{subequations}
To end up bounding $U$ and $V$ we need equation F6 \cite{babbush2018low}
\begin{equation}
\begin{split}
    \sum_{(\nu_x,\nu_y, \nu_z)\neq (0,0,0)}\frac{1}{\nu_x^2 + \nu_y^2+ \nu_z^2} \leq 4\pi \left(\sqrt{3}\frac{N^{1/3}}{2}-1\right)\\ + \int_1^{N^{1/3}}\frac{3dz}{z^2} + \int_1^{N^{1/3}}\int_1^{N^{1/3}}\frac{3dxdy}{x^2 + y^2} =  \\
    4\pi \left(\sqrt{3}\frac{N^{1/3}}{2}-1\right) + 3- \frac{3}{N^{1/3}} + \\
    \int_1^{N^{1/3}}\int_1^{N^{1/3}}\frac{3dxdy}{x^2 + y^2}.
    \label{1/(x^2 + y^2 + z^2) sum babbush2018low}
\end{split}
\end{equation}
Using this and the previous equations, it is possible to calculate the actual value of $r$, given $t$ and $\epsilon_{HS}$.

\subsubsection{Taylorization `database' algorithm}
Since the Prepare($W$) method is not specified in the main reference for this appendix \cite{babbush2018low}, we will also use the method from \cite{babbush2018encoding}. As explained in appendix \ref{app:QROM}, the cost for Prepare($W$) $6N  + 40\lceil\log_2 N\rceil + 16\lceil\log_2 \epsilon^{-1}_{SS}\rceil + 10\mu$. Notice that the cost is linear because although there are $O(N^2)$ coefficients, only $O(N)$ are independent. In any case this will be multiplied by $\lambda = O(N^2)$. 

Similarly taken from \cite{babbush2018encoding} and explained in appendix \ref{app:QROM} the cost of Select($H$) can be taken to be $12N + 8\lceil\log_2 N\rceil  + O(1)$ T gates, since the implementation proposed in both references (\cite{babbush2018encoding} and \cite{babbush2018low}) is virtually the same.

\subsubsection{Taylorization `on-the-fly' algorithm}

Finally, the main cost of the `on-the-fly' algorithm comes from the Sample($W$) operations that compute \eqref{W Taylor babbush2018low}. This will require arithmetic operations as those indicated in table \ref{tab:arithmetic}.

The main difference here will be calculating the value of $\lambda'$, that influences the number of segments $r$.
From equation K2 in \cite{babbush2018encoding} the Hamiltonian will have the form
\begin{equation}
    H = \zeta \sum_{p,q,b,\nu,m} W_{p,q,b,\nu,m} H_{p,q,b}
\end{equation}
Similarly as in previous appendices, we take 
\begin{equation}
    \zeta = \frac{\epsilon_H}{\Gamma r}.
\end{equation}
In contrast to appendix \ref{app:Taylorization} there is no integral over any volume, so we do not include $\mathcal{V}$ in the denominator; and in contrast to appendix \ref{app:Configuration_interation} we do not sum over $\rho$ so there is no division by $\mu$. The main consequence of this form of preparing the initial state is changing the value of $\lambda$, that will now be, from eq. K5 in \cite{babbush2018low}
\begin{equation}
    \lambda' = \zeta \sum_{p,q,b,\nu,m} |W_{p,q,b,\nu,m}|,\quad W_{p,q,b,\nu,m}\in\{-1,+1\}.
\end{equation}
As a consequence, given that $m\in{0,...,M-1}$, $b$ can take values $0$ and $1$ and there are $8N$ values for $p$, $q$ and $\nu$
\begin{equation}
    \lambda' = 2M\zeta (8N)^3.
\end{equation}
Since
\begin{equation}
    M = \frac{\max_{p,q,b,\nu}|W_{p,q,b,\nu}|}{\zeta}
\end{equation}
we have that 
\begin{equation}
    \lambda' = 2(8N)^3 \max_{p,q,b,\nu}|W_{p,q,b,\nu}|
\end{equation}
As the sum of the nuclear charges is equal to the number of electrons $\sum_j \zeta_j = \eta$, we can bound $\max_{p,q,b,\nu}|W_{p,q,b,\nu}|$ as the maximum of 1 (the identity term);
\begin{subequations}
\begin{equation}
\begin{split}
    &\frac{\pi}{2\Omega k_\nu^2} - \frac{k_\nu^2}{8N} +
\frac{\pi}{\Omega }\sum_j \zeta_j \frac{\cos [k_\nu \cdot (R_j-r_p)]}{k_{\nu}^2}\leq \\
    &\frac{\pi}{2\Omega k_\nu^2} - \frac{k_\nu^2}{8N} +
\frac{\pi\eta}{\Omega k_\nu^2}\leq \frac{\pi}{2\Omega k_\nu^2} + \frac{\pi\eta}{\Omega k_\nu^2}  - \frac{k_\nu^2}{8N} \\
&= \frac{(2\eta + 1)\pi}{2\Omega k_\nu^2} - \frac{k_\nu^2}{8N};
\end{split}
\end{equation}
\begin{equation}
    \frac{\pi}{4\Omega}\frac{\cos[k_\nu \cdot r_{p-q}]}{k_\nu^2}\leq \frac{\pi}{4\Omega k_\nu^2};
\end{equation}
or
\begin{equation}
    \frac{k_\nu^2}{4N}\cos[k_\nu\cdot r_{p-q}]\leq \frac{k_\nu^2}{4N}.
\end{equation}
\end{subequations}
Since the smallest value of $|k_\nu|$ for $\nu\neq 0$ is $k_{\nu}= 2\pi/\Omega^{1/3}$, and the largest is $k_\nu^2 = 3 \times \frac{(2\pi)^2 N^{2/3}}{\Omega^{2/3}}$
\begin{equation}
\begin{split}
    &\max_{p,q,b,\nu}|W_{p,q,b,\nu}|\leq \\
    &\max\left[ \frac{(2\eta + 1)}{8 \pi\Omega^{1/3}} - \frac{\pi^2}{2N\Omega^{2/3}}, \frac{1}{8\pi\Omega^{1/3}}, \frac{6\pi^2 }{N^{1/3}\Omega^{2/3}} \right],
\end{split}
\end{equation}
Provided that the first option is the largest,
\begin{equation}
    \lambda' \leq (8N)^3\left(\frac{(2\eta + 1)}{4\Omega^{1/3} \pi} - \frac{\pi^2}{N\Omega^{2/3}}\right)\label{lambda' plane waves Taylorization on-the-fly}.
\end{equation}

Now we want to compute the number of arithmetic operations in the Prepare($w$) operation. $p = q$ case of \eqref{W Taylor babbush2018low}:
\begin{enumerate}
    \item Calculating $k_\nu$ and $r_p$ requires three multiplications each, one for each coordinate component, with $n = \lceil\log_2 N^{1/3}\rceil$. 
    \item There are three subtraction for each value of $j$ in $R_j-r_p$ and another $r_{p-q} = r_p-r_q$, with $n = \lceil\log_2 N^{1/3}\rceil$.
    \item Computing $k_\nu^2$ requires 3 multiplications and 2 additions.    
    \item Calculating the product within the cosines costs three multiplications of length $n = \lceil\log_2 N^{1/3}\rceil$, and two sums between those terms.
    \item One of the fastest ways to compute the cosine is to use the CORDIC algorithm \cite{volder1959cordic}, which requires a prefactor division (if expanded to a fixed order) and 2 sums per order since divisions by powers of two can be performed virtually.
    \item We have to sum $J$ cosine computations.
    \item We have to divide or multiply such sum of cosines by a constant, and $k_\nu^2$. Costs up to $\approx 3\cdot21\log^2 N$.
\end{enumerate}
Thus, the T-gate cost of this first calculation is $\approx J\left[\frac{35o}{2} + 63 + \frac{2o}{\log_2 N}\right]\log_2^2 N$, where $J$ is the number of values of $j$, that indexes the atoms.

For $(b=0) \land (p\neq q)$ and $(b = 1)\land (p\oplus q = 0)$:
\begin{enumerate}
    \item We can reuse the previously calculated values of $k_{\nu}$, $k_{\nu}^2$ and compute $r_q$ (3 multiplications) and $r_{p-q}$ (3 subtractions). 
    \item We can perform the dot product in the cosine with 3 multiplications and 2 sums
    \item Similarly, we have to perform a cosine calculation via the CORDIC algorithm again.
    \item Finally we perform a multiplications and a division (by $k_\nu^2$)
\end{enumerate}
To perform the case $b = 1 \land (p+q = 0 \mod 2)$ we can reuse the cosine result from the previous point, as well as the $k_\nu^2$ value, so we only need two multiplications.

\subsection{\label{app:Plane_waves_adaptation}How to adapt the Hamiltonian simulation to control the direction of the time evolution}

\subsubsection{Trotterization method}

In the Phase Estimation protocol we should be controlling such rotations depending on the control ancilla qubits. However, since they are $R_z$ rotations and $XR_z(\alpha) X = R_z(-\alpha) $ we can actually use a formulation similar to \cite{babbush2018encoding} where the mapping is $\ket{1}\ket{\phi}\rightarrow e^{i\phi}\ket{1}\ket{\phi}$ and $\ket{0}\ket{\phi}\rightarrow e^{-i\phi}\ket{0}\ket{\phi}$ (except for the first segment, but this is a minor cost). To control between both rotations we use C-Nots which change the direction of the Z rotation \cite{whitfield2011simulation}. 

\subsubsection{Taylorization methods}

Adapting the Hamiltonian simulation for its use in Quantum Phase Estimation can be done as in appendix \ref{app:Taylorization_adaptation}. The cost can be therefore calculated in the same way.

\section{\label{app:SHC_ Trotter}Trotter simulation: tighter bounds}

In the previous appendix \ref{app:Plane_wave} we have explained how to perform Trotter simulation in plane waves. However, the bounds provided by \eqref{loose trotter bounds} are somewhat loose, so the number of steps needed to achieve the same error are lower than required. Similarly happens for the methods covered in appendix \ref{app:qDRIFT}. In this appendix we give tighter bounds for the second order Hamiltonian simulation deterministic Trotter operator. We aim to approximate $e^{iH\delta_t}$, for $H= \sum_{\gamma=1}^{\Gamma}w_\gamma H_\gamma$, with
\begin{equation}
    \mathcal{S}_2(H;\delta_t)= \left(\prod_{\gamma=1}^{\Gamma}e^{\frac{it}{2}w_\gamma H_\gamma}\right)\left(\prod_{\gamma=\Gamma}^{1}e^{\frac{it}{2}w_\gamma H_\gamma}\right).
\end{equation}
This general expression will reduce, for the plane wave basis, to
\eqref{Trotter segment Hamiltonian simulation in plane waves}
\begin{equation}
\begin{split}
    e^{-iH\delta_t} &\approx
    e^{-i(U+V)\delta_t/2}\cdot\\
    &FFFT^\dagger e^{-i(\delta_t/2)\sum_{\nu,\sigma}k_{\nu}^2 a_{\nu,\sigma}^\dagger a_{\nu,\sigma}}FFFT\\
    &\cdot e^{-i(U+V)\delta_t/2} + O(\delta_t^3).
\end{split}
\end{equation}
The error in this expression will be
\begin{equation}
    \|e^{iH\delta_t} - \mathcal{S}_2(H;\delta_t)\| \leq W_2 \delta_t^3,
\end{equation}
for $\delta_t = t/r$ and $W_2$ a commutator expression. Since in plane waves, operators $U$ and $V$ commute, in a Hamiltonian $H = T+U+V$ we only have to care about commutators $[[T, U+V], T]$ and $[[T, U+V], U+V]$. This can be better seen in the dual basis, where
\begin{equation}
    V=\sum_{p\neq q}V_{pq}n_p n_q
\end{equation}
and 
\begin{equation}
    U = \sum_p U_p n_p = \sum_p U_p n_p n_p,
\end{equation}
for $n_p$ the occupancy fermionic operator. Since the $n_p$ operators commute with each other, so do $U$ and $V$. Consequently, Ref. \cite{su2021nearly} proposes to write $H = T+ \Bar{V}$ with
\begin{equation}
\Bar{V} := U+V =  \sum_{p,q}\Bar{V}_{p,q}n_p n_q.
\end{equation}

One additional insight to bound the commutator $W_2$ as tightly as possible is to restrict our space to the space of $\eta$ electrons. Usually, the error has been described in terms of the spectral norm distance, that is, in other words $\|H\|_2 = \max_{\psi}\|\braket{\psi|H|\psi}\|$. However, this takes into account states $\psi$ that do not live in the subspace of $\eta$ electrons, potentially leading to a higher norm and a looser bound. To remedy this, one can instead use the `fermionic seminorm', defined as
\begin{equation}
    \|H\|_\eta = \max_{\phi,\psi\in \mathcal{H}_\eta}\|\braket{\phi|H|\psi}\|,
\end{equation}
for $\mathcal{H}_\eta$ the Hilbert subspace with $\eta$ electrons. While this seminorm fulfills many properties of norms such as the triangle inequality, it is not a norm because some operators can evaluate to $0$ without being operator $0$ in the full Hilbert space, for example $\|n_p n_q\|_{\eta=1} = 0$.

Using the fermionic seminorm, we express the commutator error bound $W_2$ as \cite{mcardle2022exploiting}
\begin{equation}
W_2\leq \frac{1}{12}\|[[T, U+V], T]\|_{\eta}+\|[[T, U+V], U+V]\|_{\eta}. 
\end{equation}
Furthermore, it is possible to bound each of the two terms independently as (\cite{su2021nearly} and appendix A in \cite{mcardle2022exploiting}):
\begin{align}
\|[[T, \bar{V}], T]\|_{\eta}&\leq 4\|T\|_2^2 \|\bar{V}\|_{\max} \eta(4\eta+1)\\
\|[[T, \bar{V}], \bar{V}]\|_{\eta}&\leq 12\|T\|_2 \|\bar{V}\|_{\max}^2 \eta^2(2\eta+1),
\end{align}
collectively known as the SHC bound, which scales as $O(N^3)$ with the number of basis functions $N$. Other bounds exist too (see sections 3 and 4, and table 1 in \cite{mcardle2022exploiting}), and shall be included in future updates to the library.

From equation 8 in \cite{babbush2018low}, we also know that
\begin{equation}
\|U+V\|_{\max} \leq \frac{4\pi}{\Omega}  \frac{\Omega^{2/3}}{4\pi^2} \sum_{i}\zeta_i = \frac{\Omega^{1/3}\eta}{\pi},
\end{equation}
while $\|T\|_2$ can be bounded as we did in \eqref{bound T babbush2018low}. 
From this, and the implementation cost of \eqref{Trotter segment Hamiltonian simulation in plane waves} that we discussed in appendix \ref{app:Plane_wave_cost}, we can obtain an even lower cost of the Trotter simulation.

\section{\label{app:Sparsity_low-rank}Sparsity and low rank factorization}
\subsection{\label{app:Low_rank_method}Method explanation}
In the previous appendix we have seen that using carefully crafted Prepare and Select operators, it is possible to lower the complexity of the Quantum Phase Estimation. However, this came at the cost of having to use plane waves or similar basis sets. The method proposed in this appendix allows to leverage QROM techniques while working in arbitrary basis \cite{berry2019qubitization,babbush2018encoding}. The other main consideration of this article is how to leverage the sparsity and a low rank factorization of the Hamiltonian to lower the complexity of the algorithm.

Let us start by the second aspect, the low rank tensor factorization. We know that we can write the Hamiltonian in the second quantization in the following form
\begin{equation}
    \begin{split}
H &=\sum_{\sigma \in \{\uparrow, \downarrow\}} \sum_{p,q = 1 }^{N/2}h_{pq}a_p^\dagger a_q\\
&+ \frac{1}{2}\sum_{\alpha,\beta \in \{\uparrow, \downarrow\}} \sum_{p,q,r,s = 1 }^{N/2}h_{pqrs}a_{p,\alpha}^\dagger a_{q,\beta}^\dagger a_{r,\beta} a_{s,\alpha}\\
&=\sum_{\sigma \in \{\uparrow, \downarrow\}} \sum_{p,q = 1 }^{N/2}T_{pq}a_p^\dagger a_q\\
&+ \sum_{\alpha,\beta \in \{\uparrow, \downarrow\}} \sum_{p,q,r,s = 1 }^{N/2}V_{pqrs}a_{p,\alpha}^\dagger a_{q,\alpha}a_{r,\beta}^\dagger a_{s,\beta}
    \end{split}
    \label{Hamiltonian second quantization}
\end{equation}

The coefficients $h_{pq}$ and $h_{pqrs}$ are efficiently computable integrals. On the previous equation, the ordering $a^\dagger a^\dagger a a$ is called the `physics notation' whereas the second ordering, $a^\dagger a a^\dagger a$ follows the chemists convention and will be the one we will use because it allows us to perform the factorization. Notice that $T_{pq}$ and $V_{pqrs}$ are real and have symmetries $p\leftrightarrow q$, $r\leftrightarrow s$ and $pq\leftrightarrow rs$. Notice also that the one-body operator changes as a result of the swapping of $a_p$ and $a_p^\dagger$ and their anticommutation in the two-body term, and so does the sign of the latter.

Since $V$ is a 4-rank tensor, with indices ranging from $0$ to $N/2-1$, we can transform it to a $N^2/4\times N^2/4$ matrix called $W$, with composite indices $pq$ and $rs$, and symmetric and positive definite. 
Diagonalizing $W$ we get,
\begin{equation}
    W g^{(l)} = w_l g^{(l)}; \quad W = \sum_{l=1}^L w_l g^{(l)} (g^{(l)})^T,
\end{equation}
where $g^{(l)}$ denotes the $l$-th eigenvector, with eigenvalue $w_l$, and entries $g^{(l)}_{pq}$.

Let us denote the rank with $L$. If $W$ where full rank, $L = N^2/4$. However, in most cases and due to Coulomb interaction being a two-body interaction, the rank will be $L = O(N)$. Now, we can rewrite
\begin{equation}
\begin{split}
        &\sum_{\alpha,\beta \in \{\uparrow, \downarrow\}} \sum_{p,q,r,s = 1 }^{N/2}V_{pqrs}a_{p,\alpha}^\dagger a_{q,\alpha}a_{r,\beta}^\dagger a_{s,\beta}\\
        &= \sum_{l=1}^L w_l \left( \sum_{\sigma \in \{\uparrow, \downarrow\}}\sum_{p,q = 1 }^{N/2} g^{(l)}_{pq}a_{p,\sigma}^\dagger a_{q,\sigma} \right)^2.
\end{split}
\end{equation}

From the right-hand side of the equation we can see that there are $O(LN^2)= O(N^3)$ independent coefficients. In fact, due to the symmetry $p\leftrightarrow q$ there are $1/2\cdot N/2(N/2-1)$ terms off diagonal, and when $p = q$ there are $N/2$ additional free coefficients. Therefore, in total there are $N^2/8 + N/4$ independent terms for each value of $l$. Further factorization is possible \cite{motta2018low,von2020quantum,lee2020even}, but this work is not covered in this appendix.

As in the previous article, we do not attempt to perform phase estimation over $e^{\pm iH}$ but rather over $e^{\pm i\arccos (E_k/\lambda)}$, which is the phase produced by one step of the qubitization quantum walk. Also as in the previous article, this method uses Jordan-Wigner mapping too.

We have to explain how to perform operators Prepare and Select. Let us start with the former. The state we want to prepare is the following
\begin{equation}
\begin{split}
\ket{\psi} = \ket{0}\ket{+}\ket{0}\sum_{p,q,\sigma}\sqrt{\frac{|T_{pq}|}{\lambda}}\ket{\theta^{(0)}_{pq}}\ket{0}\ket{p,q,\sigma}\ket{0}+\\
+\sum_l \sqrt{\frac{w_l}{\lambda}}\ket{l}\ket{+}\ket{+}\otimes\\
\otimes\sum_{p,q,r,s,\alpha,\beta}\sqrt{|g_{pq}^{(l)}g_{rs}^{(l)}|}\ket{\theta^{(l)}_{pq}}\ket{\theta^{(l)}_{rs}}\ket{p,q,\alpha}\ket{r,s,\beta}.
\end{split}
\end{equation}
Here, $\theta_{pq}^{(l)}$ indicates the sign of each term, and are defined as
\begin{equation}
    \begin{split}
        \theta^{(l)}_{pq} = \left\{ \begin{array}{lc}
0,  & T_{pq} > 0, \\
1,  & T_{pq} < 0, \\
\end{array}
\right. \qquad
\theta^{(l)}_{pq} = \left\{ \begin{array}{lc}
0,  & g^{(l)}_{pq} > 0. \\
1,  & g^{(l)}_{pq} < 0. \\
\end{array}
\right.
    \end{split}
\end{equation}

We can see that the first register selects between the $T$ terms (for state $\ket{0}$) and each of the $L$ terms for $g^{(l)}$. The second and third register use $\ket{+}$ to select between $\mathbf{1}$, and $Z_{p,\sigma}$, $Z_{p,\alpha}$ and $Z_{q,\beta}$, whenever $p = q$ or $r = s$ respectively. Additionally, depending on whether $p>q$ or $p<q$ we apply $X_{p,\sigma}\vec{Z}X_{q,\sigma}$ or $Y_{p,\sigma}\vec{Z}Y_{q,\sigma}$ respectively.

The number of coefficients to fix is $(L+1)(N^2/8+N/4)$, so the complexity will be $O(N^3 + \log_2\epsilon^{-1}_{\mu})$, where the $\mu$ appears due to the use of Subprepare techniques from \cite{babbush2018encoding}. To perform the preparation, we follow this steps
\begin{enumerate}
    \item Starting from the state $\ket{0}$, prepare a superposition over the first register
    \begin{equation}
        \begin{split}
            \left( \ket{0}\sqrt{\sum_{p,q}\frac{2|T_{pq}|}{\lambda}} + 2\sum_l \sqrt{\frac{w_l}{\lambda}}\ket{l} \sum_{p,q}|g_{p,q}^{(l)}| \right)\otimes\\
            \otimes \ket{0}\ket{0}\ket{0}\ket{0}\ket{0}\ket{0}.
        \end{split}
    \end{equation}
    If we allow for error $\epsilon_{SS}$, the complexity of this step, in terms of T-gates using the QROM is $4L +4\mu + 14\lceil \log_2 L \rceil + 8 \lceil\log_2\epsilon^{-1}_{SS}\rceil$ \cite{babbush2018encoding}. The $\epsilon_{SS}^{-1}$ dependence is due to the Uniform operator preparation, that requires to use two controlled Z rotations, at cost $4\lceil\log_2\epsilon^{-1}_{SS}\rceil$ each. On the other hand, the Uniform preparation requires $10\lceil \log_2 L \rceil$ T gates as can be seen from figure 12 in \cite{babbush2018encoding}, which has to be added to $4\lceil \log_2 L \rceil$ T-gates due to the controlled-swap operations in Subprepare. The value of $\mu$ can be taken from equation 36 in \cite{babbush2018encoding}.
    
    \item Perform a Hadamard in the second register and another on the third, controlled on the first register being $\ket{l>0}$.
    \begin{equation}
        \begin{split}
&\left( \ket{0}\ket{+}\ket{0}\sqrt{\sum_{p,q}\frac{2|T_{pq}|}{\lambda}}\right.\\
&\left.+ 2\sum_l \sqrt{\frac{w_l}{\lambda}}\ket{l}\ket{+}\ket{+} \sum_{p,q}|g_{p,q}^{(l)}| \right)\otimes\\
&\otimes \ket{0}\ket{0}\ket{0}\ket{0}.
        \end{split}
    \end{equation}
    The cost of this step is negligible compared with the following one, and can be performed using a multicontrolled Hadamard.
    
    \item Prepare a superposition over register six with amplitudes $\sqrt{|T_{pq}|}$ if $\ket{l = 0}$ or $\sqrt{|g_{pq}^{(l)}|}$ if $\ket{l>0}$.
    \begin{equation}
        \begin{split}
\ket{0}\ket{+}\ket{0}\sum_{p,q,\sigma}\sqrt{\frac{|T_{pq}|}{\lambda}}\ket{0}\ket{0}\ket{p,q,\sigma}\ket{0}+\\
  +\sqrt{2}\sum_l \sqrt{\frac{w_l}{\lambda}}\ket{l}\ket{+}\ket{+}\otimes\\
  \otimes \sum_{p,q,\alpha}\sqrt{|g_{p,q}^{(l)}|}\sqrt{\sum_{r,s}|g_{r,s}^{(l)}|} \ket{0}\ket{0}\ket{p,q,\alpha}\ket{0}.
        \end{split}
    \end{equation}
    This step and the following have the largest complexities, since we need to use the unary iterator and Subprepare circuit of \cite{babbush2018encoding}. We have to iterate over $L$, $p$, and $q$, and that gives a Toffoli complexity of $(L+1)N^2/4-1$ plus the cost of the comparison and the controlled swaps from Subprepare.

    \item For $\ket{l>0}$, prepare weights $\sqrt{|g_{rs}^{(l)}|}$ in register 7.
     \begin{equation}
        \begin{split}
\ket{0}\ket{+}\ket{0}\sum_{p,q,\sigma}\sqrt{\frac{|T_{pq}|}{\lambda}}\ket{0}\ket{0}\ket{p,q,\sigma}\ket{0}+\\
+\sum_l \sqrt{\frac{w_l}{\lambda}}\ket{l}\ket{+}\ket{+}\otimes\\ \otimes\sum_{p,q,r,s,\alpha,\beta}\sqrt{|g_{p,q}^{(l)}g_{r,s}^{(l)}|} \ket{0}\ket{0}\ket{p,q,\alpha}\ket{r,s,\beta}.
        \end{split}
    \end{equation} 
    In this step the Toffoli complexity is also $LN^2/4$ plus the cost of the compare and controlled swaps.

    \item Finally, use the QROM to output $\ket{\theta^{(l)}_{pq}}$ and $\ket{\theta^{(l)}_{rs}}$ in registers four and five.
     \begin{equation}
        \begin{split}
\ket{0}\ket{+}\ket{0}\sum_{p,q,\sigma}\sqrt{\frac{|T_{pq}|}{\lambda}}\ket{\theta^{(0)}_{pq}}\ket{0}\ket{p,q,\sigma}\ket{0}+\\
+\sum_l \sqrt{\frac{w_l}{\lambda}}\ket{l}\ket{+}\ket{+}\otimes\\
\otimes\sum_{p,q,r,s,\alpha,\beta}\sqrt{|g_{pq}^{(l)}g_{rs}^{(l)}|}\ket{\theta^{(l)}_{pq}}\ket{\theta^{(l)}_{rs}}\ket{p,q,\alpha}\ket{r,s,\beta}.
        \end{split}
    \end{equation}   
\end{enumerate}

To alleviate the cost of this procedure we follow three procedures:
\begin{enumerate}
    \item Leverage the $p\leftrightarrow q$ symmetry in $T_{pq}$ and $g^{(l)}_{pq}$, which divides the cost by half. This can be done preparing initially 
    \begin{equation}
        \sqrt{2}\sum_{p>q}\sqrt{|g^{(l)}_{pq}|}\ket{p,q,\alpha} + \sum_p\sqrt{|g^{(l)}_{pp}|}\ket{p,p,\alpha}.
    \end{equation}
    Then, one can use the second register, in state $\ket{+}$ to swap $\ket{p}$ and $\ket{q}$ when $p \neq q$ or to apply either $\mathbf{1}$ or $Z_{p,\sigma}$ when $p=q$. This means that in step 3 we will have to prepare $(L+1)(N^2/8+N/4)$ entries, and in step 4, $L(N^2/8+N/4)$.
    
    \item We can also reduce the preparation cost in the QROM by performing the comparison between the probability $\ket{\text{keep}_j}$ and an ancilla in uniform superposition, at the same time for all $l\in(0,..., L)$. The controlled swap between the register $\ket{j}$ and $\ket{\text{alt}_j}$ can also be performed for all values of $l$ simultaneously.
    
    \item The dominant cost is outputting $(2L+1)(N^2/8+N/4)$ qubits using the QROM \cite{babbush2018encoding}. The outputs will have a size $M = \lceil\log_2 N^2\rceil  + \lceil\log_2 \epsilon^{-1}_{QPE}\rceil + O(1)$ where $\lceil\log_2 N^2\rceil$ is the size of $\ket{\text{alt}}$ and $\mu = \lceil\log_2 \epsilon^{-1}_{QPE}\rceil + O(1)$ $\ket{\text{keep}}$, the size of the probability register. The key aspect of this third point is substituting the QROM of \cite{babbush2018encoding} by another from \cite{low2018trading} which allows to trade some gate complexity by space complexity. We will call it QROAM. Calling also $d =(2L+1)(N^2/8 + N/4)$ the number of entries we must look in the QROAM (including steps 3, 4 and 5), and $k = 2^n$ an arbitrarily chosen power of 2. Then the complexity of computing the QROAM is $\lceil d/k_c \rceil  + M(k_c-1)$ uncomputing it in Prepare$^\dagger$ is $\lceil d/k_u \rceil + k_u$, where the $k_c$ and $k_u$ in compute and uncompute respectively can be different.
    
    As an aside, we can indicate that if we were to use dirty ancillae (anciallae that is already being used for other purposes) the cost would be $2\lceil d/k\rceil + 4M(k-1)$ and $2\lceil d/k\rceil + 4k$ for compute and uncompute respectively. 
    
    Since the largest bottleneck is in the number of Toffolis required, we will focus on minimizing that variable. This means taking $k \approx \sqrt{d/M}$ for compute and $k\approx \sqrt{d}$ for the uncompute step, what means a cost of $2\sqrt{dM}$ and $2\sqrt{d}$ respectively, giving a total cost of $2\sqrt{d}(\sqrt{M} + 1)$. Since we have chosen $d\approx LN^2/8$ and $M \approx \lceil\log_2(N^2)\rceil + \mu$, this means an overall cost 
    $\sqrt{LN^2(\lceil\log_2(N^2)\rceil+\mu)/2}$ and half as many ancillae. Since $L = O(N)$, the number of Toffolis is $O(N^{3/2}\sqrt{\lceil \log_2 N \rceil + \mu})$.

\end{enumerate}

A technical detail is that since the QROAM requires a continous output register, we will compute a single continous register for $(l,p,q)$
\begin{equation}
    s' = l(N^2/8 +N/4)  +p(p+1)/2 +q
    \label{continous register mapping berry2019qubitization}
\end{equation}

The second operator we have to explain is Select, which is decomposed in two, Select$_1$ and Select$_2$ \cite{berry2019qubitization}, performed again similarly as is done in appendix \ref{app:QROM} \cite{babbush2018encoding}. The cost of this procedure is not dominant, as it will have complexity $O(N)$. From the representation of Select$_1$ in Figure 1 of \cite{berry2019qubitization}, we can see that we need two QROM applications, as well as 2 equality comparisons.

Apart from the implementation of Prepare and Select, some other minor costs to have in mind are
\begin{itemize}
    \item The cost in Select of each ranged operation is $N$, and each inequality test is $\lceil\log_2 N\rceil$. Since these operations have to be performed twice for $(p,q)$ and again twice for $(r,s)$, the total cost is $4N +4\lceil\log_2 N\rceil $.
    
    \item In the Prepare operator we have to initially prepare superpositions over $l\leq L$, $q\leq p <N/2$, $s\leq r <N/2$. We propose doing this by using the Uniform routine from the previous appendix (figure 12 in \cite{babbush2018encoding}). The initial uniform superposition over $l$ requires $8 \log_2 L  + 8\log_2 \epsilon^{-1}_{SS}$ T gates. Enforcing an uniform superposition (in the Subprepare method) where $p\geq q$ requires a different method. We will slightly modify the suggestion of \cite{berry2019qubitization} to control the number of Amplitude Amplification steps. We do this by implementing the Uniform protocol both for $p$ and $q$ independently, and then adding an ancilla to check whether $p\geq q$. The success probability will be $\frac{N^2/8 + N/4}{N^2/4}$ which approaches $1/2$ from above. Since we cannot straightforwardly amplify that we add a second ancilla with success amplitude $\frac{1}{2}\sqrt{\frac{N^2/4}{N^2/8 + N/4}}$. As a consequence, the product of success probabilities will be $1/2$ that corresponds to a Grover's $\theta = \pi/6$ which can be amplified to amplitude 1 with a single step. So this step requires 2 Uniform$_{N/2}$ procedures and 1 ancilla rotation, to be performed thrice: preparation and twice for Grover step.
    This second procedure has to be repeated twice to account for $r$ and $s$ too. 
    
    \item The inequality test in state preparation has cost $\mu$ Toffolis due to the $\mu$ bits in $\ket{\text{keep}}$; and the same number of gates as qubits needed in the swap gate. We have to perform swap gates in the preparation procedure in the QROM where the register $\ket{l,p,q}$ has size $\lceil\log_2 L\rceil + 2\lceil\log_2 (N/2)\rceil$. Then, we have to perform the same swap for $\ket{r,s}$ controlling on $l>0$, with registers of size $2\lceil\log_2 (N/2)\rceil$. This means a Toffoli cost $\mu + \lceil\log_2 L\rceil  + 4\lceil\log_2 N/2\rceil$. Here $\mu \approx \left\lceil \log \left( \frac{2\sqrt{2}\lambda}{\epsilon_{QPE}} \right) \right\rceil$.
    
    \item For state preparation remember that we only prepare those states that have $p>q$ and then use a controlled swap. These controlled swap for $(p,q)$ and $(r,s)$ cost $2\lceil \log_2 N/2 \rceil$ Toffolis.
    
    \item The arithmetic operations for computing \eqref{continous register mapping berry2019qubitization} require $2(\lceil \log_2 N/2 \rceil)^2$ Toffolis gates.

\end{itemize}

In any case, the leading cost of the model is $\sqrt{LN^2(\log(N^2)+\mu)/2}$ Toffoli gates due to the QROAM. We can further reduce the cost by increasing the sparsity of the $V$ operator, by zeroing all the terms $|V_{p,q,r,s}|<c$. Choosing $c$ should be done in a way that does not affect the final error $\Delta E$, as will be done choosing $L$ too. To do that, the main aspect is substituting the QROM indexing
\begin{equation}
    \frac{1}{\sqrt{d}}\sum_{j=1}^d \ket{j}\ket{\text{alt}_j}\ket{\text{keep}_j}
\end{equation}
by another
\begin{equation}
    \frac{1}{\sqrt{d}}\sum_{j=1}^d \ket{j}\ket{\text{ind}_j}\ket{\text{alt}_j}\ket{\text{keep}_j}
\end{equation}
where ind$_j$ indicates the $j$-th non-zero index, and $d$ the number of non-zero terms in each case. This means that the swapping must now be performed between ind$_j$ and alt$_j$. In this case we cannot simplify $\lceil d/k_c\rceil +(k_c-1)M+\lceil d/k_u\rceil +k_u$ with $M = \mu  + 2\log_2 N+2 \approx \log_2 (N^2) + \mu$, directly to $\sqrt{LN^2(\log(N^2) + \mu)/2}$. The $2$ in $M$ is because we have to choose between $T_{pq}$, $g^{(l)}_{pq}$ and $g^{(l)}_{rs}$.

\subsection{\label{app:Sparsity_low-rank_cost}How to compute its cost}

Notice that in contrast to other appendices, this calculation was already present in the original article \cite{berry2019qubitization} so the method to compute the cost is not our contribution. Only the automatisation of the computation is.
\begin{enumerate}
    \item Steps 1 and 2 in state preparation can be performed using a QROM for $L$ values and a multicontrolled-Hadamard gate respectively.

    \item The largest cost in each step is the use of the QROAM for steps 3, 4 and 5, that as we saw is $\left\lceil d/k_c \right\rceil + M(k_c-1) - \left\lceil d/k_c \right\rceil + k_c$ Toffolis. It takes into account both the Prepare and Prepare$^\dagger$ operators. We also use this step to prepare step 5, registers $\ket{\theta_{pq}}$ and $\ket{\theta_{rs}}$.
    \item Here 
    \begin{itemize}
        \item $d = (2L+1)(N^2/8 + N/4)$, as we take into account both steps 3 and 4 at the same time,
        \item $L$ is the rank of $W$. If $W$ were full rank, $L = N^2/4$, but since $W$ has a lot of structure $L = O(N)$.
        \item $k_c \approx \sqrt{d/M}$ (closest power of 2),
        \item $k_u \approx \sqrt{d}$ (closest power of 2),
        \item $M = \log_2 N^2  + \mu$,
        \item and $\mu\approx \left\lceil \log \left( \frac{2\sqrt{2}\lambda}{\epsilon_{QPE}} \right) \right\rceil$,
    \end{itemize}
    \item Each step requires to use Select once, at cost $4N + 4\lceil\log_2 N\rceil$.
    \item At each Prepare we have to use Uniform three times: for $l$ (accounted for in point 1 of this list), and two copies of $\tilde{s}$ for $(p,q)$ and $(r,s)$.
    \item Other minor contributions of the Subprepare circuit (see \cite{babbush2018encoding}) include a $\mu$-bit comparison and a $\log_2(LN^2/4)$-bit controlled swap.
    \item The calculation of \eqref{continous register mapping berry2019qubitization}, which is carried out for pairs $(p,q)$ and $(r,s)$ can be done from the value of $\tilde{s}$ with 2 multiplications and three multiplications.
    \item We need to perform amplitude amplification to prepare Uniform superposition over $p\geq q$ and $r\geq s$. This requires 6 Uniform$_{N/2}$ (for $p$ and $q$ and the three times of Amplitude Amplification), thrice an arbitrary rotation of the ancilla, thrice comparison between registers $\ket{p}$ and $\ket{q}$, and 2 Multi-controlled Z; and similarly for $r$ and $s$ respectively.
    
\end{enumerate}

\section{\label{app:Interaction_picture}Interaction picture}
\subsection{\label{app:Interaction_picture_method}Method explanation}

Although in previous appendices we have explored both the plane wave and Gaussian basis, there are two characteristics we have maintained constant over all the previous methods: all simulations were done in the Schrödinger picture and second quantization. These changes in later articles \cite{low2019hamiltonian,babbush2019quantum, su2021fault}, and in this appendix, we present how the interaction picture can help make more efficient Hamiltonian Simulation algorithms \cite{low2019hamiltonian}.

Let us recall that the Schrödinger picture time evolution is dictated by the solution to the Schrödinger equation
\begin{equation}
    \partial_t \ket{\psi(t)} = -iH(t)\ket{\psi(t)}
\end{equation}
what implies that
\begin{equation}
    \ket{\psi(t)}_S= e^{-iHt/\hbar}\ket{\psi(0)},
\end{equation}
whenever the Hamiltonian is time independent. We can see that in this case it is the state the one that evolves in time. 

On the other hand we have the Heisenberg picture, where the dynamics are included in the operators. As such we have
\begin{equation}
    \frac{d}{dt}A(t) =\frac{i}{h}[H, A(t)]+ \left(\frac{\partial A}{\partial t}\right)_{H}.
\end{equation}
If the Hamiltonian is time independent this becomes
\begin{equation}
    A(t)_H = e^{iHt/\hbar}A(0)e^{-iHt/\hbar}.
\end{equation}

An intermediate option is to choose the interaction or Dirac picture, where both the state and the operators become time dependent. In this case we divide the Hamiltonian in two parts $H_S = H_{S,0} + H_{S,1}$, where $H_{S,1}$ carries the complexity and time dependence of the Hamiltonian. Then, the quantum state will evolve as
\begin{equation}
    \ket{\psi}_I = e^{iH_{S,0}t/\hbar}\ket{\psi(0)}
\end{equation}
and the operators will evolve as
\begin{equation}
    A(t)_I = e^{iH_{S,0}t/\hbar} A(0)e^{-iH_{S,0}t/\hbar}.
\end{equation}
In particular
\begin{equation}
    H(t)_I = e^{iH_{S,0}t/\hbar} H_{S,1}e^{-iH_{S,0}t/\hbar}.
\end{equation}

\begin{comment}
This last equation can be seen from the Schrödinger equation
\begin{equation}
\begin{split}
    i\partial_t \ket{\psi(t)} = (H_{S,0} + H_{S,1})\ket{\psi(t)}\rightarrow\\
    i\partial_t \ket{\psi_I(t)} = e^{iH_{S,0}t/\hbar} H_{S,1}e^{-iH_{S,0}t/\hbar} \ket{\psi_I(t)}.
\end{split}
\end{equation}
\end{comment}

If the Hamiltonian is time-independent, we can evolve the state using $e^{-iHt}$, but if it is time-dependent, there is no closed expression in general. The time evolution operator is
\begin{equation}
    U(t) = \lim_{r\rightarrow \infty}\prod_{j = 1}^r e^{-iH(t(j-1)/r)\tau}:= \mathcal{T}e^{-i\int_0^t H(s)ds}.
\end{equation}

The authors of \cite{low2019hamiltonian} explore two topics. In the first place, they build a time-dependent Hamiltonian simulation algorithm that is based on synthesizing a Dyson series.
The second part of the article analyses how to apply the previous algorithm to simulate a Hamiltonian in the interaction picture. In particular, this allows us to simulate $e^{-i(H_{S,0} +H_{S,1})t}$ using
\begin{equation}
    O(\lambda_1 t \text{poly}\log ((\lambda_0+\lambda_1)t/\epsilon))
\end{equation}
queries to an oracle \begin{equation}
    (\bra{0}_a\otimes \mathbf{1}_s) O_1 (\ket{0}_a\otimes \mathbf{1}_s) = \frac{H_{S,1}}{\lambda_1},
    \label{qubitization oracle O_B}
\end{equation}
and a similar amount of $e^{-iH_{S,0}\tau}$ queries with $\tau = O(\lambda_1^{-1})$. Here we were taking $\lambda_0\geq ||H_{S,0}||$ and $\lambda_1\geq ||H_{S,1}||$. If we had used the Schrödinger picture we would have instead needed
\begin{equation}
    O((\lambda_0 + \lambda_1) t \text{poly}\log ((\lambda_0+\lambda_1)t/\epsilon_{HS}))
\end{equation}
queries to oracles $O_0$ and $O_1$ of the form of \eqref{qubitization oracle O_B}. If $||H_{S,0}||\gg ||H_{S,1}||$, and the complexity of applying $e^{-iH_{S,0}t}$ is similar to $O_1$, the interaction picture algorithm is advantageous.

Finally, the article applies the algorithm to the generalized Hubbard model and the electronic Hamiltonian, with a final complexity $\tilde{O}(N^2 t)$.

\begin{figure}[htp]
    \centering
    \includegraphics[width=.5\textwidth]{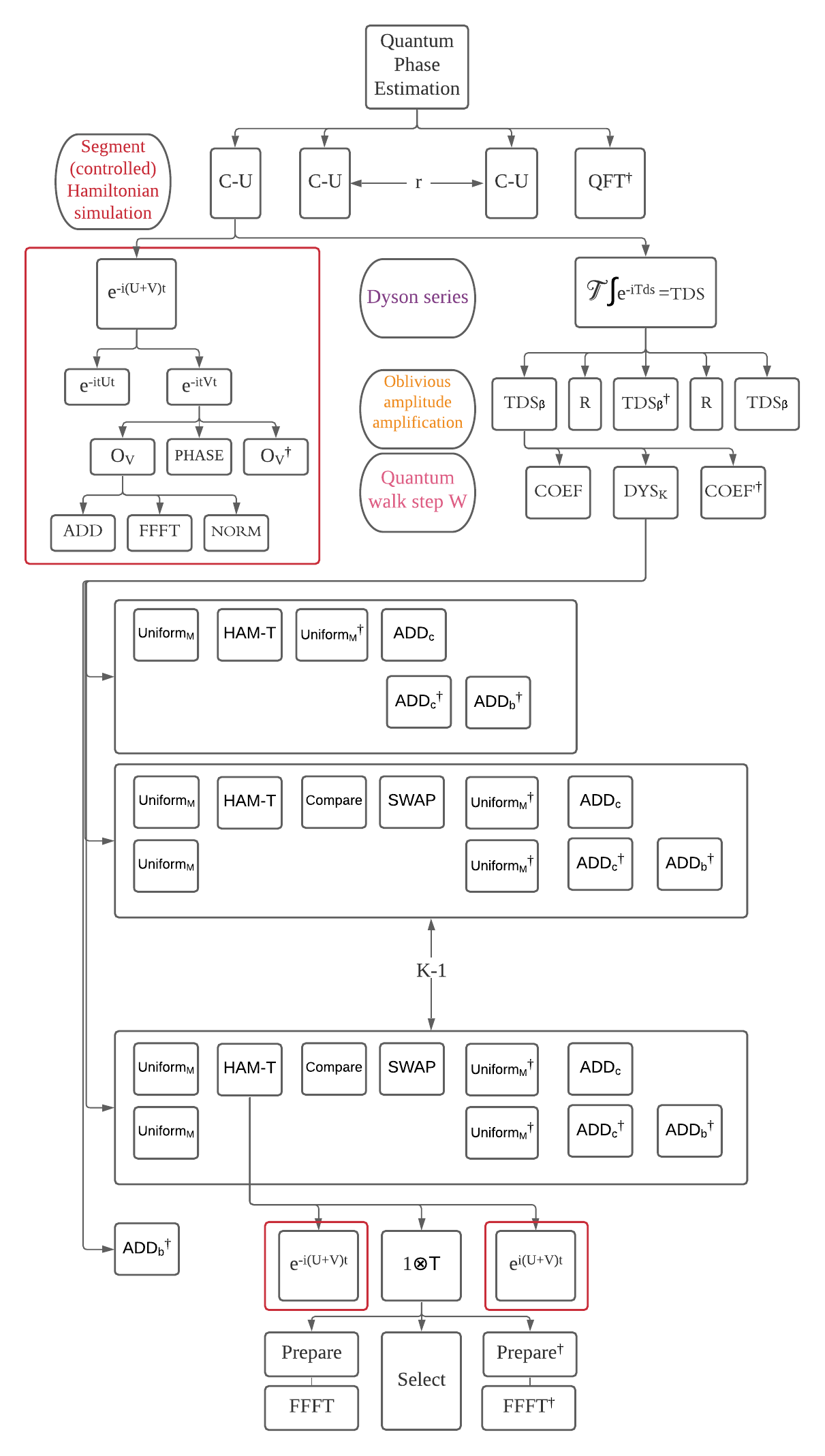}
    \caption{Abstraction level decomposition of the interaction picture protocol of \cite{low2019hamiltonian}. The boxes in red represent the same protocol, only decomposed for one of them.}
    \label{fig:Interaction_picture}
\end{figure}

In contrast with previous algorithms, we cannot approximate $U(t)$ with a Taylor series unless $[H(t), H(t')] = 0$. The alternative is the Dyson series that converges absolutely whenever $t>0$ and bounded $||H(t)||$:
\begin{equation}
\begin{split}
    U(t)&= \mathbf{1} - i \int_{0}^t H(t_1)dt_1 -\int_{t_2}^t\int_0^{t_2} H(t_2) H(t_1) dt_1 dt_2\\
    &+ i\int_{t_3}^t \int_{t_2}^{t_3}\int_{0}^{t_2} H(t_3)H(t_2)H(t_1)dt_1 dt_2 dt_3 + ...
\end{split}
\end{equation}
We can rewrite the previous expression using the time ordering operator
\begin{equation}
\begin{split}
    U(t) = \mathcal{T}[e^{-i\int_0^t H(s)ds}] = \sum_{k = 0}^\infty (-i)^k D_k.\\
    D_k = \frac{1}{k!}\int_{0}^t...\int_{0}^t\mathcal{T}[H(t_k)...H(t_1)]d^k t.
\end{split}
\end{equation}

As we did for the Taylor series, we have to truncate the series to order $K$ such that the error remains lower than target $\epsilon_{HS}$. We will see that $K$ will be logarithmic in the corresponding precision.

Let us now focus on the input model. We need two definitions. The first is the usual block encoding
\begin{equation}
    \text{HAM} = \begin{pmatrix}
H/\lambda & \cdot\\
\cdot & \cdot
\end{pmatrix}
\Rightarrow (\bra{0}_a\otimes \mathbf{1}_s)\text{HAM}(\ket{0}_a\otimes \mathbf{1}_s) = \frac{H}{\lambda}
\label{HAM}
\end{equation}
where we decompose HAM as in previous appendices
\begin{equation}
    HAM = (\text{Prepare}^\dagger\otimes \mathbf{1}_s)\text{Select}(\text{Prepare} \otimes \mathbf{1}_s)
\end{equation}
 
For a time dependent Hamiltonian we similarly define HAM-T as substituting $H$ in the matrix form of HAM in \eqref{HAM} with
\begin{equation}
\quad H = \text{Diagonal}[H(0),H(t/M),...,H(1-t/M)].
\end{equation}
In other words:
\begin{equation}
\begin{split}
    (\bra{0}_a\otimes \mathbf{1}_s)\text{HAM-T}(\ket{0}_a\otimes \mathbf{1}_s)\\
    =
    \sum_{l=0}^{M-1}\ket{m}\bra{m}\otimes \frac{H\left(\frac{mt}{M}\right)}{\lambda}.
\end{split}
\end{equation}

Having defined the main constructions for our algorithm, HAM and HAM-T, we now need the main theorem for simulating a time-dependent Hamiltonian for a short time segment:
\begin{Theorem} \cite{low2019hamiltonian}\label{Theorem 3 low_2019}
Let $H(s)$ be a time-dependent Hamiltonian such that $\max_s ||H(s)||\leq \lambda$ and $\braket{||\dot{H}||}$ the average value of its time derivative. Let $M\in O\left(\frac{t^2}{\epsilon_{HS}}(\braket{||\dot{H}||} + \max_s ||H(s)||^2)\right)$. 
Then, for all $t\in [0,\frac{1}{2\lambda}]$ and $\epsilon_{HS}>0$, exists $W$ such that $||W-\mathcal{T}[e^{-i\int_0^t H(s)ds}]||\leq \epsilon_{HS}$ with probability $1-O(\epsilon_{HS})$, and $K = O\left(\frac{\log \epsilon_{HS}^{-1}}{\log \log \epsilon_{HS}^{-1}}\right)$ queries to HAM-T.
\end{Theorem}

The proof is given in Appendix B \cite{low2019hamiltonian}. The key idea is that we want to approximate the time evolution operator with $W=$ TDS, the oblivious amplitude amplification of TDS$_\beta = \sum_{k = 0}^K \frac{(-it)^k}{M^k\beta} B_k$. As customary to require a single step of oblivious amplitude amplification, one takes $\beta = 2$.

The general strategy for TDS$_\beta$ is similar to the Prepare Select Prepare$^\dagger$ scheme.
For the Select operator we first construct a sequence of $K$ unitaries $U_1...U_K$ block-encoding matrices $H_1...H_K$:
\begin{equation}
    (\bra{0}_a\otimes \mathbf{1}_s)U_k(\ket{0}_a\otimes \mathbf{1}_s) = H_k; \quad ||H_k||\leq 1.
    \label{U_k block encoding H_k}
\end{equation}
The consecutive applications of such matrices, $H_k ... H_1\propto B_k$, the $k$-th term in the Dyson series. DYS$_K$ is the Select-like unitary that will apply this sequence $U_1...U_k$ controlled on index $\ket{k}$,
\begin{equation}
    (\bra{0}\otimes \bm{1})DYS_K (\ket{0}\otimes \bm{1}) = \sum_{k=0}^{K}\ket{k}\bra{k}\otimes \gamma_k B_K,
\end{equation}
where $\gamma_k = M^{-k}$ will be a weighting coefficient of the Dyson series. Constructing such $U_k$ operators is explained in the appendix B \cite{low2019hamiltonian}. 

The second ingredient needed is Prepare-like operators $COEF$ and $COEF'^\dagger$, the difference between them being the phase in the Dyson series term. This allows us to perform the TDS$_\beta$ operator (see fig. \ref{fig:Configuration_interaction}). Uniform$_M$, needed in the implementation of DYS$_K$ can be implemented as suggested in the main reference for this appendix \cite{babbush2018encoding}, while the rest are arithmetic operations, and the implementation of HAM-T discussed later on.

To extend Theorem \ref{Theorem 3 low_2019} to longer time periods one can just apply it multiple times with the corresponding scaled error, as given by Corollary 4 of \cite{low2019hamiltonian}. Since Theorem 1 indicates that the maximum simulation time for a single segment is $\tau = t/ \lceil 2\lambda t\rceil $ with $\max_s ||H(s)||\leq \lambda$, the number of segments is $r = \lceil 2\lambda t\rceil$, and the error allowed for each segment $\delta = \epsilon_{HS}/r$. Then Lemma 5 in \cite{low2019hamiltonian} states that the constants $K$ and $M$ for the simulation of a single segment to error $\delta$ are 
\begin{subequations}
\begin{equation}
    K= \left\lceil -1 + \frac{2\log(2r/\epsilon_{HS})}{\log\log(2r/\epsilon_{HS})+1}\right\rceil
    \label{K low2019}
\end{equation}
and
\begin{equation}
    M = \left\{\frac{16\tau^2}{\delta}(\braket{||\dot{H}||} + \max_s ||H(s)||^2), K^2\right\}.
\label{M low2019}
\end{equation}\label{K and M low2019}
\end{subequations}

The next step is to use this framework to simulate time-independent Hamiltonians in the interaction picture
\begin{equation}
    H_I(t)= e^{iH_{S,0}t}H_{S,1}e^{-iH_{S,0}t}.
    \label{H_I interaction Hamiltonian}
\end{equation}
The advantage of simulating in this frame will happen when the norm of $||H_I(t)|| = ||H_{S,1}||\ll ||H_{S,1}||+||H_{S,0}||$. We can apply this formalism to the Hamiltonian in the dual wave basis, equation \eqref{diagonal dual plane wave Hamiltonian}, where $H_{S,0} = U+V$ and $H_{S,1} = T$.

To simulate our time-independent Hamiltonian we use the Theorem \ref{Theorem 3 low_2019} and Corollary 4 in \cite{low2019hamiltonian}. 
As
\begin{equation}
\begin{split}
    \ket{\psi_S(t)} &= e^{-iH_{S,0}t}\ket{\psi_I(t)} \\ &=e^{-iH_{S,0}t}\mathcal{T}[e^{-i\int_0^t H_I(s)ds}]\ket{\psi(0)}
\end{split}
\end{equation}
we can divide the evolution in $r$ segments, $\tau = t/r$,
\begin{equation}
\begin{split}
    \ket{\psi(t)} &=(e^{-i(H_{S,0}+H_{S,1})\tau})^r\ket{\psi(0)}\\
    &= \left(e^{-iH_{S,0}\tau}\mathcal{T}[e^{-i\int_0^\tau H_I(s)ds}]\right)^r\ket{\psi(0)},
    \label{eq 19 low_2019}
\end{split}
\end{equation}
and simulate it in a similar way as suggested by Corollary 4 in \cite{low2019hamiltonian}. We have already explained ow to perform $\mathcal{T}[e^{-i\int_0^\tau H_I(s)ds}]$. However, we have yet to specify how implement HAM-T, which can be done as
\begin{equation}\label{HAM-T}
\begin{split}
    \text{HAM-T} = \left(\sum_{m=0}^{M-1}\ket{m}\bra{m}_d\otimes \mathbf{1}_a\otimes e^{iH_{S,0}\tau m/M}\right)\cdot \\
    \cdot \left(\mathbf{1}_d\otimes H_{S,1}\right)\cdot\left(\sum_{m=0}^{M-1}\ket{m}\bra{m}_d\otimes \mathbf{1}_a\otimes e^{-iH_{S,0}\tau m/M}\right).
\end{split}
\end{equation}

We also have to explain how to implement $e^{-i(U+V)t}$. Notice that $U$ and $V$ commute because they are diagonal in the dual wave basis and can therefore be fast-forwarded. However, to avoid the $O(N^2)$ cost in the $V$ term we will
\begin{enumerate}
    \item Define the Fourier transform of $V$ coefficients, $\tilde{V}(\vec{k}) = \sum_{\vec{x}}V(\vec{x}) e^{2\pi i \vec{x}\cdot \vec{k}/ N^{1/d}}$, and of the operators, $\tilde{\chi}_{\vec{k}} = \frac{1}{\sqrt{N}}\sum_{\vec{x}} e^{-2\pi i \vec{x}\cdot \vec{k} /N}\sum_{\sigma}n_{\vec{x},\sigma}$;
    and assuming $V$ is real and symmetric, equation 39 from \cite{low2019hamiltonian} writes
    \begin{equation}
    \begin{split}
        V &= \sum_{(\vec{x},\sigma)\neq (\vec{y},\sigma')}V(\vec{x} - \vec{y})n_{\vec{x},\sigma}n_{\vec{y},\sigma'}\\
        &=\sum_{\vec{k}}\tilde{V}(\vec{k})\tilde{\chi}_{\vec{k}}\tilde{\chi}_{\vec{k}}^\dagger  + \sum_{\vec{p},\sigma}\left(\sum_{\vec{k}}\tilde{V}(\vec{k})\right)n_{\vec{p},\sigma}.
    \end{split} 
    \end{equation}
    \item Use a binary oracle $O_A$ such that $O_A\ket{j}\ket{0}_o\ket{0}_{garb} = \ket{j}\ket{A_j}_o\ket{g(j)}_{garb}$. Then we can implement the phase operator
    \begin{equation}
    \begin{split}
        \ket{j}\ket{0}_o\ket{0}_{garb}\rightarrow_{\tilde{O}_A}\ket{j}\ket{A_j}_o\ket{g(j)}_{garb}\ket{0}\rightarrow_{PHASE}\\
        e^{-iA_jt}\ket{j}\ket{A_j}_o\ket{g(j)}_{garb}\ket{0}\rightarrow_{\tilde{O}_A^\dagger}
        e^{-iA_jt}\ket{j}\ket{0}_o\ket{0}_{garb}
    \end{split}
    \end{equation}
\end{enumerate}
We want to implement the oracle $O_V$ to calculate the Fourier Transform of $V$ (omitting garbage registers), so this oracle can be decomposed as
\begin{equation}
    \begin{split}
        \left(\bigotimes_{x,\sigma}\ket{n_{x,\sigma}}\right)\ket{0}\rightarrow_{ADD} \bigotimes_x \ket{\sum_\sigma n_{x,\sigma}}\rightarrow_{FFT}\bigotimes_k \ket{\tilde{\chi}_k}\\
        \rightarrow_{|\cdot|^2} \bigotimes_k \ket{|\tilde{\chi}_k|^2}\rightarrow_{\times V_k}\bigotimes_{k}\ket{V_k|\tilde{\chi}_k|^2}.
    \end{split}
\end{equation}
Notice that $\ket{n_x}$ indicates the occupancy of the corresponding orbital, and as we are working with fermions, the FFT is the Fermionic Fast Fourier Transform.

\subsection{\label{app:Interaction_picture_cost}How to compute its cost}

To be able to count the complexity of the circuit it is useful to first indicate the size of each of the registers that appear in the algorithm, and more in particular in the analysis of the TDS operator in appendix B \cite{low2019hamiltonian}: 
\begin{enumerate}
    \item Register $s$ is the register containing the state. In second quantization it has size $N$.
    \item Register $a$ has a size given by the block encoding. Using \cite{low2018hamiltonian} this can be bound by the logarithm of the number of unitary terms in Hamiltonian that are summed, $n_a = \lceil \log_2 \Gamma\rceil$.
    \item Register $b$ has $n_b = \log_2 (K+1)$ qubits.
    \item Register $c$ has $n_c = 1 + \log_2 (K+1)$ qubits.
    \item Registers $d$ and $e$ require $\log_2 M$ qubits.
    \item Register $f$ only requires $1$ qubit.
\end{enumerate}

Secondly, we have to specify the value of $K$ and $M$ in \eqref{K and M low2019}. For $K$ we already mentioned that $\delta = \epsilon_{HS}/r$, while in the usual definition of $r$ we will take $\lambda = \lambda_1 = ||H_{S,1}|| = ||T||$. Instead of taking $\tau = \frac{1}{2\lambda_1}$ \cite{low2019hamiltonian}, we may take it slightly higher, $\tau = \frac{\ln 2}{\lambda_1}$ as this limit comes from the oblivious amplitude amplification technique \cite{berry2015simulating}, and we will do so to carry out similar treatment between the algorithms. Then, since $t = \frac{\pi}{\epsilon_{QPE}}$ this implies that $r:= t/\tau= \lceil||T|| t\rceil = \left\lceil \frac{ \pi ||T|| }{\epsilon_{QPE}\ln 2}\right\rceil$.

Additionally, we need to obtain the value of $M$. Since $\max_s ||H_I(s)||\leq ||H_{S,1}||$ and $\braket{||\dot{H}||} = ||[H_{S,0},H_{S,1}]||\leq 2||H_{S,0}||\cdot ||H_{S,1}||$, substituting $ \tau = \ln 2/\lambda_1$ and $ \delta = \epsilon_{HS}/r = \epsilon_{HS}t/\tau$ in the value of $M$
\begin{equation}
\begin{split} \label{M value low2019hamiltonian}
    M &= \max\biggl\{\frac{16\tau^2}{\delta}(\braket{||\dot{H}||} + \max_s ||H(s)||^2), K^2\biggr\}\\
    & =  \max\biggl\{\frac{16t\ln 2 }{\lambda_1 \epsilon_{HS}} (2||H_{S,1}||||H_{S,0}|| + ||H_{S,1}||^2 ), K^2\biggr\} \\
    & = \max\biggl\{\frac{16t\ln 2 }{ \epsilon_{HS}}(2||H_{S,0}|| + ||H_{S,1}||), K^2\biggr\}.
\end{split}
\end{equation}

To finish giving a description of the algorithm we need to particularize HAM-T for the time-independent Hamiltonian that we want to use, as given in Lemma 7 in \cite{low2019hamiltonian}
\begin{equation}
\begin{split}
    &\text{HAM-T}= \left(\sum_{m=0}^{M-1}\ket{m}\bra{m}\otimes \mathbf{1}_a\otimes e^{i(U+V)\tau m/M}\right)\cdot\\
    &\cdot\left(\mathbf{1}\otimes O_T\right)\cdot\left(\sum_{m=0}^{M-1}\ket{m}\bra{m}\otimes \mathbf{1}_a\otimes e^{-i(U+V)\tau m/M}\right).
\end{split}
\end{equation}
Here, 
$$\left(\sum_{m=0}^{M-1}\ket{m}\bra{m}\otimes \mathbf{1}_a\otimes e^{i(U+V)\tau m/M}\right)$$
can be implemented with $\lceil \log_2 M \rceil$ controlled-rotations of the kind
$e^{i(U+V)\tau/M}$, $e^{i(U+V)\tau 2/M}$, $e^{i(U+V)\tau 4/M}$...

Lastly, performing $O_T$ can be done using
$O_T = (\text{Prepare}_T^\dagger \otimes \text{FFFT}^\dagger)\text{Select}_T(\text{Prepare}_T\otimes \text{FFFT})$, where 
\begin{subequations}
\begin{equation}
    \text{Prepare}_T\ket{0_a}=\sum_p 
\sqrt{\frac{\tilde{T}(p)}{\lambda_T}}\ket{\vec{p}},
\end{equation}
\begin{equation}
    \text{Select}_T = \sum_p \ket{\vec{p}}\bra{\vec{p}}\otimes n_{p},
\end{equation}
\end{subequations}
and FFFT applied using $O(N\log N)$ gates, as we already discussed in appendix \ref{app:Plane_wave}.

Finally, let us highlight that there is a way to avoid the extra cost posed by Amplitude Amplification. The key idea is to implement a block encoding of $\sin (H \tau )= \frac{e^{+iH\tau}-e^{-iH\tau}}{2i}$, so that the qubitization walk operator will implement $e^{-i\arcsin \mu_j /\lambda'} = e^{-i\arcsin (\sin E_j\tau) /\exp (\lambda_1 \tau)} \approx e^{-i E_j \tau /\exp(\lambda_1\tau)}$ \cite{su2021fault}, for $\mu_j = \sin E_j\tau$ and $\lambda' = \exp (\lambda_1 \tau)$ \cite{kieferova2019simulating}. If the segment time length of the amplitude amplified algorithm is $\tau \approx \ln 2/\lambda_1$, then the adjusted segment length is $\tau_{\text{eff}} \approx \ln 2/2\lambda_1$, and if one increases $\tau\approx 1/\lambda_1$, then $\tau_{\text{eff}}\approx 1/(e\lambda_1)$. This means that each step of the algorithm does not need to be amplified, but the number of time segments increases from $\lambda_1/(\epsilon_{QPE}\ln 2)$ to $e\lambda_1/\epsilon_{QPE}$ \cite{su2021fault}. In this case we also aim to perform Hamiltonian simulation over $H-\tilde{E}_0$, where $\tilde{E}_0$ is an approximation to the ground state energy, so that we operate on the linear regime of the sine and arcsine functions, and the error introduced is small.

\subsection{\label{app:Interaction_picture_adaptation}How to adapt the Hamiltonian simulation to control the direction of the time evolution.}

We will use the trick of setting the controlled evolution of phase estimation $\ket{1}\ket{\phi}\rightarrow e^{i\phi}\ket{1}\ket{\phi}$ and $\ket{0}\ket{\phi}\rightarrow e^{-i\phi}\ket{0}\ket{\phi}$. This will be reflected in the Dyson expansion, where we will have to add a $i$ factor to the coefficients in $COEF$ conditional on the phase estimation ancilla being on state $1$; and in the sign of the exponential $e^{-i(U+V)\tau}$. 

\end{document}